\documentclass[preprint]{aastex}

\tighten

\newcommand{\asec}{\hbox to 1pt{}\rlap{$^{\prime\prime}$}.\hbox to 2pt{}}
\newcommand{\amin}{\hbox to 1pt{}\rlap{$^{\prime}$}.\hbox to 2pt{}}
\newcommand{\myr}{{\rm Myr}}
\newcommand{\mpc}{{\rm Mpc}}
\newcommand{\kms}{{\rm km~s^{-1}}}
\newcommand{\Msun}{M_\odot}
\newcommand{\Lsun}{L_\odot}
\newcommand{\Ser}{S\' ersic\ }

\slugcomment{Submitted to {\it The Astrophysical Journal}}
\shortauthors{Lauer et al.}
\shorttitle{The Most Massive Black Holes}

\begin{document}

\title{The Masses of Nuclear Black Holes in Luminous Elliptical Galaxies
and Implications for the Space Density of the Most Massive Black Holes. 
\footnote{Based on observations made with the NASA/ESA 
{\it Hubble Space Telescope}, obtained at the Space Telescope Science Institute,
which is operated by the Association of Universities for
Research in Astronomy, Inc., under NASA contract NAS 5-26555. These
observations are associated with GO and GTO proposals
\# 5236, 5446, 5454, 5512, 5943, 5990, 5999, 6099, 6386, 6554, 6587, 6633,
7468, 8683, and 9107.}}

\author{Tod R. Lauer}
\affil{National Optical Astronomy Observatory\footnote{The National Optical
Astronomy Observatory is operated by AURA, Inc., under cooperative agreement
with the National Science Foundation.},
P.O. Box 26732, Tucson, AZ 85726}

\author{S. M. Faber}
\affil{UCO/Lick Observatory, Board of Studies in Astronomy and
Astrophysics, University of California, Santa Cruz, CA 95064}

\author{Douglas Richstone}
\affil{Department of Astronomy, University of Michigan, Ann Arbor, MI 48109}

\author{Karl Gebhardt}
\affil{Department of Astronomy, University of Texas, Austin, TX 78712}

\author{Scott Tremaine}
\affil{Princeton University Observatory, Peyton Hall, Princeton, NJ 08544}

\author{Marc Postman}
\affil{Space Telescope Science Institute, 3700 San Martin Drive, Baltimore, MD
21218}

\author{Alan Dressler}
\affil{The Observatories of the Carnegie Institution of Washington,
Pasadena, CA 91101}

\author{M. C. Aller}
\affil{Department of Astronomy, University of Michigan, Ann Arbor, MI 48109}

\author{Alexei V. Filippenko}
\affil{Department of Astronomy, University of California, Berkeley,
CA 94720-3411}

\author{Richard Green}
\affil{LBT Observatory, University of Arizona, Tucson, AZ 85721}

\author{Luis C. Ho}
\affil{The Observatories of the Carnegie Institution of Washington,
Pasadena, CA 91101}

\author{John Kormendy}
\affil{Department of Astronomy, University of Texas, Austin, TX 78712}

\author{John Magorrian}
\affil{Department of Physics, University of Durham, Durham, United Kingdom,
DH1 3LE}

\author{Jason Pinkney}
\affil{Department of Physics and Astronomy, Ohio Northern University,
Ada, OH 45810}

\vfill

\begin{abstract}

Black hole masses predicted from the $M_\bullet-\sigma$
relationship conflict with those predicted from the
$M_\bullet-L$ relationship for the most luminous galaxies,
such as brightest cluster galaxies (BCGs).
This is because stellar velocity dispersion, $\sigma,$ increases
only weakly with luminosity for BCGs and other giant ellipticals.
The $M_\bullet-L$ relationship predicts
that the most luminous BCGs may harbor black holes with $M_\bullet$
approaching $10^{10}M_\odot,$ while the $M_\bullet-\sigma$ relationship
always predicts $M_\bullet<3\times10^9M_\odot.$
Lacking direct determination of $M_\bullet$ in a sample of the
most luminous galaxies, we advance arguments that
the $M_\bullet-L$ relationship is a plausible or
even preferred description for BCGs and other galaxies of similar luminosity.
Under the hypothesis that cores in central stellar density
are formed by binary black holes, the inner-core cusp radius,
$r_\gamma,$ may be an independent witness of $M_\bullet.$
Using central structural parameters derived from a large sample
of early-type galaxies observed by {\it HST},
we argue that $L$ is superior to $\sigma$ as an indicator of
$r_\gamma$ in luminous galaxies.  Further, the observed  $r_\gamma-M_\bullet$
relationship for 11 core galaxies with measured $M_\bullet$ appears to be
consistent with the $M_\bullet-L$ relationship for BCGs.
BCGs have large cores appropriate for their large
luminosities that may be difficult to generate with the more modest
black hole masses inferred from the $M_\bullet-\sigma$ relationship.
$M_\bullet\sim L$ may be expected to hold for BCGs,
if they were formed in dissipationless mergers, which should preserve
ratio of black hole to stellar mass.
This picture appears to be consistent
with the slow increase in $\sigma$ with $L$ and the more rapid
increase in effective radii, $R_e,$ with $L$ seen
in BCGs as compared to less luminous galaxies.
If BCGs have large BHs commensurate with their high luminosities,
then the local black hole mass function
for $M_\bullet>3\times10^9M_\odot$ may be nearly an order of magnitude
richer than what would be inferred from the $M_\bullet-\sigma$ relationship.
The volume density of the most luminous QSOs at earlier epochs
may favor the predictions from the $M_\bullet-L$ relationship.

\end{abstract}


\section{The Most Luminous Galaxies $\Longleftrightarrow$ The Most Massive
Black Holes} 

Nearly every elliptical galaxy and spiral bulge
has a black hole at its center \citep{mag}.
The masses of the black holes, $M_\bullet,$ are
related to the $V$-band luminosity, $L,$ and average
stellar velocity dispersion, $\sigma,$
of their host galaxies \citep{d89, k93, kr, mag, fm, g00, tr02, hr}.
The $M_\bullet-\sigma$ and $M_\bullet-L$ relationships are powerful tools
as they allow the prediction of black hole masses --- which are
difficult to measure directly --- from readily available galaxy
parameters.

The black hole population in the most massive galaxies has yet to be assayed,
however, which means that estimates of $M_\bullet$ in these objects are
based on extrapolations of relationships defined by smaller galaxies.
The current record for largest black hole mass measured directly is
$M_\bullet\sim3\times10^9M_\odot$ in M87 \citep{m87},
yet M87 is only the {\it second}-ranked galaxy in a cluster of modest richness.
Brightest cluster galaxies (BCGs) in nearby Abell clusters
are typically $\sim3\times$ more luminous \citep{pl} and may host
proportionately more massive BHs.  Testing this hypothesis through
measurements of stellar dynamics requires both high sensitivity
and high spatial-resolution, given the low central surface brightnesses and
relatively large distances of BCGs.
Such observations were not possible with the {\it Hubble Space Telescope}
({\it HST}) even before the failure of the {\it Space Telescope
Imaging Spectrograph;}
they are only now becoming
feasible with the advent of adaptive optics spectroscopy on 10m class
telescopes.

A number of arguments suggest that black holes with
$M_\bullet>3\times10^9M_\odot$ do exist, even if this conclusion
is not universal (e.g. \citealt{mclure}).
\citet{net} argues that
some QSOs have $M_\bullet>10^{10}M_\odot$ based on an empirical
relationship between $M_\bullet,$ broad-line width and nuclear
luminosity for AGN.  \citet{bechtold} and \citet{vester1} also argue that
some QSOs have black holes approaching this mass.
Of particular relevance for BCGs is the hypothesis
that cluster cooling flows are inhibited by AGN heating from the
central galaxy \citep{bt, chur}.  Recent {\it Chandra} observations
support a picture in which episodic AGN outbursts in BCGs heat
the intra-cluster medium \citep{vd};
the energetics required to terminate cooling flows
imply $M_\bullet>10^{10}M_\odot$ for many clusters \citep{fab}.

Arguments for such massive black holes
appear to be in conflict, however, with the expectations
from the $M_\bullet-\sigma$ relationship applied to the local galaxy
velocity-dispersion distribution function.
\citet{tr02} find
\begin{equation}
\log (M_\bullet/M_\odot)=(4.02\pm0.32)\log(\sigma/200
{\rm\ km\ s^{-1}})+8.19\pm0.06,
\label{eqn:msig}
\end{equation}
for $H_0=70~{\rm km~s^{-1}Mpc^{-1}}$ (which we will use throughout this paper).
The \citet{sheth} local velocity dispersion function shows a strong cut-off
at $\sigma\approx400{\rm\ km~s^{-1}},$ which implies that
galaxies harboring black holes with $M_\bullet>3\times10^9M_\odot$
would be extremely rare.  \citet{bern3} have identified
a handful of galaxies with $\sigma>400{\rm\ km~s^{-1}},$
but their results do not alter this conclusion.

Extrapolation of the $M_\bullet-\sigma$
relationship to galaxies more massive than M87 assumes
that $\sigma$ (and not galaxy mass) is the fundamental
parameter for determining $M_\bullet.$
The uncertainty in such an extrapolation is
underscored by \citet{wyithe}, who argues that the $M_\bullet-\sigma$
relationship is curved rather than linear in log-log space,
in the sense that, at the high-$\sigma$ end, the
``log-quadratic'' relationship predicts higher $M_\bullet$
than does equation (\ref{eqn:msig}). 
The Wyithe $M_\bullet-\sigma$ relationship, implies that the
space density of black holes with $M_\bullet>5\times10^9M_\odot$
may be substantially higher than that implied
by equation (\ref{eqn:msig}) (although the exact difference is highly
sensitive to both the details of the velocity dispersion
distribution function, and the assumed level of cosmic 
scatter in the $M_\bullet-\sigma$ relationship).

In this paper we point out that the $M_\bullet-L$ relationship
applied to the most luminous galaxies predicts
$M_\bullet$ values that are significantly larger than those predicted
by either the \citet{tr02} or \citet{wyithe} $M_\bullet-\sigma$ relationships.
This difference arises because BCGs do not follow the \citet{fj} relationship
between $L$ and $\sigma.$
The relationship between $L$ and $\sigma$
``plateaus'' at large $L$ in the sense that BCGs have relatively low $\sigma$
for their high $L$ 
(\citealt{ho}; see also \citealt{bmq}, who have seen this effect
in simulations.)

Resolution of which of the $M_\bullet-L$ or $M_\bullet-\sigma$ relationships
is most representative of the black hole population in the most massive
galaxies will only be possible when black hole masses can be
measured in such galaxies.  In advance of such work, however,
we can advance a number of arguments that suggest that the $M_\bullet-L$
is a plausible and perhaps even preferred description for such systems.

The first set of arguments are based on the central structure of
BCGs and other luminous elliptical galaxies that
have cores in their central brightness profiles \citep{l95, laine}.
A core is evident as a radius at which the steep envelope of the
galaxy ``breaks'' and transitions to an inner cusp with a shallow slope
in logarithmic coordinates.
The favored theory for core formation
posits that cores are formed when stars are ejected
from the galaxy's center by the decay of a binary BH created in a merger
\citep{bbr, ebi, f97, qh, mnm}.
The size of the core then reflects the total mass ejected, which should
be a function of $M_\bullet.$  The size of the core may thus be
an independent witness of $M_\bullet.$  In BCGs and other galaxies
of similar luminosity, galaxy luminosity is more closely related
to the physical scale of the cores than $\sigma,$ and the
observed core size $M_\bullet$ relationship for galaxies with
cores and directly measured black hole masses appears to
be consistent with the $M_\bullet-L$ relationship.

A second set of arguments come from considering the formation
of BCGs.  If BCGs are formed in ``dry'' mergers, then the ratio
of black hole to stellar mass should be preserved over mergers,
leading to the observed $M_\bullet-L$ relationship.  In contrast,
$\sigma$ may change little over such mergers, and no longer track
black hole mass as well it does for the less luminous galaxies
from which the $M_\bullet-\sigma$ has been determined.

Lastly, we consider the relative predictions of the
$M_\bullet-L$ and $M_\bullet-\sigma$ relationships for the
volume mass distribution function of black holes and which we compare
to the predictions from QSO luminosity functions.  A decisive
discrimination between the two relationships is not possible
without a better understanding of the cosmic scatter in both
relationships, but the \citet{tr02} version of the $M_\bullet-L$
relationship probably predicts too few high mass black holes to support the
QSO luminosity function.

\section{A Large Sample of Early Type Galaxies With Central Structure
Characterized by {\it HST}}

We start by comparing the two predictions $M_\bullet(\sigma)$
($M_\bullet$ predicted from the $M_\bullet-\sigma$ relationship) and
$M_\bullet(L)$ ($M_\bullet$ predicted from the $M_\bullet-L$ relationship)
for a sample of 219 galaxies for which we have central structural parameters
derived from {\it HST} imagery \citep{l06}.  We then present
the separate relationships between
core structure versus $\sigma$ and $L.$
This leads in turn to two separate predictions for how core size
should be related to $M_\bullet,$ which can be compared to the observed
relationship between core size and $M_\bullet$ for 11 core
galaxies that have direct $M_\bullet$ determinations.

The galaxy sample combines several different
{\it HST} imaging programs that all used
the Nuker-law parameterization \citep{l95} to
characterize the central starlight distributions.
The properties and definition of this sample are presented in detail
in \citet{l06}, but briefly, we combine surface photometry presented in
\citet{l95}, \citet{f97}, \citet{laine}, \citet{rest}, \citet{rav},
\citet{quil}, and \citet{l05}.
This diverse source material has been transferred to a common photometric
system ($V-$band) and a common distance scale, adopting
$H_0=70$ km s$^{-1}$ Mpc$^{-1}.$
The primary source of distances is the SBF survey of \citet{ton},
but when possible we use the
group memberships in \citet{f89} and average SBF distances
over the group.  As the \cite{ton} SBF scale is consistent with
$H_0=74,$ we scale up their SBF distances by 6\%.
The treatment of galaxies not in the SBF survey is detailed in \citet{l06}.
The sample is listed in Table \ref{tab:glob}.
It comprises 120 core galaxies, 87 power-law galaxies, and 12 intermediate
galaxies.

The most important Nuker-law parameter for the present analysis
is the break radius, $r_b,$ which is used to calculate the cusp radius,
$r_\gamma,$ which in turn is used to represent the physical
scale of the core (this parameter is discussed in detail in
$\S\ref{sec:rgam1}$ and Appendix \ref{sec:rgam2}).
The average error in $r_\gamma$ is 30\%, based on comparison
of Nuker parameters to non-parametric estimates of the same parameters.

Central velocity dispersions are provided by the ``Hyperleda''
augmentation of the \citet{ps} compendium
of published velocity dispersions;
no values were available for 30 of the total of 219 galaxies.
We adopt a 10\%\ typical error in $\sigma.$
The $M_\bullet-\sigma$ relationship as initially presented
by \citet{g00} used the average luminosity-weighted velocity dispersion
measured in a slit along the major axis interior to the effective radius.
Velocity dispersion profiles are unfortunately not available for the
bulk of the galaxies; however, \citet{g00} showed that the central values
are likely to be within 5\%\ of the radial averages.

\subsection{Galaxy Luminosities\label{sec:gallum}}

The sources of the present galaxy luminosities are discussed in detail in
\cite{l06}.  Most of the magnitudes
are derived from $V_T$ or $B_T$ values drawn from the RC3 \citep{rc3}.
Bulge luminosities are given
for S0 and spiral galaxies based on bulge/disk decompositions in the literature.
Absolute luminosities were calculated using the \citet{sfd} Galactic
extinction values; we assume a typical $M_V$ error of 10\%.

The accuracy of the BCG luminosities is of special concern
as we will argue that they
imply higher $M_\bullet$ than would be inferred from the $\sigma$
values for the same galaxies.  The present BCG luminosities
are based on fitting $r^{1/4}$ laws to the inner portions ($r<50$ kpc)
of the R-band \citet{pl} brightness profiles,
limiting the fits to radii that are well matched by this function.
\citet{graham} show that BCG brightness profiles are better described
by \Ser profiles with \Ser $n>4,$  which is also true of
giant elliptical galaxies in general (e.q. \citealt{f06,k06}).
However, BCGs with \Ser $n>4,$ typically have
extremely large effective radii that are factors of several larger
than the actual radial limit of the surface photometry;
this in turn implies unrealistically large total luminosities.
The $r^{1/4}$ laws give a conservative lower limit
for BCG total luminosities.
Even so, the derived luminosities are systematically much larger than
those provided by the Sloan Digital Sky Survey (SDSS).
We resolve this issue in Appendix \ref{sec:bcg} with a demonstration that
the SDSS BCG luminosities are strongly biased to low values
by excessive sky subtraction.
The NIR apparent magnitudes provided by the 2MASS Extended Source
Catalogue \citep{xsc, lga}
have also been used to provide BCG total luminosities \citep{bat};
however in Appendix \ref{sec:2mass} we show that the 2MASS apparent
magnitudes are also likely to be underestimates.

A separate issue raised by a number of our colleagues is that BCG luminosities
may need to be ``corrected'' for intracluster light (ICL).
One such treatment of ICL assumes that the BCG is coincident with the center
of the cluster potential, and that the composite BCG+ICL can be
modeled as two superimposed $r^{1/4}$ laws (cf. \citealt{gzz}).
The ICL component is then subtracted to yield the ``true''  BCG luminosity.
A key feature of such models is that the ICL profile is assumed
to continue to rise in brightness at radii well interior to where it dominates,
thus implying a substantial contribution at even small radii.
There is little physical justification for a correction of this form, however.
As noted above, giant elliptical galaxies in general (not just BCGs)
have \Ser $n>4.$  Further, the presumption that BCGs sit exactly at the
center of their clusters is an idealization that is actually realized
in only a small fraction of systems.
\citet{pl} show that BCGs are typically displaced
from the geometric cluster center by $\sim90$ kpc in projection and
$\sim260~{\rm km~s^{-1}}$ in velocity.  \citet{patel} showed that
BCGs are typically displaced from the centroid of cluster X-ray emission
by 129 kpc, consistent with the \citet{pl} analysis.
Lastly, the presumption that the
ICL follows an $r^{1/4}$ law into small radii is not uniquely demanded,
and is probably inconsistent with the large velocity dispersion
of stars truly not bound to the BCG.  Again, BCGs are well described
over a large radial range by \Ser laws; in no case in the \citet{graham}
sample are there any profiles that have a distinct feature that objectively
supports a two component model.  This is not to say that ICL is not present,
but the surface brightness at which it dominates even in the two component
models are well outside the radii
at which we measure the $r^{1/4}$ laws
used to estimate total luminosity (typically less than 50 kpc).
The \citet{zib} models of ICL show that it begins to dominate the BCGs
at $r\sim80$ kpc from the BCG centers, corresponding to $\mu_r\sim26.$
We conclude that a strong correction to our BCG luminosities
for ICL is poorly justified.

\section{A Contradiction Between the $M_\bullet-\sigma$ and
$M_\bullet-L$ Relationships\label{sec:mpred}}

The $M_\bullet-L$ relationship emerged from the first attempts
to relate black hole mass to properties of the host galaxy \citep{d89, k93, kr}.
Much of the recent work on this problem, however, has focused on the
$M_\bullet-\sigma$ relationship due to its apparent smaller scatter
(although see \citealt{novak} on the significance of this),
as well as arguments that $\sigma,$ rather than galaxy luminosity
is the more fundamental parameter that determines how galaxies were formed
(e.g., \citealt{wl}).
While $L$ and $\sigma$ are related by the \citet{fj}
relationship, since the discovery that galaxies lie on
a ``fundamental-plane'' determined by $L,$ $\sigma,$ and the
effective radius, $R_e,$ \citep{dd, d87}, we know that
neither $L$ nor $\sigma$ alone is sufficient to codify
the full range of galaxy properties.
The $M_\bullet-L$ relationship thus may contain information
that is not a trivial projection of the $M_\bullet-\sigma$ relationship.

\begin{figure}
\plotone{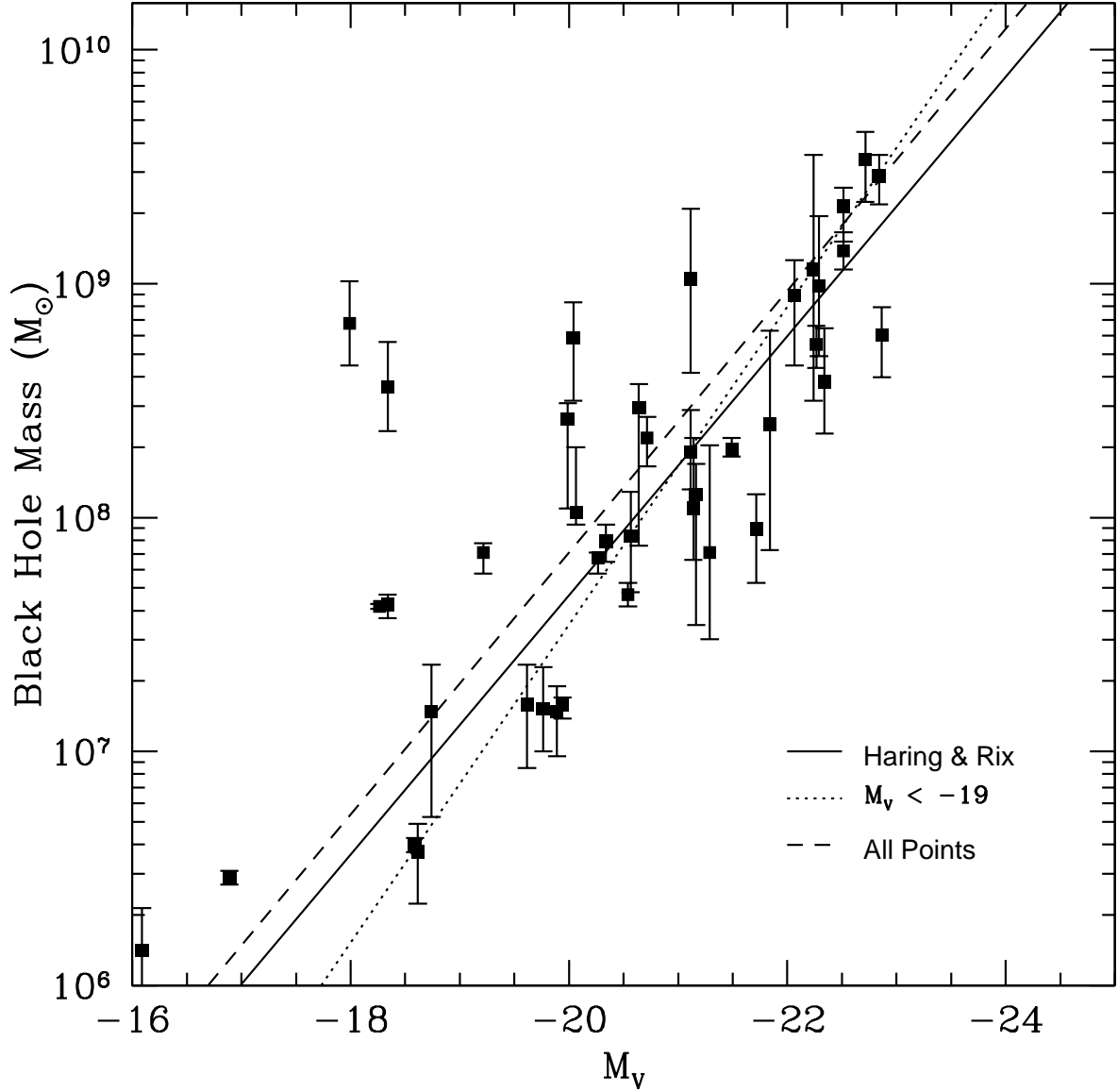}
\caption{Black hole masses for all black holes with direct mass determinations
are plotted as a function of $M_V.$  Galaxies are drawn from the \citet{tr02}
sample with augmentations as described in the text.
The solid line is the \citet{hr} relationship (equation (\ref{eqn:ml_hr}))
between $M_\bullet$ and galaxy mass transformed to luminosity
using $M/L_V\sim L_V^{0.23}$ with zeropoint $M/L_V=6$ at $M_V=-22.$
A symmetrical least-squares fit to all data points is shown as the dashed line
(equation \ref{eqn:ml_all}), and a fit to just the galaxies with $M_V<-19$
is shown as the dotted line (equation \ref{eqn:ml_bright}).}
\label{fig:ml}
\end{figure}

The relationship between $M_\bullet$ and $L$ is shown
in Figure \ref{fig:ml}.  Most of the galaxies shown are those presented
in \citet{tr02},\footnote{We augment the \citet{tr02} sample with $M_\bullet$
determinations in NGC 1399 \citep{n1399}, NGC 3031 \citep{n3031},
NGC 3998 \citep{n3031}, NGC 4374 \citep{n4374}, NGC 4486B \citep{n4486b},
NGC 4945 \citep{n4945}, NGC 5128 \citep{n5128}, NGC 7332 \citep{n7332},
and Cygnus A \citep{cyga}.} transformed to $H_0=70~{\rm km~s^{-1}~Mpc^{-1}}.$
Due to the large scatter of the data points in Figure \ref{fig:ml},
estimating a mean $M_\bullet-L$ relationship is likely to be sensitive
to the fitting algorithm.
We have elected to use the ``symmetric'' least-squares algorithm of
\citet{press} throughout this analysis.  This technique allows
for errors in both variables being fitted, and finds the best slope
and intercept parameters without assigning either parameter
as the independent or dependent variable.
As a way of bracketing uncertainties in the mean $M_\bullet-L$ relationship,
we performed one fit using all the data points, but for a second fit we used
only galaxies with $M_V<-19,$ because they appear to have less scatter.
The fit to all data points gives
\begin{equation}
\log (M_\bullet/M_\odot)=(1.40\pm0.17)(-M_V-21)/2.5+8.41\pm0.11, 
\label{eqn:ml_all}
\end{equation}
which is shown as the dashed line in Figure \ref{fig:ml}.
Just fitting galaxies with $M_V<-19$ gives
\begin{equation}
\log (M_\bullet/M_\odot)=(1.70\pm0.22)(-M_V-21)/2.5+8.22\pm0.08, 
\label{eqn:ml_bright}
\end{equation}
which is shown as the dotted line in Figure \ref{fig:ml}.
Both relationships agree well for $-23<M_V<-19;$ their differences in slope
cause them to diverge slightly when extrapolated to more luminous galaxies.
Both relationships also agree well with the 
\citet{hr} relationship between $M_\bullet$ and galaxy {\it mass}
transformed back to luminosity, which we consider as a third
$M_\bullet-L$ relationship.
\citet{novak} found that the $M_\bullet$-mass relationship was not
significantly less tight than the $M_\bullet-\sigma$ relationship,
given the errors of the various samples.
If so, then the reduced scatter
in the $M_\bullet$-mass relationship means that it should serve
well as a relationship between $M_\bullet$ and $L;$ we transform it
by adopting $M/L_V\approx6\times10^{-0.092(M_V+22)}M_\odot/L_\odot,$
based on the $M/L$ estimates given in \citet{g03}; this gives
\begin{equation}
\log (M_\bullet/M_\odot)=(1.38\pm0.07)(-M_V-22)/2.5+8.78\pm0.10. 
\label{eqn:ml_hr}
\end{equation}
This is shown in Figure \ref{fig:ml} as the solid line; within errors
it is essentially identical to equation (\ref{eqn:ml_all}) for $-25<M_V<-23,$
the interval over which we will be extrapolating the
$M_\bullet-L$ relationship to the most luminous galaxies in the sample.

\begin{figure}
\plotone{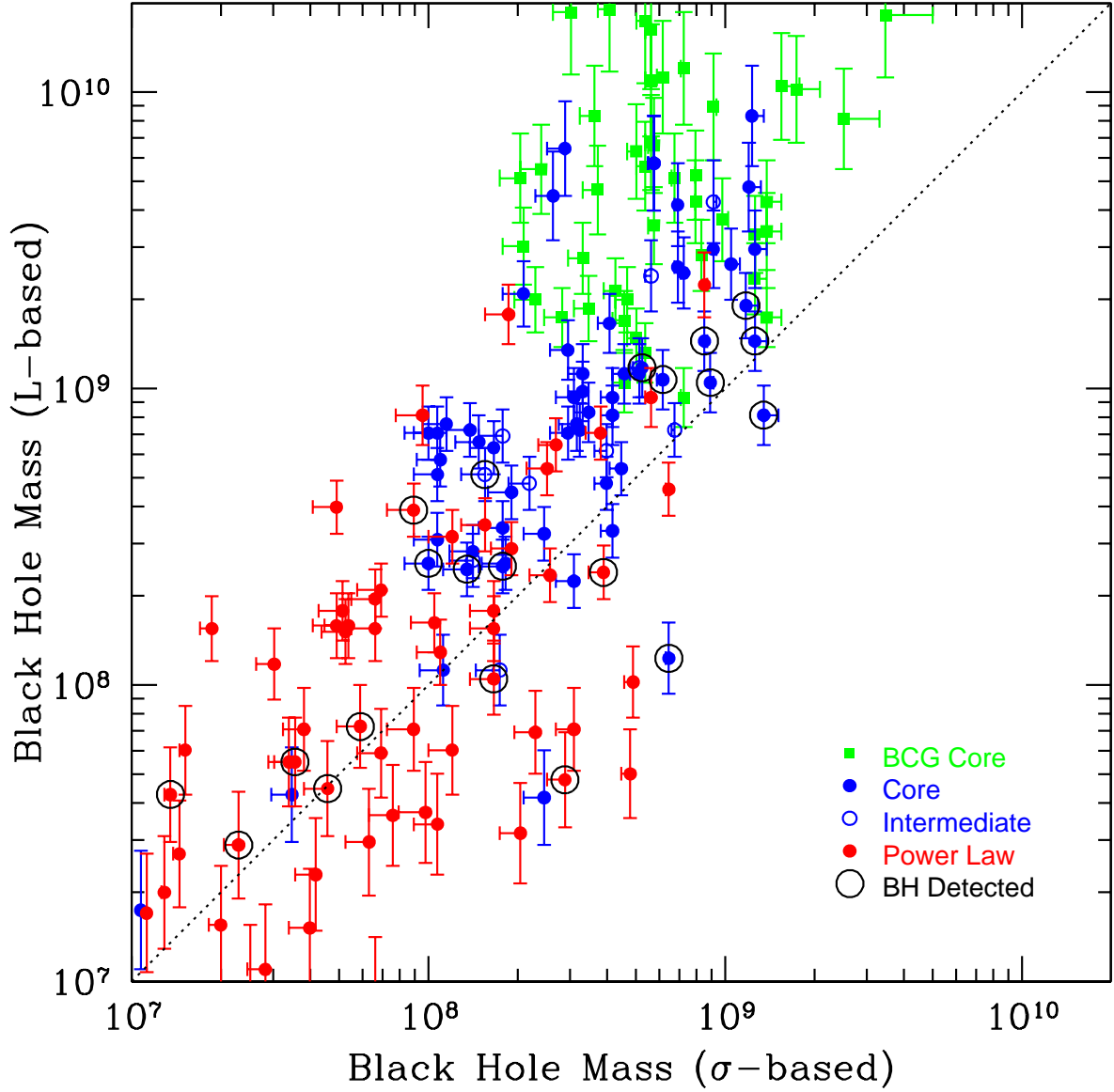}
\caption{$M_\bullet(L)$ versus $M_\bullet(\sigma)$
the sample galaxies that have $\sigma$ measurements.
Power-law galaxies are plotted as red dots, core galaxies are blue dots,
``intermediate'' galaxies are plotted as small open circles,
and BCGs with resolved cores are plotted as green squares.
Galaxies with large circles have directly determined black hole
masses; however, the predicted rather than observed $M_\bullet$
values are still plotted.
The $M_\bullet-\sigma$ relationship is that of Tremaine et al.\ (2002).
The asymmetric error bar in the horizontal direction shows the
change in predicted $M_\bullet$ if the \citep{wyithe} log-quadratic
$M_\bullet-\sigma$ relationship is used instead.
$M_\bullet(L)$ is the average of the minimum and
maximum predictions for a given $L$ from the three $M_\bullet-L$ relationships
in Figure \ref{fig:ml}, with the error bars showing the range of the
predictions.}
\label{fig:bh_pred}
\end{figure}

Figure \ref{fig:bh_pred} shows $M_\bullet(L)$ based on a combination of the
three relationships presented in Figure \ref{fig:ml} plotted against
$M_\bullet(\sigma)$ from equation (\ref{eqn:msig}) for the sample.
The error bars along the $M_\bullet(L)$ axis reflect the minimum and maximum
predictions of $M_\bullet$ given by the three relationships shown
in Figure \ref{fig:ml}; the central values plotted are the
mean of the minimum and maximum predicted $M_\bullet.$ 
The $L$ and $\sigma$ predictors diverge at large $L,$ with all three
$M_\bullet-L$ relationships predicting $M_\bullet\sim10^{10}M_\odot$
for the most luminous galaxies, while equation (\ref{eqn:msig}) predicts
no values of $M_\bullet$ larger than $\sim3\times10^9M_\odot.$
The errors in $M_\bullet(L)$ increase somewhat with galaxy luminosity
but are much smaller than the differences between $M_\bullet(L)$
and $M_\bullet(\sigma),$ which approach an order of magnitude
for some of the most luminous galaxies.\footnote{The error bars
in Figure \ref{fig:bh_pred} do not include the systematic errors
associated with the uncertainties in the individual relationships
themselves.}

The differences between $M_\bullet(L)$ and $M_\bullet(\sigma)$
cannot be reconciled by the \citet{wyithe} log-quadratic
$M_\bullet-\sigma$ relationship.  The asymmetric error bars in the
$\sigma$-based predictions of $M_\bullet$ shown in Figure \ref{fig:bh_pred}
reflect the implied change in predicted $M_\bullet$ if the \citet{wyithe}
relationship is used instead of the \citet{tr02} log-linear $M_\bullet-\sigma$
relationship.  The \citet{wyithe} relationship predicts slightly
larger $M_\bullet$ only for the largest $\sigma$ values ($\sim30\%$), but still
does not match the even larger $M_\bullet(L)$ for the same
galaxies.  As expected, $M_\bullet(L)$ and $M_\bullet(\sigma)$
do agree on average for the sample galaxies
that actually have direct $M_\bullet$ determinations,
since it was this subset of galaxies that defined the relationships
in the first place.

\begin{figure}
\plotone{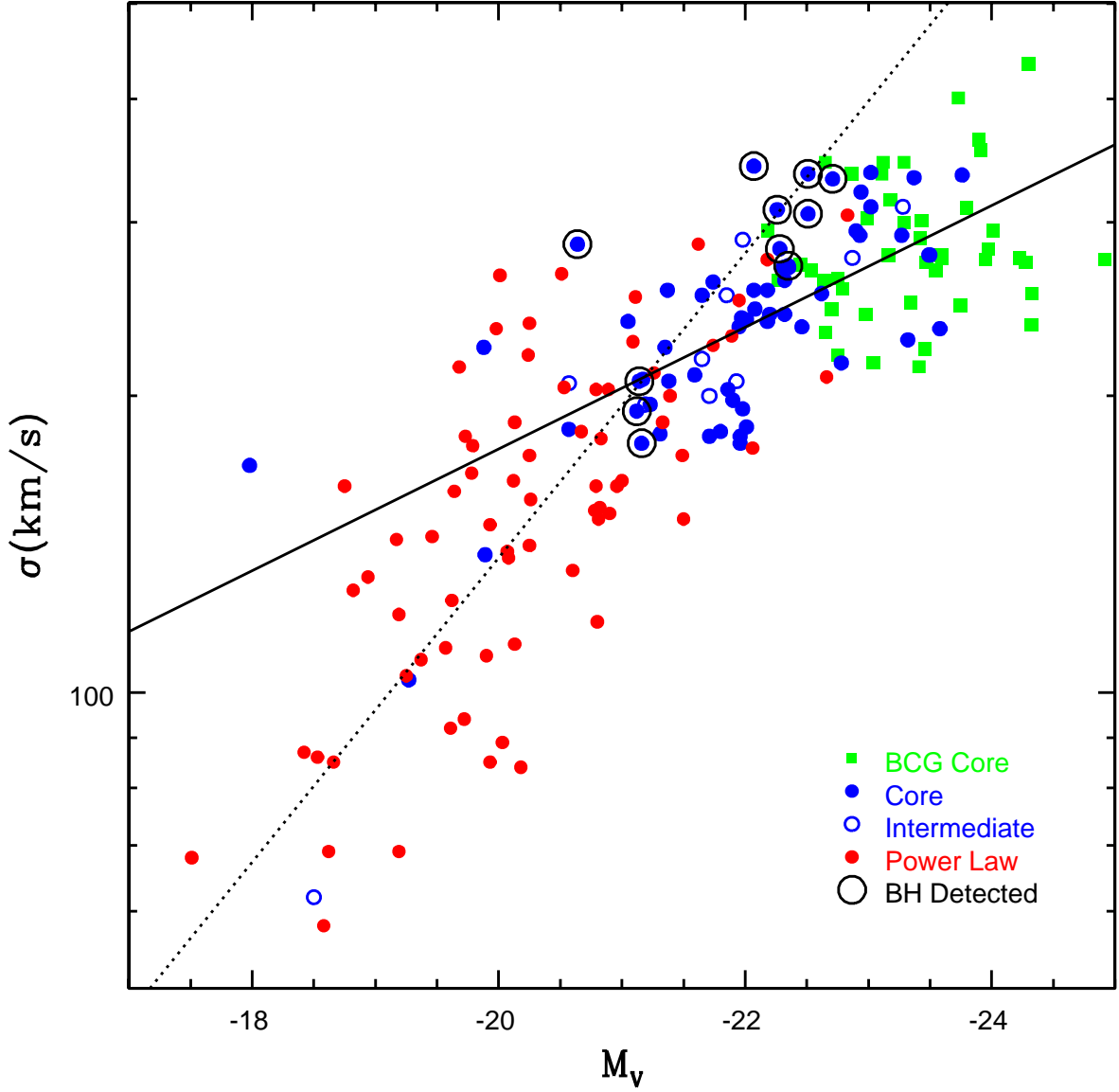}
\caption{The relationship between central velocity dispersion, $\sigma,$
and $L$ for the sample is plotted.
A fit to just the core galaxies and BCGs (solid line; equation \ref{eqn:fj})
gives $L\sim\sigma^7,$
a much steep relationship then the standard $L\sim\sigma^4$ Faber-Jackson
relationship, and $L\sim\sigma^2$ for the power-law galaxies alone
(dashed line; equation \ref{eqn:fj_pl}).
It is this change in slope that leads to conflicting
predictions for $M_\bullet$ from the $M_\bullet-L$ and $M_\bullet-\sigma$
relation for the most luminous galaxies.  Core galaxies with directly
measured black hole masses are circled.}
\label{fig:mv_sig}
\end{figure}

The disagreement of the two $M_\bullet$ predictors for the larger
set of galaxies lacking direct $M_\bullet$ determinations can be traced to
changes in the form of the $L-\sigma$ relationship as a function
of galaxy luminosity.  Figure \ref{fig:mv_sig}
shows this relationship for the sample galaxies.
The typical $\sigma$ value appears to level off for large $L$;
indeed, there appears to be little relationship between $\sigma$ and
$L$ for galaxies with $M_V<-22.$
While most of the galaxies in this luminosity range are BCGs, other
bright ellipticals show the same behavior.
Put simply, the high luminosities of BCGs and other similarly bright ellipticals
are not matched by similarly large velocity dispersions.
The $M_\bullet-\sigma$ relationship thus predicts unexceptional black
hole masses for these exceptionally luminous galaxies.

This ``saturation'' in $\sigma$ at BCG luminosities
was noted in the BCG velocity dispersion study of \citet{ho}, but it appears
only weakly in the SDSS study of \citet{bern2}.
We suggest that this may be due to the use of different BCG luminosities,
based on the analysis of the SDSS magnitudes of BCGs presented in Appendix
\ref{sec:bcg}.  For the core galaxies, we find $L\sim\sigma^7,$
a much steeper relationship than the classic $L\sim\sigma^4.$
Specifically, a symmetrical least-squares fit \citep{press}
to the 99 core galaxies with $M_V<-21$ and having a $\sigma$ value produces:
\begin{equation}
M_V=-2.5~(6.5\pm1.3)\log(\sigma/{\rm 250~km~s^{-1}})-22.45\pm0.18.
\label{eqn:fj}
\end{equation}
However, since the $L-\sigma$ relationship appears
to be nonlinear, even this fit may not be the best approximation
for the most luminous galaxies.
This result also contrasts with the relationship measured for
power-law galaxies alone,
\begin{equation}
M_V=-2.5~(2.6\pm0.3)\log(\sigma/{\rm 150~km~s^{-1}})-20.30\pm0.10.
\label{eqn:fj_pl}
\end{equation}

The distribution of points with $M_\bullet$ measurements shows
what appears to be a bias in the BH sample: galaxies with $M_V\sim-22.5$
with measured $M_\bullet$ have a higher-than-average $\sigma$ than typical
galaxies at this luminosity --- or conversely have low luminosities for
their $\sigma$ values (see also \citealt{b06c}).
The 7 galaxies with measured $M_\bullet$ at $M_V\sim-22.5$ have
average $\sigma=311\pm25{\rm~km~s^{-1}},$ while equation (\ref{eqn:fj})
predicts only $\sim250{\rm~km~s^{-1}}$ at $M_V\sim-22.5$ in agreement with
the average $\sigma$ at this luminosity for the SDSS sample \citep{bern2}.
If $\sigma$ is the best predictor of $M_\bullet,$ then the black holes
in these galaxies should be on average $(314/250)^4\approx2.4\times$ more
massive than is typical for galaxies with $M_V\sim-22.5.$
The $M_\bullet-L$ relationship in turn would be biased at
the high luminosity end, and the large black hole masses predicted
from $L$ shown in Figure \ref{fig:bh_pred} will be over-estimates.
Conversely, if $L$ is the better predictor of $M_\bullet,$ then
then the $M_\bullet-\sigma$ relationship would be biased to predict
lower $M_\bullet$ than would be correct.

The possibility that the galaxies with measured $M_\bullet$ are
a biased sampling of the $L-\sigma$ relationship is echoed
in Figure \ref{fig:bh_pred}.  For $M_\bullet>10^8M_\odot,$ $M_\bullet(L)$ 
is on average greater that $M_\bullet(\sigma)$ for galaxies in the present
sample.  Lowering $M_\bullet(L)$ by the bias factor inferred above, or
increasing $M_\bullet(\sigma)$ by a similar factor would bring the
average predictions into excellent agreement, however.  Note the
galaxies with measured $M_\bullet$ in Figure \ref{fig:bh_pred},
are presently in excellent agreement, since these are the very systems
used to define the $M_\bullet-\sigma$ and $M_\bullet-L$ relationships.

Figure \ref{fig:bh_pred} also shows, however, that the large
$M_\bullet(L)$ predicted for the most luminous galaxies still deviate from
$M_\bullet(\sigma)$ by a much larger factor than this putative bias.
The strong curvature in $L-\sigma$ relationship leads to the
upward curvature in $M_\bullet(L)$ versus $M_\bullet(\sigma)$
well in excess of the selection biases implied by Figure \ref{fig:mv_sig}.
Any luminosity-based predictor of $M_\bullet$ calibrated
for $M_V>-22$ would still predict $M_\bullet$ in excess of the
$M_\bullet-\sigma$ relationship for $M_V<-22,$ since $\sigma$ for
the brightest galaxies does not increase with luminosity.

\section{Core Structure as an Independent Witness of $M_\bullet$}

\subsection{The Cusp Radius\label{sec:rgam1}}

Resolving whether $L$ or $\sigma$ is the best predictor for
$M_\bullet$ for galaxies with $M_V<-23$ will only be possible when real
$M_\bullet$ determinations can be made in this luminosity regime.
Lacking this, we can attempt to obtain preliminary
information by considering whether the central
structure of galaxies may provide an independent witness to $M_\bullet.$
We characterize the physical scale of the core by
the ``cusp radius,'' $r_\gamma,$ which is the radius at which
the negative logarithmic-slope of a galaxy's surface brightness profile
reaches a pre-specified value $\gamma'.$
This measure of core size was first proposed
by \citet{carollo}; we will discuss it in detail in Appendix \ref{sec:rgam2}.
The core is also characterized by
the cusp brightness, $I_\gamma,$ the local surface
brightness at $r_\gamma$ ($\mu_\gamma$ is $I_\gamma$ expressed in
magnitude units).
In terms of the Nuker-law parameters, for $\gamma\leq\gamma'\leq\beta,$
\begin{equation}
r_\gamma\equiv r_b\left({\gamma'-\gamma\over\beta-\gamma'}\right)^{1/\alpha};
\end{equation}
$I_\gamma$ is then found directly from the fitted Nuker-law,
\begin{equation}
I_\gamma=2^{(\beta-\gamma)/\alpha}I_b\left({r_b\over r_\gamma}\right)^{\gamma}
\left[1+\left({r_\gamma\over r_b}\right)^\alpha\right]^{(\gamma-\beta)/\alpha}.
\label{eqn:rgam}
\end{equation}
\citet{carollo} advocated use of $r_\gamma$ with $\gamma'=1/2$
as a core scale-parameter.
We show in Appendix \ref{sec:rgam2} that
using $r_\gamma$ with $\gamma'=1/2,$
indeed gives tighter correlations with other galaxy parameters
than the choice of $r_b$ as a scale parameter.  While the Nuker-law
$r_b$ is still used to calculate $r_\gamma,$ we no longer use it
directly as a measure of core size, in contrast to the analysis
presented in \cite{f97}.  Lastly, we emphasize that since $r_\gamma$ is
generally well interior to $r_b$ it is not meant to describe the actual
complete extent of the core; it is just a convenient representative scale.

\subsection{Core Structure and Galaxy Parameters}

\begin{figure}
\plotone{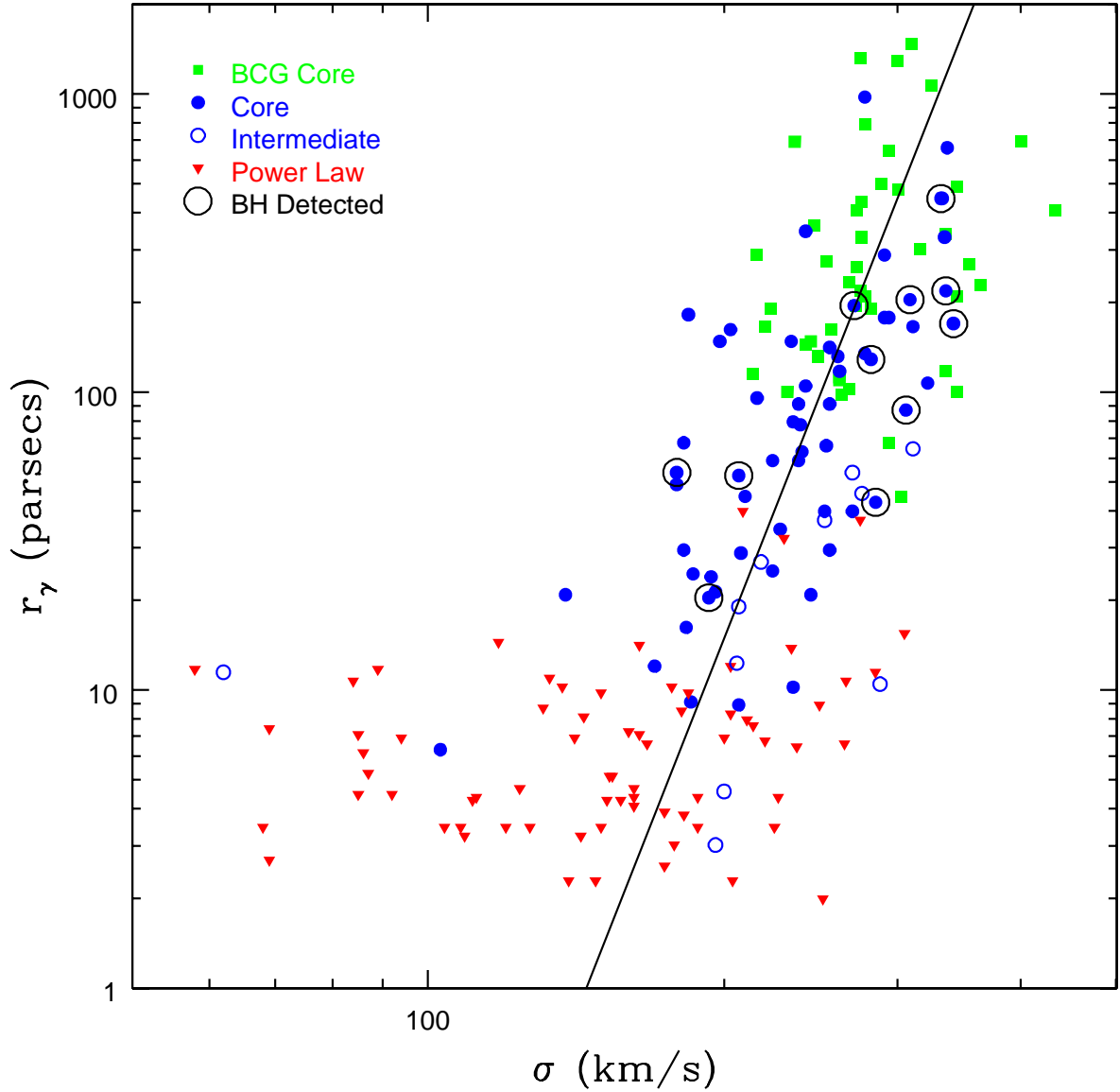}
\caption{Cusp radius, $r_\gamma,$ is plotted as a function of stellar velocity
dispersion.  The power-law galaxies are now plotted as triangles
to indicate that their cusp radii are only upper limits.  The
solid line is the fitted relationship (equation \ref{eqn:r_sig})
between $r_\gamma$ and $\sigma$ for core galaxies.
The figure shows that $r_\gamma$ is a steep function of $\sigma.$
If $M_\bullet\sim\sigma^4$ as equation (\ref{eqn:msig}), the observed
empirical relationship between $r_\gamma$ and $\sigma$ (solid line)
implies that $r_\gamma\sim M_\bullet^{2.1\pm0.4}.$}
\label{fig:sigrc}
\end{figure}

It has long been known that the physical scale of cores in early-type
galaxies is correlated with galaxy luminosity \citep{l85, k85}.
This relationship may be
due to the action of central black holes on the central distribution
of stars (e.g., \citealt{f97}).  Figures \ref{fig:sigrc} and \ref{fig:mvrc} show
the relationships between $\sigma,$ $L,$ and cusp radius,
$r_\gamma,$ for the present sample.  The $r_\gamma-\sigma$
relationship is particularly steep, as $r_\gamma$ varies by over
two orders of magnitude, while $\sigma$ changes by only a factor of two.
For core galaxies with $M_V<-21,$ a symmetrical least-squares fit gives
\begin{equation}
\log(r_\gamma/{\rm pc})=(8.4\pm1.6)\log(\sigma/{\rm 250~km~s^{-1}})+1.99\pm0.09,
\label{eqn:r_sig}
\end{equation}
while the $r_\gamma-L$ relationship is
\begin{equation}
\log (r_\gamma/{\rm pc})=(1.32\pm0.11)(-M_V-23)/2.5+2.28\pm0.04.
\label{eqn:r_mv}
\end{equation}
Of the two relationships, $L$ is the better predictor of $r_\gamma,$
with only 0.31 rms scatter in $\log r_\gamma,$ while the scatter
of $\log r_\gamma$ versus $\sigma$ is 0.63.
Note that BCGs and non-BCG core galaxies appear to follow the same
relationships between $r_\gamma$ and $\sigma$ or $L.$
For the non-BCG core galaxies,
\begin{equation}
\log(r_\gamma/{\rm pc})=(7.4\pm1.2)\log(\sigma/{\rm 250~km~s^{-1}})+2.00\pm0.07,
\end{equation}
and
\begin{equation}
\log (r_\gamma/{\rm pc})=(1.94\pm0.29)(-M_V-22)/2.5+1.77\pm0.06,
\end{equation}
while for the BCGs,
\begin{equation}
\log(r_\gamma/{\rm pc})=(15.2\pm7.5)\log(\sigma/{\rm
300~km~s^{-1}})+2.80\pm0.31,
\end{equation}
and
\begin{equation}
\log (r_\gamma/{\rm pc})=(1.24\pm0.17)(-M_V-23)/2.5+2.28\pm0.05.
\end{equation}
While the slopes of the relationships are different from those
for the entire sample of core galaxies, there is no
significant difference between the relationships within the parameter
ranges in which BCGs and non BCGs overlap.
A critical result that is evident in Figure \ref{fig:mvrc} is that
while BCGs have larger cores than less luminous core galaxies,
they are completely consistent with the larger total luminosity of BCGs.
In contrast, there is
essentially no correlation between $r_\gamma$ and $\sigma$
for $r_\gamma>300$ pc, as is evident
in Figure \ref{fig:sigrc}; luminosity is a much better predictor
of core size in BCGs than $\sigma.$

The core is characterized by a surface brightness as well as a physical scale,
thus one could also explore the relationships between $I_\gamma$ and
$\sigma$ or $L,$ but as we show in Figure \ref{fig:rc_ic}, $I_\gamma$
and $r_\gamma$ are so closely related that they can be regarded as
interchangeable.
The fitted relationship between the two parameters for core
galaxies with $M_V<-21$ is
\begin{equation}
\mu_\gamma=2.5~(1.05\pm0.07)\log(r_\gamma/{\rm100~pc})+16.23\pm0.10,
\label{eqn:ic_rc}
\end{equation}
where $\mu_\gamma$ is $I_\gamma$ in units of $V-$band magnitudes
per square-arcsecond.

Lastly, $I_\gamma$ and $r_\gamma$ can be combined to estimate the
stellar mass of the core interior to the cusp as $M_\gamma=\pi I_\gamma
r_\gamma^2(M/L_V),$ again using the conversion between mass and light
given in the context of equation (\ref{eqn:ml_hr}).  Symmetrical
fits give the relationships between $M_\gamma$ and $L$ or $\sigma$ as;
\begin{equation}
\log(M_\gamma/M_\odot)=(1.35\pm0.13)(-M_V-22)/2.5+9.17\pm0.05,
\label{eqn:mc_l}
\end{equation}
and
\begin{equation}
\log(M_\gamma/M_\odot)=(8.6\pm1.0)\log(\sigma/200{\rm km~s^{-1}})+8.55\pm0.13.
\label{eqn:mc_sig}
\end{equation}

\subsection{Core Scouring and Black Hole Mass}

\subsubsection{Black Hole Mass and $r_\gamma$}

The existence of the $r_\gamma-\sigma$ and $r_\gamma-L$ relationships implies
empirical relationships between $r_\gamma$ and $M_\bullet,$
given the separate $M_\bullet-\sigma$ and $M_\bullet-L$ relationships.
By combining equation (\ref{eqn:msig}) with
equation (\ref{eqn:r_sig}) we find $r_\gamma\sim M_\bullet^{2.1\pm0.4},$ or
more precisely,
\begin{equation}
M_\bullet(\sigma)\Longrightarrow\quad
\log(r_\gamma/{\rm pc})=(2.1\pm0.4)\log (M_\bullet/10^9M_\odot) + 2.7\pm0.2.
\label{eqn:mbh_rs}
\end{equation}
At the same time, we can also combine
equations (\ref{eqn:r_mv}) and (\ref{eqn:ml_hr}) to find
$r_\gamma\sim M_\bullet^{1.0\pm0.1},$ or
\begin{equation}
M_\bullet(L)\Longrightarrow\quad
\log(r_\gamma/{\rm pc})=(0.96\pm0.09)\log (M_\bullet/10^9M_\odot) + 1.9\pm0.1.
\label{eqn:mbh_rl}
\end{equation}
Equations (\ref{eqn:mbh_rs}) and (\ref{eqn:mbh_rl}) are inconsistent.
The conflict between $M_\bullet(L)$ and $M_\bullet(\sigma)$
leads in turn to contradictory predictions
for how the physical scale of cores is related to black hole mass.

Comparison of the observed $r_\gamma-M_\bullet$ relationship to
the two inferred relationships presented above may offer a path to
determining which of $M_\bullet(\sigma)$ or $M_\bullet(L)$ is more
accurate for the most massive galaxies.
In the ``core-scouring'' scenario, cores are created by the orbital decay of a
massive binary black hole, which would be formed during the merging of two
galaxies.  As the merger progresses, black holes in the nuclei of the progenitor
galaxies are brought to the center of the merged system by dynamical friction.
While the center of the merger
may initially be highly concentrated \citep{mnm}, as is the case
for power-law galaxies, central stars interacting with the binary black
hole are ejected from the center as the binary hardens.
The ejection of stars erodes the steep central stellar density profile,
creating a shallow cusp, or break from the steeper profile that still
persists at larger radii.
A core is the region of the galaxy interior to the break (cf. \citealt{l95}).

Under this hypothesis, the relationship
between core scale and $M_\bullet$ ought to be more fundamental
than either of the $r_\gamma-\sigma$ or $r_\gamma-L$ relationships alone.
The action of the black hole mass on stellar orbits at the galaxy center
{\it creates the core structure directly,} and the $r_\gamma-\sigma$
and $r_\gamma-L$ relationships are then merely consequences
of the separate $M_\bullet-\sigma$ and $M_\bullet-L$ relationships.
According to this logic, we would conclude that the
larger cores of BCGs are evidence of higher BH masses.

A major caveat standing in the way of this conclusion is that
core scouring may not lead directly
to a clean relationship between $r_\gamma$ and $M_\bullet.$
The binary BH ejects a total mass of stars, $M_{ej},$ that is expected to be
proportional to the total merged $M_\bullet$ \citep{q06, mnm, m06}.
However, the resultant $r_\gamma$ would depend on the radii over
which stars are ejected from the center.
Further, \citet{m06} presents simulations that show that core
formation should be a cumulative process.
Cores formed in one merger event will
be depleted even further in subsequent mergers, presumably leading to even
larger increases in $r_\gamma$ that would reflect not only the total BH mass,
but the integrated merger history as well.
Under this hypothesis, cores resulting from successive dry mergers
would be abnormally large compared to their BH masses, potentially
explaining the extra-large cores of BCGs, which are thought to
be formed by such multiple mergers.  Under scrutiny, 
however, this explanation seems difficult to support, since the cores of BCGs
show no excess {\it compared to the luminosities of their host galaxies,}
and it is this latter quantity that is probably
the best indicator of the amount of dry merging that any massive
elliptical has experienced.
In other words, the core masses of BCGs galaxies are the same
fixed fraction of their total light as in other galaxies, not
some amplified value driven by multiple mergers.
Thus, we seem to be driven back to the basic explanation that the larger cores
of BCGs are due simply to larger BH masses.

Can we use actual core data to identify
the correct $r_\gamma-M_\bullet$ relationship?
Figure \ref{fig:bh_rc} tries this by plotting $r_\gamma$ versus $M_\bullet$ for
the 11 core galaxies for which there are direct determinations of $M_\bullet.$
A symmetric fit to $r_\gamma$ and $M_\bullet$ for these galaxies has the form
\begin{equation}
\log(r_\gamma/{\rm pc})=(0.83\pm0.25)\log (M_\bullet/10^9M_\odot) + 2.20\pm0.10.
\label{eqn:mbh_r_fit}
\end{equation}
This equation is essentially consistent with equation (\ref{eqn:mbh_rl}),
the relationship inferred from $M_\bullet(L),$ rather than
equation (\ref{eqn:mbh_rs}), which inferred from $M_\bullet(\sigma).$
At the same time, the scatter in Figure \ref{fig:bh_rc} is large, thus
this result is sensitive to how the $r_\gamma-M_\bullet$ relationship
is fitted.  For example, if $r_\gamma$ is treated as the independent
variable in an attempt to predict $M_\bullet,$ given $r_\gamma,$ then
\begin{equation}
\log(r_\gamma/{\rm pc})=(1.5\pm0.8)\log (M_\bullet/10^9M_\odot) + 2.20\pm0.11
\label{eqn:mbh_r_ind}
\end{equation}
(although we express $r_\gamma$ as the dependent variable for comparison
with the relationships above).  The slope of this relationship is
intermediate between that in equations (\ref{eqn:mbh_rs})
and (\ref{eqn:mbh_rl}).
For completeness, if $M_\bullet$ is treated as the independent variable,
which corresponds to the scouring hypothesis that $M_\bullet$ determines
$r_\gamma,$ then
\begin{equation}
\log(r_\gamma/{\rm pc})=(0.59\pm0.18)\log (M_\bullet/10^9M_\odot) + 2.19\pm0.10
\label{eqn:mbh_m_ind}
\end{equation}

These three fits do not in fact suffice to identify the ``correct"
$r_\gamma-M_\bullet$ relation for four reasons: 1)
the various slopes differ considerably because the native scatter in the data
is large; 2) we are seeking the ``true" underlying
relationship (i.e., the ``theorist's'' question of \citealt{novak}), but
without a knowledge of cosmic scatter and its separate contribution to both
$M_\bullet$ and $r_\gamma,$ we cannot fit the data properly to find it;
3) the slopes in equations (\ref{eqn:mbh_rs}) and (\ref{eqn:mbh_rl})
were likewise meant to embody ``true" relations, but they were
derived from prior fits that themselves suffered a similar ambiguity;
and 4) the sample of core galaxies with measured $M_\bullet$
is potentially biased in some way that is not understood
(cf. Figures \ref{fig:mv_sig} to \ref{fig:rc_ic}), and
any new fit based on these galaxies might therefore not be representative.
On this last point, we emphasize caution. While the galaxies with
measured $M_\bullet$ may on average have offsets in the parameter plots
shown, this does not mean {\it a priori} that the directly
observed $r_\gamma-M_\bullet$ relationship is biased.  The small number
of core galaxies with measured $M_\bullet$ plus the number of parameters
in play means that understanding any biases must await a richer sample.

Likewise, the sample of core galaxies with measured $M_\bullet$ will
have to be increased considerably before it can be used to convincingly
discriminate between the $M_\bullet(L)$ and $M_\bullet(\sigma)$ relations.
Nevertheless, we may be able to obtain some guidance in advance
of such observations by comparing $M_\bullet$ estimated from
$r_\gamma$ to values estimated from $L$ or $\sigma.$
Figure \ref{fig:bh_rgpred} shows the results of using either
equation (\ref{eqn:mbh_r_fit}) or (\ref{eqn:mbh_r_ind}) to predict $M_\bullet$
from $r_\gamma,$ in analogy to Figure \ref{fig:bh_pred}, which compared
predictions of $M_\bullet$ based on $\sigma$ versus $L.$
Both versions of the $r_\gamma-M_\bullet$ relationship predict larger
$M_\bullet$ than would be inferred from $M_\bullet(\sigma).$
The symmetrically-fitted $r_\gamma-M_\bullet$ in equation (\ref{eqn:mbh_r_fit})
appears to be consistent with $M_\bullet(L),$ also predicting
$M_\bullet\sim10^{10} M_\odot$ for the most massive galaxies.

Presently, the large scatter in the observed $r_\gamma-M_\bullet$
relationship and the attendant uncertainties in any empirical relationship
derived from it does not decisively favor $M_\bullet(L)$
over $M_\bullet(\sigma).$  Equations (\ref{eqn:mbh_r_fit}) and
(\ref{eqn:mbh_r_ind}) however, on average predict greater $M_\bullet$ than
would be inferred from $M_\bullet(\sigma),$ while equation (\ref{eqn:mbh_r_fit})
is consistent with the larger black hole masses implied by $L$
for the most massive galaxies.
At this early stage the $r_\gamma-M_\bullet$ relationship thus may favor
consistency with the $M_\bullet(L)$ relationship.
The fact that the scatter of $r_\gamma$ on $L$
is smaller than that on $\sigma$ (as would be expected if $r_\gamma$
is produced directly by black hole scouring and $M_\bullet$ correlates
more closely with $L$) as well as the
fact that $M_\bullet(L)$ plausibly explains the large core of BCGs
as being due to more massive black holes, whereas $M_\bullet(\sigma)$
seems to provide no ready explanation this, may offer additional
support that $M_\bullet(L)$ is more appropriate for the most massive galaxies.

\subsubsection{Black Hole Mass and Core Mass}

An alternative approach to explore the relationship between core
structure and black hole mass is to compare the core mass,
$M_\gamma=\pi I_\gamma r_\gamma^2(M/L_V)$, rather than $r_\gamma,$
to $M_\bullet.$   Although, as we noted earlier, $I_\gamma$ and $r_\gamma$
are closely related, so the relationship between $M_\gamma$ and
$M_\bullet$ will contain information similar to the $r_\gamma-M_\bullet$
relationship, core mass should be a more direct indicator of the
amount of core scouring and its relationship to black hole mass.
If cores are created from power-law galaxies by core scouring
following a dry merger, one might expect that the core mass would be
approximately proportional to the black-hole mass. This conjecture is
supported by N-body calculations by \citet{m06}, who argues that
the core mass produced by scouring in a single merger is $\simeq
fM_\bullet$, where $M_\bullet$ is the mass of the merged black hole
and $f\simeq 0.5$, largely independent of the mass ratio of the
merging black holes; he also argues that the total core mass after $N$
dry mergers should be given by $f\simeq0.5N$.
Direct estimation of the mass ejected from the core
by scouring is much more difficult
observationally than theoretically, because we do not know the state
of the galaxy before the merger. Thus we will simply use $M_\gamma$ as an
``indicative" core mass, recognizing
that the factor $f$ relating indicative core mass to black-hole mass
is very uncertain, but should be approximately independent of galaxy
luminosity for core galaxies.

Figure \ref{fig:corem} shows the relationships between $M_\gamma$
and $M_\bullet$ as derived from the $M_\bullet-\sigma$ and
$M_\bullet-L$ relationships.  By combining  $M_\bullet(\sigma)$
(equation \ref{eqn:msig}) with the $M_\gamma-\sigma$ relationship
(equation \ref{eqn:mc_sig}), we find
\begin{equation}
M_\bullet(\sigma)\Longrightarrow\quad
\log(M_\gamma/M_\odot)=(2.2\pm0.3)\log (M_\bullet/10^9M_\odot) + 10.29\pm0.18,
\label{eqn:corems}
\end{equation}
while the combinations of $M_\bullet(L)$ (equation \ref{eqn:ml_hr})
with the $M_\gamma-L$ relationship (equation \ref{eqn:mc_l}) gives
\begin{equation}
M_\bullet(L)\Longrightarrow\quad
\log(M_\gamma/M_\odot)=(1.0\pm0.1)\log (M_\bullet/10^9M_\odot) + 9.39\pm0.11.
\label{eqn:coreml}
\end{equation}
The relation between indicative core mass and black-hole mass
predicted by the $M_\bullet-L$ relation is essentially linear, as
expected, while the
relation predicted by the  $M_\bullet-\sigma$ relation is twice as steep.
Moreover the ratio of indicative core mass to black-hole
mass is $\sim2.4$ at $M_\bullet=10^9M_\odot$ from the $M_\bullet-L$
relation, not far from the value of order unity that we might expect,
while the corresponding value from the $M_\bullet-\sigma$
relation is $\sim19$. It is
difficult to devise dynamical models in which core scouring could be
efficient enough to create cores with mass so much bigger than the
black-hole mass.

\section{The Growth of the Most Massive Galaxies and Why
$M_\bullet(L)$ Might be Favored Over $M_\bullet(\sigma)$}

Having found suggestive but not conclusive
arguments to prefer one relation over the other, we turn
now to physical arguments for additional guidance.
We stress again that the tension between $M_\bullet(L)$ and $M_\bullet(\sigma)$ 
arises due to the breakdown, or curvature, in the $L-\sigma$
relationship at high galaxy masses.
It is appropriate to inquire at least
briefly into physical reasons why this might happen,
and whether this offers insight into which of $M_\bullet(L)$ and
$M_\bullet(\sigma)$ might be preferred for the most massive galaxies.

Curvature in a gravitational scaling relation may signal a breakdown
in perfect homology in galaxy formation, which in turn may reflect a change
in the relative importance of different physical processes as 
a function of galaxy size.  One such effect, suggested some time
ago, is an increase in the importance of dissipationless
(i.e., ``dry") merging in forming the most massive ellipticals
\citep{bend92, f97, naab}.
A second effect, following logically from the first, is a change
in the nature of dissipationless mergers with galaxy
mass.  As hierarchical clustering proceeds, clusters of galaxies become
more massive, and previously formed elliptical
galaxies at the centers of these clusters merge.  Each new round
of merging thus increases the galaxy stellar mass along with
the cluster dark halo mass in which it is embedded.  The largest ellipticals
are thus produced by dry mergers at the centers of the largest
clusters.  These are BCGs.  The above scenario
is supported by the steep environmental dependence 
among bright ellipticals, the brightest ones being found in 
the densest environments \citep{hogg}.  

There are at least two trends that might
contribute to a breakdown in perfect homology 
for dry merging to produce the observed curvature
in the $L-\sigma$ relationship.  The first is a change
in the typical orbital eccentricity of central merging pairs
as cluster mass grows.  \citet{bmq}
have suggested that head-on collisions
may become more frequent when massive clusters merge, and their
N-body simulations indicate less loss of energy from
stars to dark matter in such collisions.  The resultant 
stellar remnants are puffed up in radius and have significantly 
lower stellar velocity dispersions compared to encounters
with normal orbital geometry.  A second effect, not considered by them,
is the fact that the ratio of cluster velocity dispersion
to internal galaxy velocity dispersion also rises as clustering proceeds.
This appears to happen because gas cooling is reduced in large dark-matter
halos (e.g., \citealt{birn}), which means that the baryonic masses
of central galaxies grow more slowly than their dark-matter halos.
This is why the stellar velocity dispersions within BCG galaxies in large
clusters are so much lower than those of their surrounding clusters, whereas
the same is not true of ellipticals in small groups
(cf. Figure 2 of \citealt{cdm}).
The net result is that central merging pairs will
approach each other at relatively higher speeds, with the potential
of injecting more orbital kinetic energy into the final stellar remnant.
This would also cause the remnant to puff up and have smaller final velocity
dispersion.  

The relative importance of these two effects
can only be decided using realistic two-component N-body
simulations containing both stars and dark matter
that are appropriately embedded in a cosmological clustering scenario.   
It seems probable that both effects will be 
found to play a role.  The point
for now is that there are at least two reasons
to expect non-homology in dissipationless mergers, and thus two reasons for
curvature in the $L-\sigma$ relationship.

If this logic is correct, it points towards
$M_\bullet(L)$ as being the proper scaling law for massive galaxies.
That is because the major growth in black hole mass during dissipationless
merging occurs by merging black holes as the galaxies themselves merge.
With little mass accretion
directly onto black holes during this stage and no attendant star formation,
black hole mass should increase in proportion to galaxy mass.
The ratio of stellar mass to black hole mass is constant over dry merging,
consistent with $M_\bullet$ scaling linearly with galaxy mass.
Conversely, for $M_\bullet(\sigma)$ to be maintained over dry merging, given
the plateau in $\sigma$ at high galaxy luminosity, would require one of the two
merging black holes to be ejected from the galaxy as a common occurrence.
These arguments provide yet a another reason to prefer the
$M_\bullet(L)$ relation over $M_\bullet(\sigma).$

Regardless of which mechanism is dominant for determining
the velocity dispersion of the merged galaxy,
``puffing-up" of the remnant does appear to happen
in the most luminous galaxies, supporting an overall scenario in which
velocity dispersion in the largest galaxies
does not increase strongly via mergers.
The evidence for this comes from the effective radii of the largest galaxies
(see also \citealt{b06b}).
Figure \ref{fig:mvre} shows the $R_e-L$ relationship for the whole sample,
where $R_e$ is the effective radius measured from $r^{1/4}$-law fits to
those galaxies that have ground-based surface photometry
extending to large radii \citep{l06}.
For low galaxy luminosity, the 
the mean $R_e-L$ relationship is relatively shallow.  We find
\begin{equation}
\log(R_e/{\rm pc})=(-0.50\pm0.08)(M_V+21)/2.5+3.62\pm0.04
\label{eqn:re_pl}
\end{equation}
based on fits to just power-law galaxies.
This stands in contrast to the
steeper relationship defined by core galaxies with $M_V<-21,$
\begin{equation}
\log(R_e/{\rm pc})=(-1.18\pm0.06)(M_V+23)/2.5+4.27\pm0.02
\label{eqn:re_core}
\end{equation}
The transition between the two forms occurs at $M_V\approx-22,$
which corresponds to the luminosity at which the average central
structure changes from power-law to core \citep{l06}.
This is also the scale at which $\sigma$ starts to plateau in
Figure \ref{fig:mv_sig} --- the leveling-off of the $L-\sigma$
relationship is thus associated with a rapid increase in $R_e$
with $L$ not seen in less luminous galaxies.
This is as predicted if extra energy is injected into these
galaxies by merging: the objects will have lower $\sigma$ but larger $R_e.$

It should be stressed that
the arguments presented in support of the $M_\bullet(L)$ relation 
in this section apply only to bright ellipticals, which are
those produced by {\it dissipationless} merging.  The mass-accretion
processes that built black holes when galaxies were younger were drastically
different and might have obeyed different scaling laws.
The $M_\bullet(\sigma)$ law might be a better fit to such galaxies,
which in general will be smaller than the objects considered
in this paper.  The broader point of this discussion, however, is that 
``non-homology" processes may have affected the growth of
galaxies generally, with the result that a single black hole
scaling law with global galaxy properties might not fit all galaxies.

\section{The Space Density of the Most Massive Black Holes}

The preceding sections have presented suggestive if not
conclusive reasons to suspect
that the $M_\bullet - L$ relation might be a better predictor of black hole
mass than the $M_\bullet - \sigma$ relation for the most massive galaxies.
\begin{enumerate}

\item The velocity dispersions of the most massive
elliptical galaxies rises slowly if at all with galaxy luminosity
implying that their black holes are no larger than those of much smaller
ellipticals if $M_\bullet - \sigma$ is the governing relation ---
this seems rather surprising. 

\item The core-scouring model says that $r_\gamma$
should correlate directly with $M_\bullet$ whereas correlations
between $r_\gamma$ and $L$ or $\sigma$ should be secondary; if so,
the smaller scatter of $r_\gamma$ on $L$ validates $L$ as the more
accurate predictor of $M_\bullet$.

\item The $M_\bullet -L$ relation offers a simple explanation
for the large cores of BCGs in terms of bigger black holes whereas
the $M_\bullet - \sigma$ relation offers no such ready explanation.

\item Either of the two fits of $M_\bullet$ on $r_\gamma$ for core galaxies
with measured BHs (equations \ref{eqn:mbh_r_fit} and \ref{eqn:mbh_r_ind})
predicts large $M_\bullet$ when extrapolated to luminous galaxies,
in better agreement with $M_\bullet-L$ than with $M_\bullet - \sigma$.

\item The largest elliptical galaxies are believed to be formed by dry merging,
which predicts that black hole mass should grow in proportion to stellar mass;
the observed $M_\bullet \propto L$ relation is thus the simplest relation
predicted on these grounds.  By contrast, the saturation of black holes mass
in the largest galaxies that is predicted by the
$M_\bullet-\sigma$ relation requires that
one of the two merging BHs be ejected from the galaxy as a common
occurrence, which may not by natural.

\item The largest elliptical galaxies are likely formed by dry mergers at the
centers of massive clusters; non-homology merger arguments plausibly
explain the low velocity dispersions (and large radii) of such
galaxies, but the same arguments then imply that $\sigma$ ought not to be
a fundamental parameter for predicting black hole mass in the biggest galaxies.

\item As noted in the introduction, there is evidence from AGN physics
that $M_\bullet\sim10^{10}M_\odot$ in some systems.

\end{enumerate}

Although these lines of argument are not conclusive, they 
motivate us to consider the implications for the local BH mass 
function should the $M_\bullet-L$ relation prove correct.  The differences 
from the $M_\bullet - \sigma$ relation are large, as we shall show.  In this
section we first compute these two mass functions based on $M_\bullet-\sigma$
and $M_\bullet-L$, and then we compare the resulting mass functions to
estimates of the relic black hole mass function based on the space density of
QSOs as a function of luminosity and redshift.

\subsection{The Black Hole Cumulative Mass Distribution Functions}

We first compute the black hole mass function by combining
the $M_\bullet(\sigma)$ predictor with the
velocity-dispersion function (the space density of
galaxies as a function of velocity dispersion),
We then repeat the calculation, but then using $M_\bullet(L)$ combined
with the galaxy luminosity function.
Our analysis follows the precepts 
of \citet{yu02}, departing in the choice of dispersion functions.
Both calculations use the same formalism, thus for the sake of generality
we denote the log of either the velocity dispersion, $\sigma,$
or luminosity of the galaxy, $L,$ by $s$ and 
assume that the correlations of BH mass $M_\bullet$ with either parameter
can be formalized through the statement that the probability of finding
a galaxy a given BH mass is
\begin{equation} 
 dP(M_\bullet) = (2 \pi \Delta^2)^{-1/2}
\exp\left[ - 
{[\log(M_\bullet) - F(s)]^2 \over 2 \Delta^2}\right]d\log(M_\bullet),
\label{eqn:prob}
\end{equation} 
where  $F(s)$ is the ridge line of either
$\log(M_\bullet) - s$ relation.  
The number of BHs near a given mass is then
\begin{equation}
 {d \,  n(M_\bullet) \over 
    d\log(M_\bullet)} = (2 \pi \Delta^2)^{-1/2}
\int_{-\infty}^\infty {dn \over ds} \,
\exp\left[ -
 {[\log(M_\bullet) - F(s)]^2 \over 2 \Delta^2} \right]
ds,
\label{eqn:vdiffnm}
\end{equation}
and the cumulative distribution is
\begin{equation}
 n(M_\bullet)  =
\int_{M_\bullet}^\infty
 {d \,  n(M_\bullet) \over
    d\log(M_\bullet)}   d\log(M_\bullet).
\label{eqn:vcumnm}
\end{equation}

For the dispersion-based predictor,
we start with the \citet{sheth} SDSS-based
velocity-dispersion function.
\citet{bern3} reprocessed the SDSS data and recovered a number of
galaxies with larger dispersions than those used in \citet{sheth}.  
We use that set of high dispersion 
galaxies to compute a cumulative dispersion function above 
$\sigma = 350 \kms$ by  
\begin{equation}
n(>\sigma) = N(>\sigma)/V,  
\label{eqn:bdycounts}
\end{equation}
where $N(>\sigma)$ is the number of galaxies with dispersions 
greater than $\sigma$ and $V$ is the Sloan survey volume given by 
\citet{bern3} as $3.34 \times 10^{-7} \mpc$ 
(for $H_0=70~{\rm km~s^{-1}~Mpc^{-1}}$ and $z < 0.3$).  
This cumulative function is well approximated by a power law.  
Differentiating it gives an estimate of the dispersion function 
above $\sigma = 350 \kms$ of 
\begin{equation} 
 {d \,  n(M_\bullet) \over d\log(\sigma)}
  = 6.67 \times 10^{-6} 
  \left( {\sigma \over 350 \kms} \right)^{-10.27}
   \mpc^{-3}
\label{eqn:bdydiff}
\end{equation}
Equation \ref{eqn:bdydiff} predicts about 10 times as many 
galaxies with dispersions greater than $400 \kms$ as the 
Schechter function fit given in 
\citet{sheth}.  Above $\sigma = 350 \kms$ 
we add it to the \citet{sheth} dispersion function in 
the analysis of equation (\ref{eqn:vdiffnm}).

We think this is the best available estimate of
$dn/d log(\sigma)$ for early-type galaxies at zero redshift.
We combine the Sheth/Bernardi dispersion function
with the \citet{tr02} $M_\bullet(\sigma)$ predictor
(equation \ref{eqn:msig}) 
$F(s) = 8.19 + 4.02 \times x$ 
where $x = s -\log(200~\kms)$, in equations
\ref{eqn:vdiffnm} and \ref{eqn:vcumnm}.
As an alternative, we consider the $\sigma$ -- BH mass function
using the \citet{wyithe} predictor  $F(s) = 8.11 + 4.2 \times x
+ 1.6 \times x^2$.  
We illustrate both cumulative BH
mass functions so derived in Figure \ref{fig:bh_mf}.
Choosing the Wyithe predictor only increases the 
predicted BH number density modestly 
near $M_\bullet = 3 \times 10^9 \Msun$.

Inclusion of the cosmic
scatter in the $M_\bullet$ relationships is crucial \citep{yu02, yulu, tundo}.
The total population of galaxies at any given
$L$ or $\sigma$ will host black holes with a range in $M_\bullet.$
The final BH mass functions thus are not a simple ``relabeling'' of
the $L$ or $\sigma$ distributions with black hole mass, but are rather
a convolution of the of these distributions with an assumed distribution
of $M_\bullet$ at constant $L$ or $\sigma.$
This convolution is especially critical
at a large BH mass, where both the galaxy luminosity and dispersion
functions decline rapidly.  
As a result of cosmic scatter, most
of the high mass BHs actually come from ``modest'' galaxies
with unusually large BHs for their luminosities or dispersions,
as compared to the expected contribution of massive black holes
from the most massive galaxies \citep{l07b}.

We emphasize that $\Delta$ above is the scatter about the mean
relation due to cosmic scatter in the relation and {\it not} due to
measurement error.
\citet{tr02} notes that the $\Delta = 0.30$
scatter about their derivation of $M_\bullet(\sigma)$
might be entirely due to
measurement error, leaving no room for cosmic scatter.  In plots of BH
mass functions in Figure \ref{fig:bh_mf} we illustrate illustrate 3
values of the cosmic scatter for the Tremaine and Wyithe results:
$\Delta = 0,\ 0.15 \ {\rm and} \ 0.30$.  This scatter is probably not
larger than $0.30$, but may be considerably smaller.

We compute the the $L$-based black hole mass function by the same
approach, starting with the \citet{blanton} SDSS $g'$ galaxy
luminosity function.  We convert the $g'$ galaxy luminosity function
to $V-$band using $g' =V+0.41$ --- suitable for E galaxies at $z=0$
\citep{fuku}.  Comparison of the Blanton et al.\
galaxy luminosity function with the \citet{pl} BCG survey suggests
that the Blanton work undercounts BCGs.  We argue in Appendix
\ref{sec:bcg} that this may be due to the effects of excessive sky
subtraction on the most luminous galaxies.
To correct for this, we have added an estimate of the space density of BCGs
as a function of V-band luminosity to the Blanton et al.\ sample.
We used the \citet{pl} BCG sample, which is volume-limited,
to construct an estimate of the luminosity function in $R_C,$
transforming it to $V$ using $R_C = V - 0.55$.  Using the combined $dn/dM_V$,
we determine the number of BHs above a specified mass $M_\bullet$ from
equations \ref{eqn:vcumnm} and \ref{eqn:vdiffnm},
where we set $s = M_V$ and $F(s)$ is the right-hand
side of equation (\ref{eqn:ml_hr}) --- the \citet{hr} predictor.
Because \citet{blanton} represented their luminosity function as that observed
at $z=0.1,$ rather than the present epoch, both k-corrections and
corrections for evolutionary fading are required.  These two terms
fortuitously cancel each other:  \citet{blanton} show that their
sample dims by 0.2 mag in $g'$ from $z = 0.1$ to the present, while
the filter k-correction to transform to $z=0$ is $-0.20$ mag.
Lastly, we use $\Delta=0.25$ and 0.50, larger values than were
used for $M_\bullet(\sigma),$ given the larger scatter in $M_\bullet(L).$

The cumulative BH mass functions based on the two different methods
are shown in Figure \ref{fig:bh_mf}.
Near BH masses of $10^8 \Msun$
(luminosities near $L_*$) and lower, the $L$-based function overpredicts
BH numbers by a factor of two and larger.
In part, this is
due to the fact that the $L$-based mass function includes galaxies
that are disk-dominated, thereby overestimating the numbers and
masses of BH at lower masses, since
the $L - M_\bullet$ relationship is
based on bulges and elliptical galaxies.
At higher masses this
correction is negligible since most galaxies are ellipticals
or bulge dominated S0s.
There is also the
possibility that the $L - M_\bullet$ relationship is biased in the
sense that the galaxies with black hole mass determinations have
larger velocity dispersions than the average values for galaxies of
their luminosities (see Figure \ref{fig:ml}).
If so, this would cause the $L$-based mass
function in Figure \ref{fig:bh_mf} to shift to the left, in better
agreement with the $\sigma$-based mass function.

For $M_\bullet > 10^9 \Msun$ the $L$ and $\sigma$ mass functions
diverge.  The $L$ based mass function predicts a local density of
the most massive black holes that is about an order of magnitude
greater than would be inferred from the $M_\bullet-\sigma$
relationship for $M_\bullet > 2 \times 10^9 \Msun$.
This disagreement was foretold in Figure
\ref{fig:bh_pred} --- the present Figure \ref{fig:bh_mf} simply
recasts the disagreement between $M_\bullet-L$ and
$M_\bullet-\sigma$ in terms of the BH mass function.

\subsection{The Black Hole Distribution Function Inferred From QSOs}

There are two different approaches that can be used to infer
the present BH mass function from
quasar counts (specifically from the joint distribution of quasar
numbers as a function of redshift and luminosity).
One line, started by \citet{soltan},
relies on energy conservation.  Under that argument, the energy
emitted by the quasars at any redshift propagates through the
universe declining in co-moving density due to the redshift
of the photons as $(1+z)^{-1}$, and thus behaves exactly like a
background.  Hence the observed quasar flux translates directly to the
total energy emitted given a known redshift distribution of emitters,
and further translates into the total mass accreted by black holes,
given their radiative efficiency.  An alternate approach,
started by \citet{smbl}, is to assume that all quasars go through a phase in
which they accrete at the Eddington limit, followed by a period of
slower or intermittent accretion according to a universal model
dependent on BH mass and time.  This assumption, together with a
continuity argument (the number of BHs at a given mass changes only by
accretion and merging) permits the recovery, not merely of the local
BH density, but also of the local BH mass function.
This second approach
achieves a more detailed result than the So\l tan argument, but at
the expense of additional assumptions.

A third more limited approach, which shares some logic with Small-Blanford,
is to assume simply that the BHs of known mass accrete near the
Eddington limit for some period and to ignore their fainter growth
period --- the so called ``lightbulb model.'' In this model quasars are either
on or off.  Because the 
number of luminous quasars in the universe varies strongly with time,
the model doesn't count BHs directly, rather it counts those that are
accreting.  In what follows, we evaluate the lightbulb model at
$z=2.5$ where the top end of the quasar LF is largest.  We assume that
no BHs above $10^9 \Msun$ are destroyed by mergers since that time, so
the fall-off in the LF is due to a halt in accretion.  At earlier
times the BH mass function may be evolving.  The full-width half
maximum of the bright quasar LF is about $10^9$ yrs.  So our lightbulb
model implicitly assumes that every massive BH accretes for $10^9 f$
yrs, where $f$ is defined as the duty fraction, and then shuts off.
This idealization ignores low mass BHs and low-level accretion.
Another significant limitation is that the model provides no procedure
to identify the mass below which it fails, although that failure is
implicit within the assumptions: some of the lower luminosity quasars
must be high mass BHs accreting at less than their Eddington limit.  A
third limitation is that the model is fundamentally inconsistent: the
assignment of mass corresponding to a quasar luminosity gives the
instantaneous mass of an accreting BH, while the present day mass
function depends on its final mass.  If the quasar is ``on'' for less
than the Salpeter time (for BH mass to e-fold in Eddington-limited
accretion), then the problem is small, but if it is on for much longer
than the Salpeter time it is catastrophic: the present BH mass may be
much larger than that assigned to the quasar.

Nonetheless, the lightbulb model permits a comparison of the quasar LF
with the number density of the most massive BHs in the local universe,
with the duty fraction as a free parameter, under the assumption that
the brightest objects in the quasar luminosity function are accreting
near the Eddington limit.  This approach was used by \citet{rinuc98}
to make a crude estimate of the duty fraction of BH accretion.  We
perform a similar analysis here.  We start with the \citet{rich05}
luminosity function at $z=2.5$, where the bright quasar density is
greatest.  The \citet{rich05} fit reports number densities per
magnitude at an AB absolute magnitude at rest-frame $\lambda =
$1500\AA. We integrate their fitting function from a given luminosity
to infinity
to get a cumulative number of objects brighter than that luminosity,
we apply a bolometric correction of 5 \citep{marc},
and then we deduce a mass from the Eddington limit.
This procedure identifies the number density of
BHs greater than a given mass accreting at a given redshift.  We
compute this cumulative density at redshift $z=2.5$ where the density
of bright quasars is greatest.  We divide this result by the duty
fraction $f.$  We adopt $f=0.03$ based on the extensive work by
\citet{ste02} and \citet{adel05}.  This yields the line labeled ``lightbulb''
in Figure \ref{fig:bh_mf}.

An improvement on both the lightbulb approach and the Small-Blandford
approach is to use a physical model for the accretionary evolution of
the BHs.  One such model is the merger-induced accretion model that has
been explored in detail by \citet{spr05} and \citet{rob06}.  They
simulate the merger of disk+bulge+BH galaxies containing gas using the
GADGET code, treating the BH growth by computing the Eddington-limited
Bondi accretion rate at their smallest resolution element.  They
compute the luminosity of the accreting BH from the accretion rate
under reasonable assumptions about the radiative efficiency.  Their
simulations permit the development of a model \citep{hop06} that
predicts the X-ray background and the zero-redshift BH mass function
from the quasar LF.  We believe the Hopkins model is a profound
advance over simpler analyses.  While it might turn out that
their model does not correspond in detail to the quasar phenomenon,
the approach may have broader utility.  We summarize the salient
points of their model below.

The \citet{hop06} simulations exhibit very complex behavior of
luminosity as a function of time for a given galaxy merger, but the time
spent above a given luminosity turns out to be a universal profile over
a wide range of galaxy or merger parameters, provided it is scaled
appropriately with
the peak luminosity and relic BH mass of the merger.  For their simulations,
the lifetime near a given bolometric luminosity L can be parametrized as
\begin{equation}
dt / d\log L = t_Q^* \exp (-L/L_Q^*),
\label{eqn:dt_dlogl}
\end{equation}
where the timescale $t_Q^*$ (a crude quasar lifetime)
and luminosity scale $L_Q^*$ are
functions of the peak luminosity
$L_p$ as follows.
\begin{equation}
L_Q^* = 0.2 L_p,
\end{equation}
and
\begin{equation}
t_Q^* = 1.37 \left({L_p \over 10^{10} \Lsun} \right) ^{-0.11} {\rm
  Gyr}.
\end{equation}
The final or relic BH mass has a one-to-one correspondence with the
peak luminosity given by
\begin{equation}
M_\bullet = 1.24 \left( {L_p \over 10^{13} \Lsun} \right) ^{-0.11}
             \times M_{Edd}(L_p),
\end{equation}
where $ M_{Edd}(L_p)$ is the mass with an Eddington luminosity of $L_p$.
An ensemble of objects with the same
$L_p$ should have a luminosity distribution proportional to the $dt$
in  equation \ref{eqn:dt_dlogl}.

\citet{hop06} use the model of quasar lifetimes described above
together with a log-normal distribution of quasar birth rate
per unit time to match the
quasar luminosity function.  We use their parameterization
\begin{equation}
\dot n = {\dot n_*  \over \sigma_* \sqrt{2 \pi}} \
        exp \left(- { [\log(L_p/L_*)]^2 \over 2 \sigma_*^2} \right),
\end{equation}
where $\dot n$ is the number of quasars born per unit comoving volume
per unit time.  \citet{hop06} find a good fit to the X-ray and optical
quasar luminosity functions with
\begin{equation}
L_*(z) = L_*(0)  \exp (k \tau),
\end{equation}
where $\tau$ is the dimensionless lookback time $\tau = H_0 \int dt$
and the other parameters are presumed constant.  In what follows we
use their best fit model with $(\log L_*,\ k,\ \log \dot n_*,\ \sigma_*) = $
(9.94, 5.61, -6.29, 0.91), with  $L_*$ in solar units and $\dot n_*$
in comoving $\mpc^{-3} \myr^{-1}$.

We can compute the present day density of quasar relics by integrating
the quasar birthrate over time at any specific mass or $L_p$.
Therefore the cumulative density of BHs above a given mass $M_\bullet$ is
\begin{equation}
  n(M_\bullet) = \int_{L_p(M_\bullet)}^{-\infty}
      \left[ \int_0^{z_{max}} \dot n {dt \over dz} dz \right] d \log L_p.
\label{eqn:bhmf}
\end{equation}
Following \citet{hop06} we set $z_{max} = 3$.  We plot the result of
equation (\ref{eqn:bhmf}) in Figure \ref{fig:bh_mf}.

An important feature of the Hopkins model is that owing to the exponential
distribution of time above a given luminosity (equation
\ref{eqn:dt_dlogl}) the quasar spends only a fraction of its lifetime
accreting near the Eddington rate.  For example, a $10^9 \Msun$ relic
BH had a peak luminosity, $L_p$ of $3.03 \times 10^{13} \Lsun$ and spent the
time $t_e = t_Q^* \log e \times E_1(5/e) = 15 \myr$ above a factor of
$1/e$ of $L_p,$ where $E_1(x) = \int_x^\infty \exp(-u)/u \
du$ is the usual exponential integral. The Hopkins model guarantees that the BH
will accrete enough mass, but not too much, over its lifetime.

Figure \ref{fig:bh_mf} permits us to compare the lightbulb and Hopkins models
with the two relic BH mass functions.  The BH mass functions diverge
at about $10^9 \Msun$.  The dispersion-based predictors predict considerably
fewer BH at above $10^9 \Msun$ than the luminosity-based
predictors.  They are not consistent with the lightbulb model;
consistency with the Hopkins model is possible
with the \citet{wyithe} form of $M_\bullet(\sigma),$
but with the assumption of more cosmic scatter in the $M_\bullet-\sigma$
relationship than is probably realistic.
The luminosity-based mass function is barely consistent
with the lightbulb model, but probably overpredicts the AGN density compared
to the Hopkins model.   We thus cannot make a clear determination between the
dispersion-based and luminosity-based BH mass predictors by comparing
zero redshift BH demographics to quasar demographics; however,
the linear \citet{tr02} $M_\bullet(\sigma)$ relationship is disfavored
in all of the present models to explain the QSO population.

An important caveat is that our calculations have neglected the
effects of dry merging on the most massive galaxies {\it after} the
epoch of QSOs.  Merging might produce high mass black holes as a relatively
recent phenomenon, thereby helping to reconcile the estimates made from
$M_\bullet(L)$ with the QSO population.
Another caveat for both results is the
possibility of super-Eddington accretion among the biggest BHs
\citep{begel06}.  If common, super-Eddington accretion would make it
very hard to make any estimates of the mass function of relic BHs from
quasar LFs.  

\section{Conclusions}

The $M_\bullet-\sigma$ relationship has come to be the ``gold standard"
for predicting black hole masses from galaxy properties due to its
small scatter and its implications for illuminating the co-formation
of galaxies and their nuclear black holes.  At first sight, the
$M_\bullet-L$ relationship might be dismissed as a simple consequence
of the Faber-Jackson relationship.  With $M_\bullet\sim\sigma^4$ and
$L\sim\sigma^4,$ one would expect something like $M_\bullet\sim L.$
The larger scatter in the $M_\bullet-L$ relationship further suggests
that the $M_\bullet-\sigma$ relationship is really more fundamental.
But as galaxy luminosity increases, $\sigma$ levels off and the
basic Faber-Jackson relationship does not appear to hold.
At BCG luminosities there are no direct measurements of $M_\bullet$
and $M_\bullet(\sigma)$ versus $M_\bullet(L)$
present contradictory extrapolations.
The contradiction essentially begs the question, do these exceptionally
luminous galaxies have exceptionally massive black holes?
The $M_\bullet-L$ relationship answers this in the affirmative,
while for the $M_\bullet-\sigma$ relationship to be correct we
must accept the puzzling result that the black holes in BCGs
have relatively modest masses.
But this question leads to a broader issue, namely.
is is there a significant population
of black holes with $M_\bullet$ approaching $10^{10}M_\odot$
in the local universe?

The best way to answer these questions is to attempt to weigh
the BHs in BCGs.  With the advent of LASER-guided adaptive optics-fed
spectrographs on 10m class telescopes, it is now possible to do this.
This paper may therefore be premature.  However, given the high attention to
the $M_\bullet-\sigma$ relationship and its implications for galaxy formation,
we believe that advancing the implications of $M_\bullet-L$ relationship
offers an important alternative view that should not be overlooked.
Lacking hard measures of $M_\bullet$ in the most massive
galaxies, we have marshalled a number of less-direct
arguments that this hypothesis may be favored for the most luminous galaxies.

The first set of arguments is based on the
hypothesis that cores in the most luminous galaxies are created
in a ``core scouring'' process in which a binary BH created in the
merger of two galaxies evacuates stars from the center of the newly-merged
product.  There presently is little observational support for the
creation of binary black holes in mergers, but abundant theoretical
work shows that realistic cores can be created by binary black holes,
and the prevalence of nuclear black holes in galaxies overall strongly
argues that such binaries must be created as a natural consequence of mergers.
If so, the physical scale of cores, which we have parameterized
as $r_\gamma$ may be an independent witness of $M_\bullet,$ and thus
use the large cores in BCGs as an indicator of their black hole masses.

Based on central structural parameters derived from {\it HST} observations,
we find that the large cores in BCGs are commensurate with their
high luminosities, while $\sigma$ is a poor predictor of $r_\gamma$
for $r_\gamma>300$ pc.  The scatter in the $r_\gamma-L$ relationship
is much smaller than that in the $r_\gamma-\sigma$ relationship, again
implying that $L$ and core scale are more closely related.
The observed $r_\gamma-M_\bullet$ relationship for 11 core galaxies
with directly determined black hole masses has large scatter,
but appears to be more consistent with the $M_\bullet-L$
rather than the $M_\bullet-\sigma$ relationship.  Lastly. the core
masses in BCGs are over an order of magnitude larger than the black
hole masses estimated from $M_\bullet(\sigma),$ but are only a few
times larger than those estimated from $M_\bullet(L);$ making such
large cores with the smaller $\sigma$-based black hole masses would
be a strong challenge for the core-scouring hypothesis of core
formation.

The second set of arguments comes from considering
theoretical arguments concerning whether or not $L$
rather than $\sigma$ should predict $M_\bullet$ in BCGs.  The
favored origin of BCGs is that they are the remnants of dissipationless
purely-stellar mergers of less-luminous elliptical galaxies, augmented
by ongoing galactic cannibalism of elliptical galaxies in the rich
environments at the center of galaxy clusters.
The plateau in the $L-\sigma$ relationship plus
but the steeper $L-R_e$ relationship at high
galaxy luminosity presented here strongly favor this formation scenario.
The luminosity of a BCG is the sum of the luminosities of its progenitors.
Similarly, setting aside the possible ejection of nuclear black holes in the
final stages of a merger, the final nuclear black hole mass should
be the sum of the progenitor black holes.  Stated
more directly, the ratio $M_\bullet/L$ should be largely invariant
over dissipationless mergers, leading to $M_\bullet\sim L$ at the
high end of the galaxy LF.  In contrast, $\sigma$ appears to be
nearly constant over these mergers.  In effect, even if a relationship
between $M_\bullet$ and $\sigma$ were set up in the initial stages
of galaxy formation, it might not survive in a dissipationless merging
hierarchy.

A final argument comes from attempting to infer the $z=0$ space density
of the remnant black holes associated with the most luminous
QSOs seen at $z\sim2.$  As noted in the Introduction,
the properties of the broad-line regions
in the most luminous QSOs argue that they are powered by black
holes with $M_\bullet\sim10^{10}M_\odot.$  The heating of the
intra-cluster medium in the richest galaxy clusters may also demand
that some black holes in BCGs approach this mass.
The critical part of this analysis is understanding how to correct
the QSO space density for QSO luminosity evolution.  The remnant black
holes last forever, but the QSOs represent only those BHs
made visible during an epoch
of high mass-accretion, which presumably lasts only a small fraction
of the age of the universe.
We used the \citet{hop06} simulations to estimate the QSO lifetimes.
The resulting shape and space
density of the bright end of the QSO LF falls between the higher
space density of the most massive black holes
implied by $M_\bullet(L)$ and those implied by $M_\bullet(\sigma),$
while the simple ``lightbulb'' model of QSO duty cycles favors the
$M_\bullet(L)$ relation.  This treatment is sensitive to the assumed
amount of cosmic scatter in both $M_\bullet$ relationships; however, it
appears difficult for the log-linear
$M_\bullet-\sigma$ relationship to explain
the the observed space densities of the most luminous QSOs without
assuming that its cosmic scatter is larger than is likely to be the case.

\acknowledgments

This research was supported in part by several grants provided through STScI
associated with GO programs 5512, 6099, 6587, 7388, 8591, 9106, and 9107.
Our team meetings were generously hosted
by the National Optical Astronomy Observatory, the Observatories
of the Carnegie Institution of Washington, the Aspen Center
for Physics, the Leiden Observatory, and the University of California, Santa
Cruz Center for Adaptive Optics.  We thank Mariangela
Bernardi, Megan Donahue, Tom Jarrett, Michael Strauss, and Mark Voit
for useful discussions.  We thank Qingjuan Yu for kindly reminding us
to include cosmic scatter in calculation of the black hole mass
functions.  We also thank the referee for many
excellent suggestions.

\clearpage
\appendix

\section{The Luminosities of BCGs and Comparison to SDSS Magnitudes
\label{sec:bcg}}

Our analysis depends critically on
the accuracy of the absolute luminosities of the brightest galaxies in the
sample, such as BCGs.  This is underscored by the bright-end disagreements
of our $L-\sigma$ relationship and galaxy luminosity
function with those based on Sloan Digital Sky Survey (SDSS)
data (\citealt{bern2} and \citealt{blanton}, respectively).
We thus describe the derivation of our BCG luminosities, and
compare them to magnitudes based on the SDSS for BCGs in common.
We conclude that the SDSS BCG magnitudes are strongly affected by
sky subtraction errors, causing them to be biassed to significantly fainter
values.

The BCGs in the present sample come from the \citet{laine} {\it HST}
BCG study.  This program, in turn, was based on the volume-limited \cite{pl} BCG
sample, which provides ground-based profiles and aperture
photometry.\footnote{\citet{pl} did not actually publish their BCG surface
brightness profiles, but they were presented graphically in the BCG
profile analysis of \citet{graham}.}
As outlined in $\S\ref{sec:gallum}$, we derive apparent
magnitudes of the BCGs by fitting the classic $r^{1/4}$ form to the inner
portions of the brightness profiles, where the inner limit
of the fit was set to avoid seeing and the outer
limit was specified to avoid portions of the profile
that appeared to rise {\it above} the $r^{1/4}$ fit.  \citet{graham} showed that
the BCG profiles could be described by single-component \citet{sersic} forms,
but ones that often had index $n>4.$  The apparent magnitudes,
which were derived by integrating the $r^{1/4}$-law over radius, thus if
anything are {\it underestimates} of the total BCG fluxes.
An alternative to this procedure would be to integrate the S\' ersic
forms, however, as is shown in \citet{graham}, the S\' ersic $R_e$ and $n$
values are closely coupled, such that large $n$ is matched with large
$R_e,$  The implied total magnitude strongly diverges as $n$ increases,
and must essentially be regarded as unphysical extrapolations because
the derived $R_e$ is typically well outside the actual radial domain of
the surface brightness profile for large $n;$ this is not true for $n=4.$
A contrasting treatment that occurs in much of the literature is
based on the presumption that BCG must be completely well described by
$r^{1/4}$-laws (in contrast to other giant elliptical galaxies,
which also have $n>4$), and that
S\' ersic $n>4$ is really the signature of an intracluster light
component that must be subtracted.  We conclude that an unambiguous
procedure to derive total BCG luminosity does not presently exist.
Our procedure of deriving magnitudes from just the inner portion of
the profile that is well described by an $r^{1/4}$-law again should
give a sensible lower limit to the total luminosity.

The high luminosity end of the \citet{blanton} luminosity function falls
well below the BCG space densities measured by \citet{pl}.
The \citet{bern2} $L-\sigma$ relationship shows no plateau at its
bright limit.  These discrepancies would both be explained if the total
magnitudes for BCGs in the SDSS database are significantly underestimated.
We checked this hypothesis by examining the 25 \citet{pl} BCGs
present in the SDSS DR4 database.\footnote{Two
additional BCGs were mis-identified in the SDSS database as stars.}
In detail, we compared the total $R_C$ apparent magnitude in \citet{laine}
against the SDSS $r$ ``model magnitude,'' (which in almost all cases
is the most luminous total magnitude provided by the SDSS database)
transformed by $R_C=r-0.31.$  The results
are shown in Figure \ref{fig:sdss_comp} as a function of effective radius
(based on our fits).  A strong systematic trend is evident such
that larger galaxies have greater offsets between
the two total magnitudes.  The median $r-0.31-R_C$ value is 0.54 mag and
rises to 1.57 mag for the NGC 6166 (the BCG in A2199).
As an additional check, we also compared the SDSS $r$ magnitudes against
the maximum $R_C$ aperture magnitude published by \citet{pl}.  The
maximum aperture radius was not defined in any rigorous way and does
not correspond to any fixed fraction of the total galaxy flux,
but the magnitude is a model-independent integration of all
the flux within the published radius.
The median difference between the SDSS model $r$ magnitudes (transformed
to $R_C$) and the maximum aperture magnitude is 0.24 mag
and rises to values over a full magnitude for the largest galaxies.
This demonstrates directly that the SDSS model magnitudes for the
galaxies in question cannot be regarded as total magnitudes.

Conversations with a number of experienced users of the SDSS database for
bright galaxies warned us that the SDSS pipeline measured sky levels on
angular scales too small to accommodate bright nearby BCGs of the sort observed
by \citet{pl}, and indeed the results shown in Figure \ref{fig:sdss_comp}
strongly suggest that a sky-subtraction error affects the SDSS magnitudes.
As a check, we plot the SDSS $r$
surface brightness profiles against the \citet{pl} profiles for
three of the BCGs with the largest magnitude differences
in Figure \ref{fig:profile}.
The SDSS profiles agree well at small radii but all fall
below the \citet{pl} profiles at large radii,
consistent with large sky subtraction errors.

The large SDSS sky-subtraction errors for bright galaxies
may have important implications for the \citet{bern2}
and \citet{blanton} studies, but exactly how is not clear.
Both SDSS studies are based on galaxy
samples with higher mean redshifts than the \citet{pl} sample. Their
BCGs should be angularly smaller and thus be less
vulnerable to sky subtraction errors.  Typical BCGs in the SDSS sample
are listed in the \citet{miller} sample of galaxy clusters identified
from SDSS galaxy catalogues.  Figure \ref{fig:bcg_lf} shows a histogram of
SDSS model $r$ magnitudes (converted to $M_V$) for the BCGs identified by 
\citet{miller} compared to a histogram of all \citet{pl} BCGs with
$M_V$ based on their total $R_C$ magnitudes.  There is a clear offset
between the two samples, with the \citet{pl} BCGs appearing to be
typically one magnitude brighter than the \citet{miller} BCGs.
However, a histogram
of SDSS $r$ magnitudes for the 25 \citet{pl} BCGs in common with SDSS
agrees well with the \citet{miller} BCG histogram, yet these
are the magnitudes shown to in be error.  We conclude that the total magnitudes
of nearby SDSS BCGs are wrong.

\section{The Use of Catalogued 2MASS XSC Apparent Luminosities of BCGs
\label{sec:2mass}}

After the first version of this paper was posted on astro-ph,
\citet{bat} presented a $M_\bullet-L$ relationship for BCGs based
on apparent magnitudes extracted from the 2MASS Extended Source Catalogue (XSC)
\citep{xsc, lga}.
The implied NIR luminosity differential between BCGs and other giant
elliptical galaxies is greatly reduced from that of the present work.
As a result, the plateau in the $L-\sigma$
relationship presented here is greatly reduced in the NIR and the conflict
between $M_\bullet(\sigma)$ and $M_\bullet(L)$ is thus resolved.
\cite{bat} further suggest that the relatively higher luminosities
inferred from the optical photometry may imply that the envelopes
of the BCGs are extremely blue.

We have not conducted a complete comparison of the present photometry
with that provided by the 2MASS XSC, but a spot check of a few systems
makes it clear that the 2MASS imagery from which the catalogue
magnitudes were derived is extremely shallow compared to that of
\citet{pl}, which is the source of the $R$ band optical photometry
(transformed to $V$) used in this paper.  The most likely explanation
for the difference between the present and \citet{bat} results is that the
2MASS images are simply not deep enough to obtain accurate total luminosities
of the BCGs, at least as represented by the automatic reductions used
to generate the XSC magnitudes.

Figure \ref{fig:a0779i} shows the $J$ band 2MASS image of NGC 2832, the
BCG in A0779 compared to the central portion of the $R$ band image
obtained by \citet{pl}.  The $J$ sky level is 15.67 magnitudes per
square arcsecond versus the $R$ band sky level of 20.90.  Accounting
for the $R-J=1.68$ color in the center of the galaxy implies that the
$J$ band has a sky level effectively $26\times$ brighter.  Further, the
2MASS image is a 7.8s exposure obtained with a 1.3m telescope as compared
to the 200s $R$ image obtained with a 2.1m telescope \citep{pl}.
The $J$ image is thus considerably shallower than the $R$ image as is
evident in Figure \ref{fig:a0779i}.
The galaxy envelope in the $J$ band image
disappears into the noise at radii at which it is still clearly
present in $R.$  This problem is exacerbated in the $H$ and $K$ bands,
which have yet brighter sky levels.

Despite the much brighter $J$ sky level,
the $J$ and $R$ band profiles of the A0779 BCG are consistent,
as is evident in Figure \ref{fig:warp_2mass}, which compares
the $R$ profiles measured by \citet{pl} to $J$ profiles derived
by us from the 2MASS archive images for the three BCGs shown earlier
in Figure \ref{fig:profile}.
The final $J$ band isophote shown for A0779 falls fully $\sim6.8$ magnitudes
below the sky level, but is still in
agreement with the $R$ band profile within the large error bars, which
represent a 0.004 magnitude error in the 2MASS $J$ sky level.

An $r^{1/4}$ law fitted to the $J$ band profile of A0779 yields $m_J=9.04,$
only 0.08 magnitudes fainter
than $m_J=8.96,$ estimated by subtracting the $R-J=1.68$ color of
the central isophotes from $m_R$ obtained from the \citet{pl}
photometry.  These values are in poor agreement with the XSC isophotal
($J_{m_{k20fe}}$), and extrapolated ($J_{m_{ext}}$) $m_J$ values of
9.78 and 9.67, respectively, however.
The isophotal radius, $r_{k20fe},$ is $53\asec3,$
well interior to the limits of the surface photometry shown
in Figure \ref{fig:warp_2mass}.
The extrapolated magnitude is based on a \Ser fit to a surface
brightness profile generated by the XSC pipeline. {\it However,
even for giant elliptical galaxies the XSC \Ser index is limited to}
$n<1.5$ (\citealt{lga}; Jarrett, private communication). 
The XSC calculation of total magnitudes thus assumes that the galaxies
essentially have exponential profiles,
rather than the $r^{1/4}$ form standard for giant elliptical galaxies.
The XSC pipeline gives $n=1.17$ for A0779, for example, while
\citet{graham} find $n>10$ based on the \citet{pl} photometry.
An exponential cutoff explains both the small difference between
the XSC isophotal and total magnitudes, as well as the large deficit
of both magnitudes compared to a total magnitude estimated from
an $r^{1/4}$ law.

A similar pattern is seen for the two other BCGs shown in
Figure \ref{fig:warp_2mass}.  For A2052, the BCG has $J_{m_{k20fe}}=10.92,$
corresponding to $r_{k20fe}=38\asec4,$
and $J_{m_{ext}}=10.60,$ while an $r^{1/4}$ fit to the surface photometry
recovered from the 2MASS archive image gives $m_J=9.58,$ a full magnitude
brighter and only 0.21 magnitudes dimmer than the $m_J=9.36$ inferred
from the $R$ band imagery with $R-J=1.79.$
For A2199, the BCG (NGC 6166) has $J_{m_{k20fe}}=10.51,$
corresponding to $r_{k20fe}=50\asec0,$
and $J_{m_{ext}}=10.41,$ generated from a \Ser fit with $n=1.18$
(Jarrett, private communication); \citet{graham} find $n=6.9.$
An $r^{1/4}$ fit to the surface photometry
recovered from the 2MASS archive image gives $m_J=9.66,$ 0.75 magnitudes
brighter, but 0.41 magnitudes dimmer than $m_J=9.20$ inferred
from the $R$ band imagery with $R-J=1.82.$

To summarize, our fits to the 2MASS $J$ images for the three BCGs
shown give $m_J$ values markedly brighter than the XSC apparent magnitudes,
but that are much more consistent with the $R$ photometry of \citet{pl}.
As a check, we found that the aperture photometry in \citet{pl}
interpolated to $r_{k20fe}$ was in excellent agreement with
the $J_{m_{k20fe}}$ values, assuming the $R-J$ values derived by
comparing the inner portions of the surface photometry profiles;
however, $r_{k20fe}$ is well interior to the limits of the
optical photometry, thus the XSC isophotal magnitudes
cannot be regarded as a total apparent magnitudes.
The nearly exponential profile form used by the XSC pipeline explains why
$J_{m_{ext}}$ is only modestly brighter than $J_{m_{k20fe}}$
for the three galaxies shown, but is much dimmer than our fits,
which assume \Ser $n=4,$ the standard value for giant elliptical galaxies.

This analysis thus leads us to conclude that the extremely blue BCG
envelopes and the relatively modest NIR BCG luminosities advanced
by \citet{bat} are artifacts.
We do note that the $J$ surface photometry does fall
below the $R$ photometry at large radii in all three BCGs, but this
occurs for $J$ isophotes well within the noise and well below the even the
errors in the sky levels.  Any color gradients implied by
the profiles presented in Figure \ref{fig:warp_2mass} are not
significant, and in any case are
considerably smaller than would be implied by direct comparison
of the XSC apparent magnitudes to the parameters given in this paper.

\section{The Selection of $r_\gamma$ over $r_b$ as the Core
 Scale\label{sec:rgam2}}

The break-radius parameter in the Nuker-law has been used directly
as the indicator of physical core scale in earlier work by our group.
\citet{f97} showed that $r_b$ is correlated with $M_V$ for core galaxies.
The scatter about the $r_b-M_V$ relationship is large, however.  With
the present much larger sample of galaxies, we were able to
conduct extensive searches for other parameters that might reduce the scatter,
with the goal of better understanding how
core structure varies with galaxy properties.

Several plots and parameter fits were conducted to search for
correlations between residuals of the $r_b-L$ and $r_b-\sigma$
relationships with the Nuker profile shape parameters $\alpha,~\beta,$ 
and $\gamma.$
We also tried local values of $\gamma,$ $\gamma',$ measured at constant
fraction of $r_b$ interior to $r_b$
($\gamma'(fr_b)$ with $f<1$).
A correlation emerged between the Nuker-law
outer profile slope $\beta$ and the residuals in the $r_b-L$
relationship, as is shown in Figure \ref{fig:rbmv_bet}.
Larger than average cores correspond to galaxies with large $\beta$ and
vice versa.  While one might be tempted to use this relationship to yield
some sort of ``$\beta$-corrected'' break radii, the form of the correlation
suggests a simpler, more useful approach.  In the Nuker law,
$r_b$ marks the maximum curvature of the profile in logarithmic
coordinates, but also the location where the profile slope
reaches the average of $\gamma$ and $\beta.$  Since the range
of $\gamma$ for core galaxies is greatly restricted compared to
that of $\beta,$ this effectively means that $r_b$ will fall
at steeper portions of the profile for steeper $\beta.$
This suggests that some sort of ``cusp radius,'' $r_\gamma,$ a radius
at which the profile reaches a nominal slope value, $\gamma',$ would provide
a better description of the core scale than $r_b.$  The exact definition
of $r_\gamma$ we use is given in equation (\ref{eqn:rgam}).
\citet{carollo} had already suggested use of $r_\gamma$ with $\gamma'=1/2,$
the choice that we adopt here.

\begin{figure}
\plotone{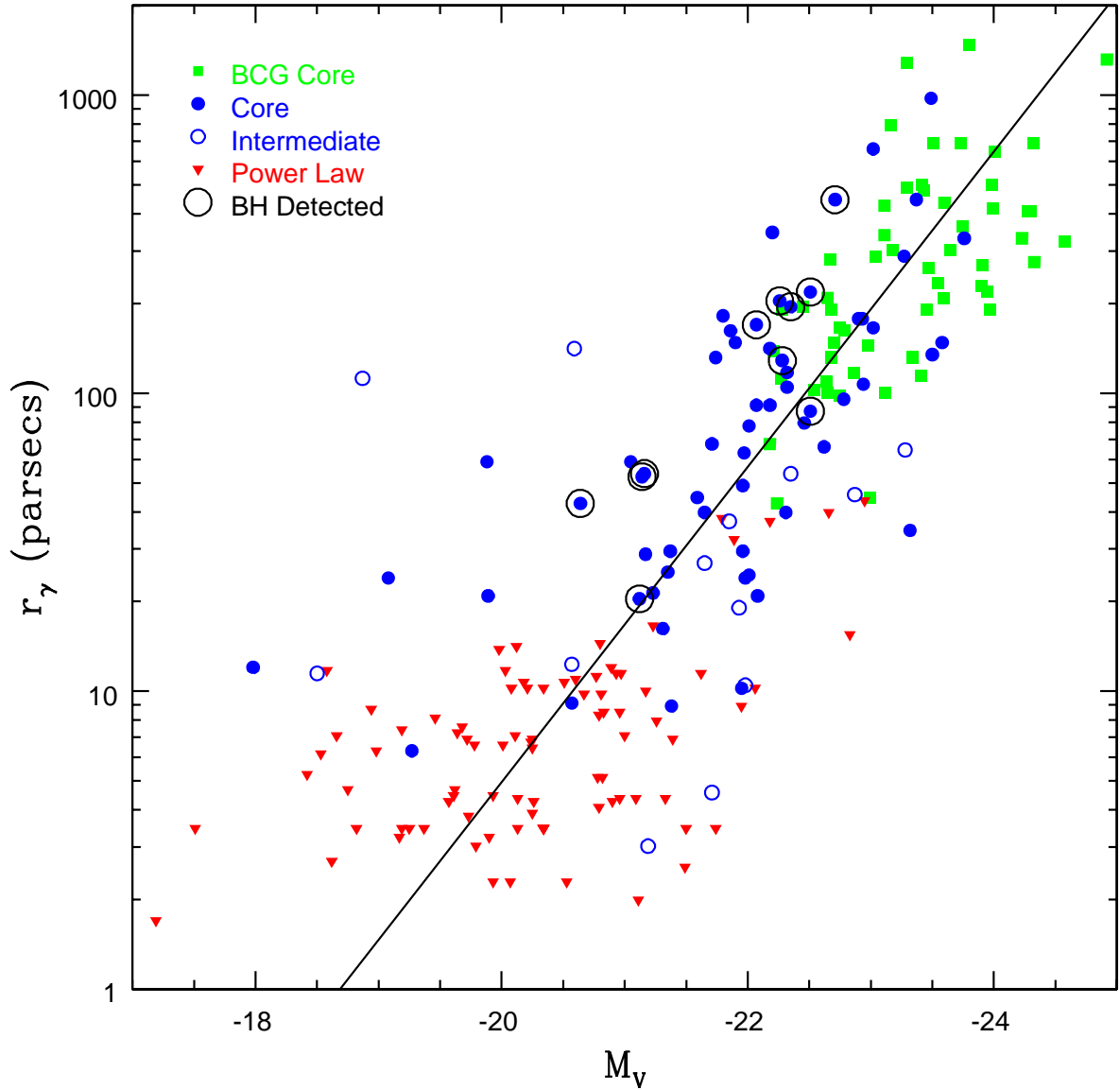}
\caption{Cusp radius, $r_\gamma,$ is plotted as a function of total
galaxy luminosity.  Power-law galaxies are plotted as $\nabla$ to
indicate that their $r_\gamma$ values are as upper limits.
The solid line shows the best-fit relationship between $r_\gamma$
and $M_V$ for core galaxies (equation \ref{eqn:r_mv}).
The figure shows that $r_\gamma$
varies nearly linearly with $L.$
If $M_\bullet\sim L_V^{1.4}$ as in equation (\ref{eqn:ml_hr}), the observed
empirical relationship between $r_\gamma$ and $L$ (solid line)
implies that $r_\gamma\sim M_\bullet^{0.96\pm0.09}.$}
Note that for core galaxies the range in $r_\gamma$ at any given $L$
is smaller than it is at a given $\sigma.$
\label{fig:mvrc}
\end{figure}

\begin{figure}
\plotone{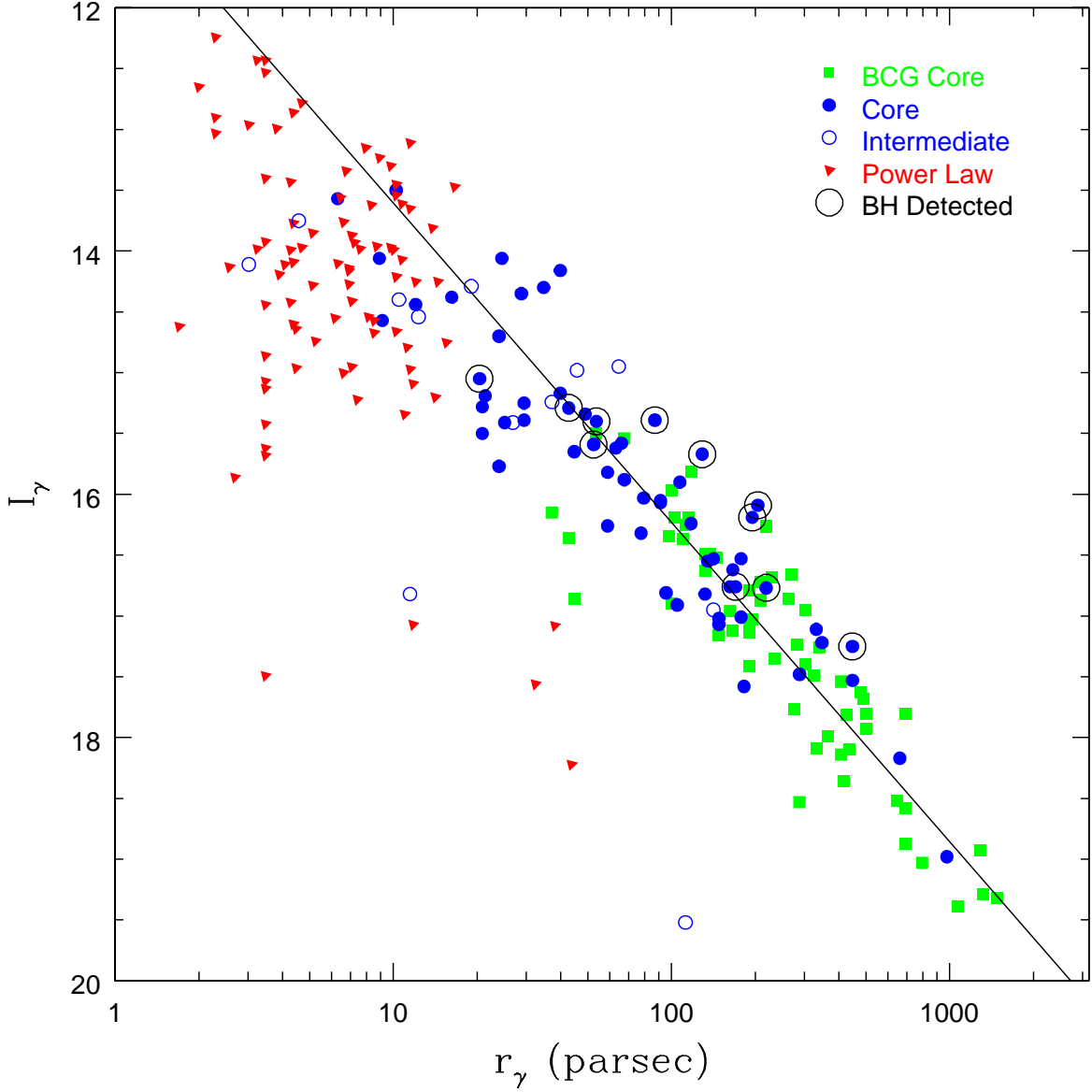}
\caption{Cusp brightness, $\mu_\gamma,$ is plotted as a function of
cusp radius, $r_\gamma.$
The power-law triangles are rotated and shown as arrows
to reflect that the points
are only upper limits for both $I_\gamma$ and $r_\gamma.$
The tight relationship between $I_\gamma$ and $r_\gamma$
(equation \ref{eqn:ic_rc}) means
that either can serve for the other in the context of relating
core structure to $M_\bullet,$ $L$ or $\sigma.$}
\label{fig:rc_ic}
\end{figure}

\begin{figure}
\plotone{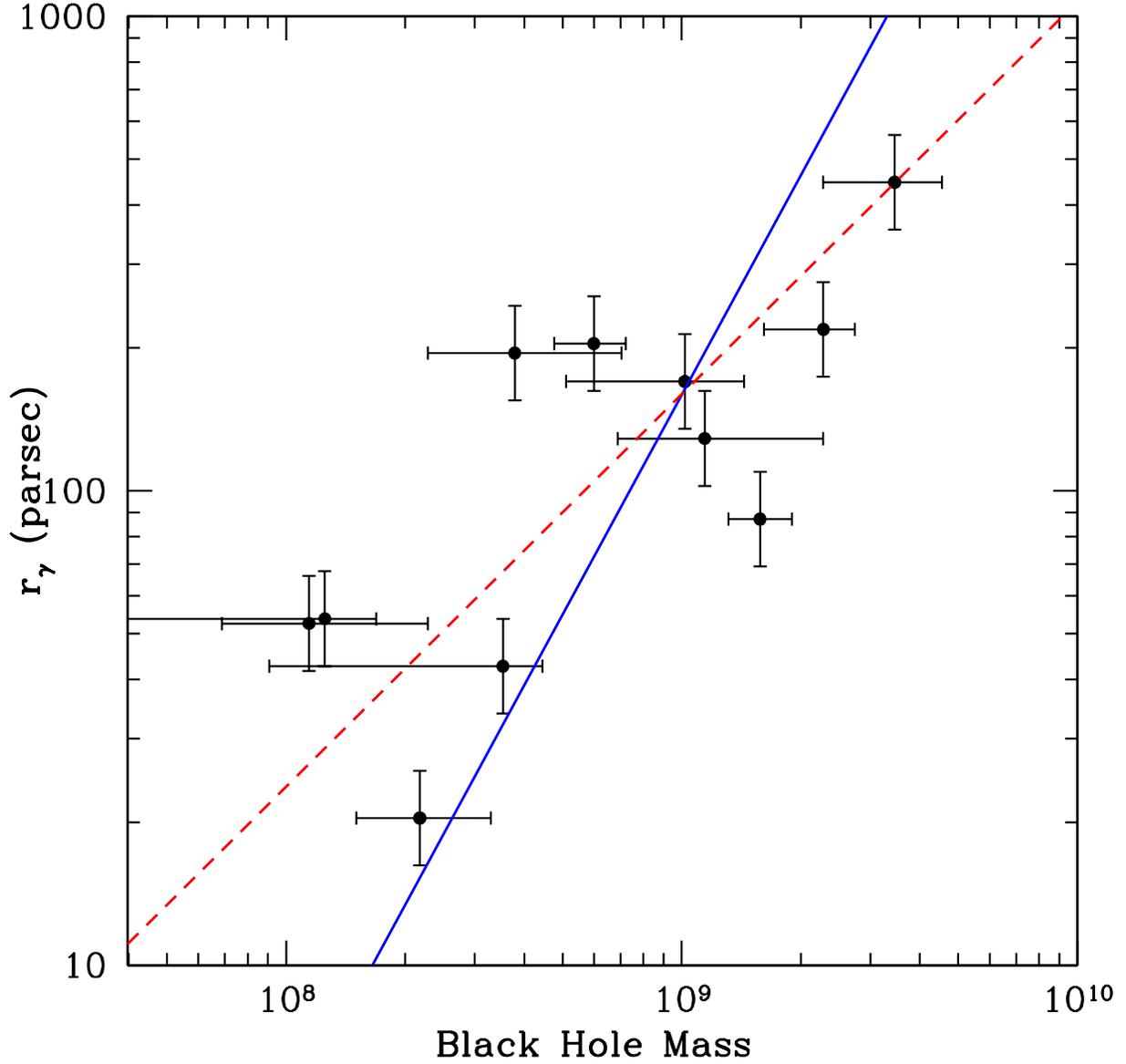}
\caption{Black hole mass versus core size, $r_\gamma,$ for the 11 core galaxies
that have $M_\bullet$ measurements.  The red line is the symmetric fit
between $M_\bullet$ and $r_\gamma$ provided by equation (\ref{eqn:mbh_r_fit}),
while the blue line gives the fit presented in equation (\ref{eqn:mbh_r_ind}),
which assumes that $r_\gamma$ is the independent variable.}
\label{fig:bh_rc}
\end{figure}

\begin{figure}
\plotone{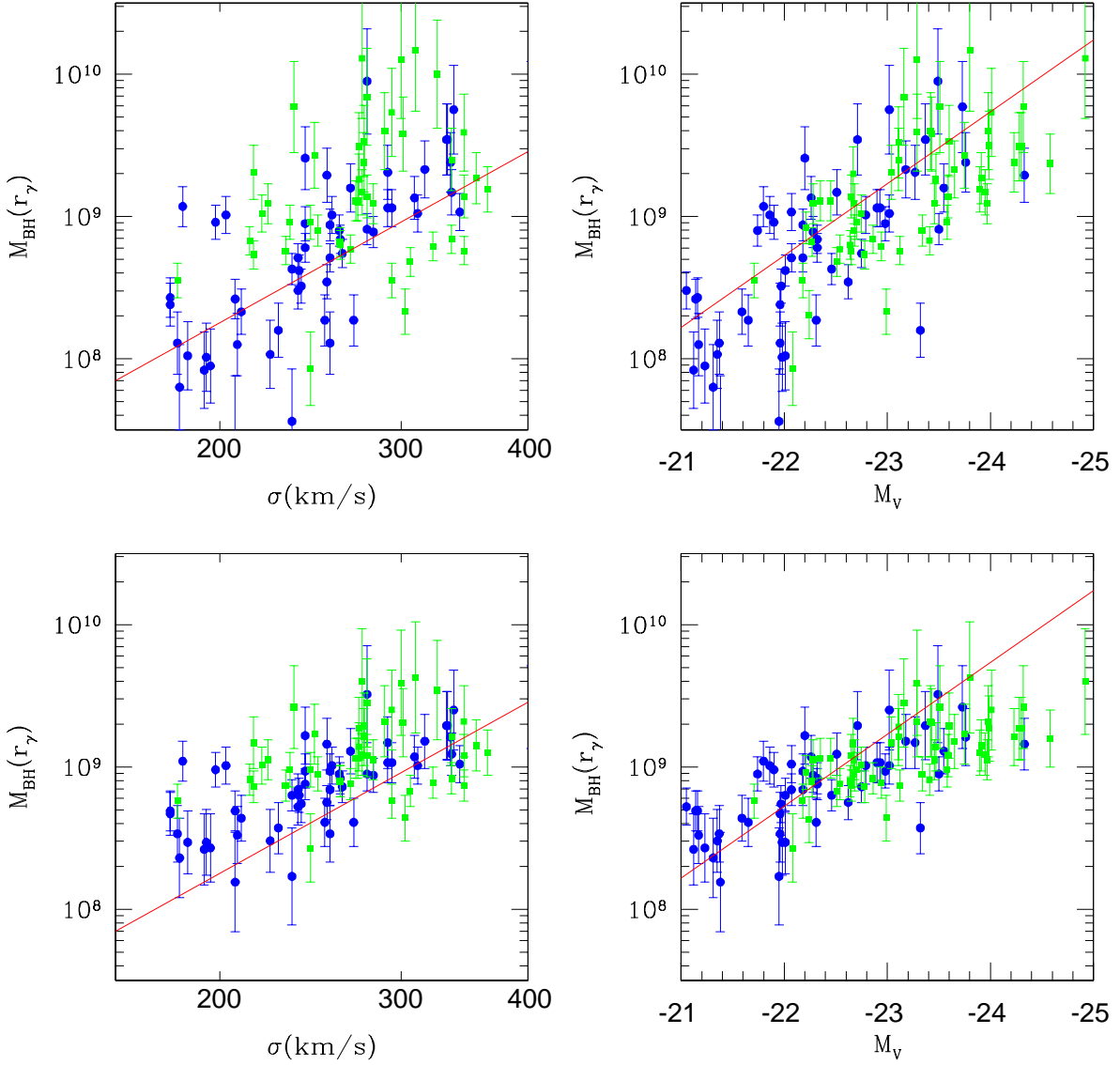}
\caption{The four panels plot $M_\bullet$ predicted from $r_\gamma$
as function of $\sigma$ and $M_V$ for core galaxies, where green
symbols are BCGs, and blue symbols are the remaining core galaxies.
The red lines give $M_\bullet$ values predicted either
from the $M_\bullet-\sigma$ relationship (equation \ref{eqn:msig})
or the $M_\bullet-L$ relationship (equation \ref{eqn:ml_hr}).
The upper panels give $M_\bullet$ predicted by $r_\gamma$
through equation (\ref{eqn:mbh_r_fit}), which was derived by a symmetrical
fit to the points in Figure \ref{fig:bh_rc}.
The bottom panels, in contrast, use equation (\ref{eqn:mbh_r_ind}),
which was derived assuming $r_\gamma$ as the independent variable.
Both equations typically predict $M_\bullet$ in excess of
the predictions from the $M_\bullet-\sigma$ relationship.}
\label{fig:bh_rgpred}
\end{figure}

\begin{figure}
\plotone{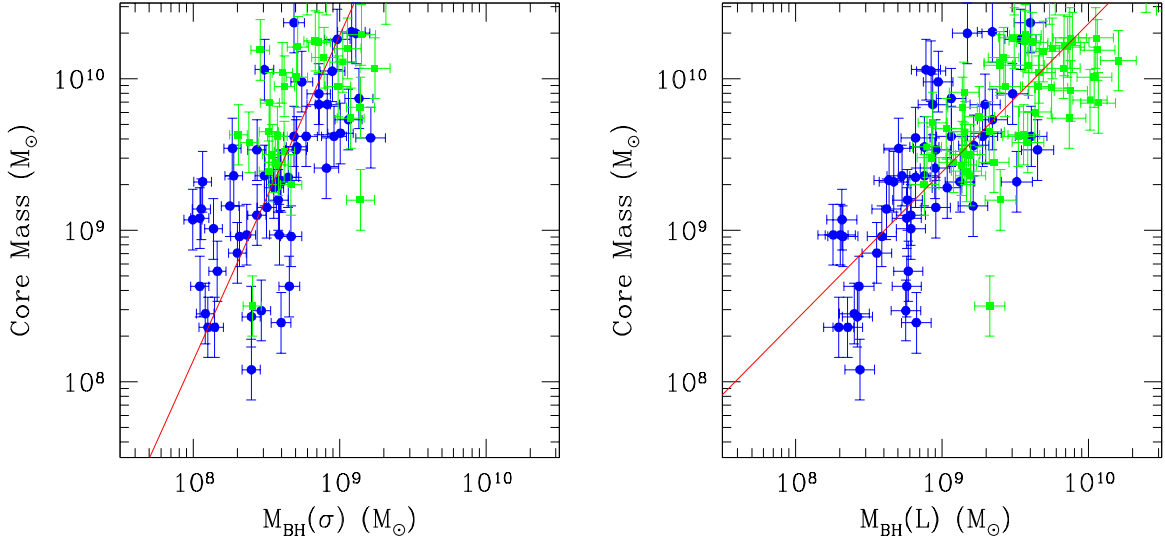}
\caption{The two panels plot core mass $M_\gamma$
as function of $M_\bullet(\sigma)$ and $M_\bullet(L)$ for core galaxies,
where green symbols are BCGs, and blue symbols are the remaining core galaxies.
The red lines give the mean $M_\gamma-M_\bullet$ relationships inferred
by combining either the $M_\bullet-\sigma$ relationship
(equation \ref{eqn:msig}) with the $M_\gamma-\sigma$
(equation \ref{eqn:mc_sig}),
or the $M_\bullet-L$ relationship (equation \ref{eqn:ml_hr})
with the $M_\gamma-L$ relationship (equation \ref{eqn:mc_l}).}
\label{fig:corem}
\end{figure}

\begin{figure}
\plotone{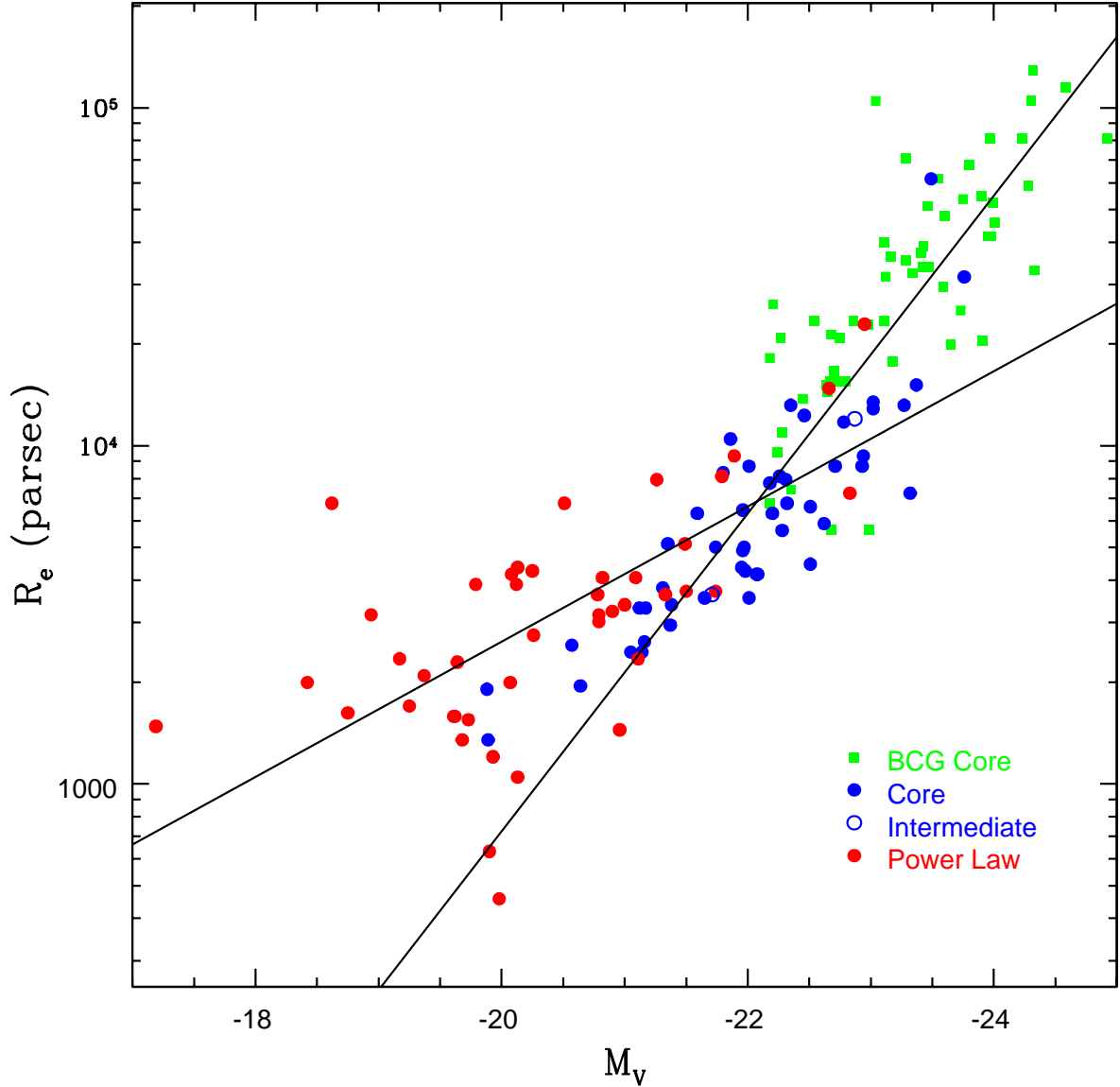}
\caption{Effective radius as a function of luminosity for the galaxy sample.
The steepening of the $R_e-L$ relationship sets in at $M_V<-22,$
similar to the luminosity at which the velocity dispersion
starts to ``plateau" in Figure \ref{fig:mv_sig}.
We attribute this to a progressive change in the character of ``dry mergers''
at higher galaxy masses.
The shallow line is defined by a fit to power-law galaxies only
(equation \ref{eqn:re_pl}), while the steep line is a fit to
core galaxies with $M_V<-21$ (equation \ref{eqn:re_core}).
Power-law galaxies and core galaxies, however, have similar $R_e$ at
$M_V\sim-21$ where the transition between the two forms takes place.}
\label{fig:mvre}
\end{figure}

\begin{figure}
\centering
\includegraphics[scale=0.6,angle=-90]{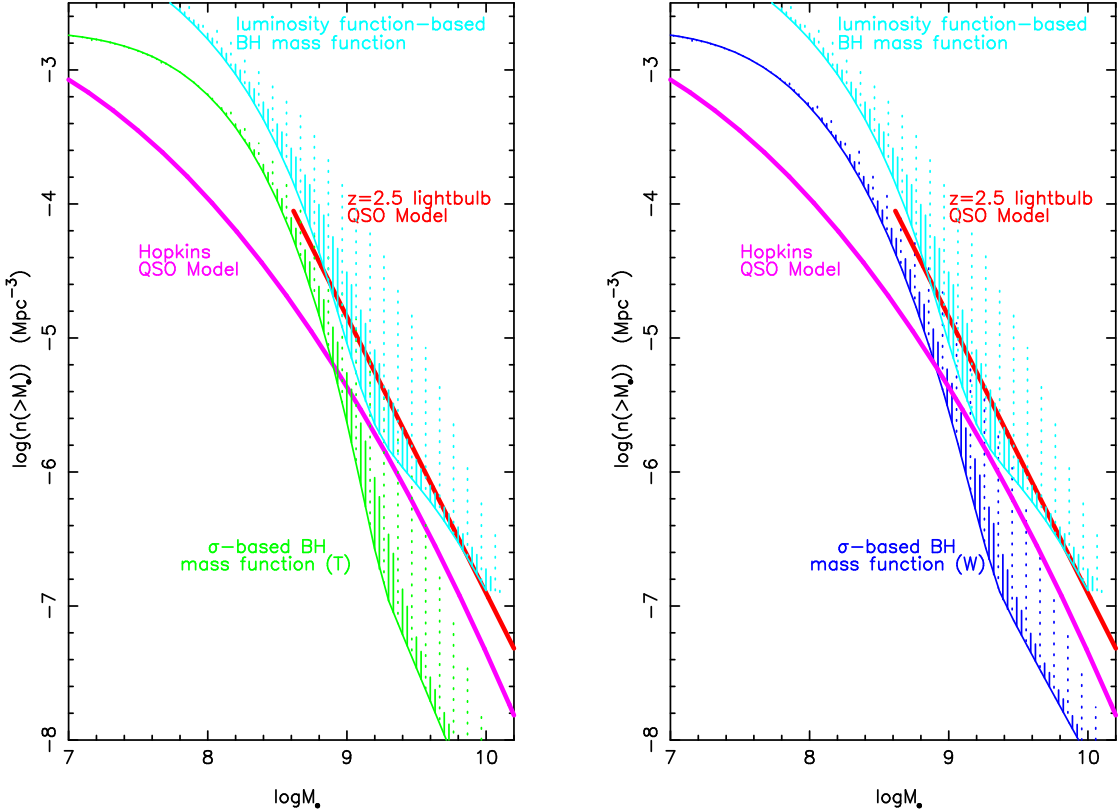}
\caption{The log cumulative density of black holes above a given mass versus
$\log M_\bullet$ for different mass functions.  The red
curve in both panels is derived from a lightbulb model for
quasars applied to the  \citet{rich05}   best-fit quasar
luminosity function from SDSS, evaluated at $z=2.5$.
The model assumes they radiate at their Eddington Luminosity
with a duty fraction $f = 0.03$ (see text).
The pink curve
in both panels is the BH mass function produced by the
\citet{hop06} model for quasar, described in detail in the text.
The cyan curve in both panels is the BH mass function obtained
by augmenting the \citet{blanton} best-fit \citet{schechter}
luminosity function for SDSS galaxies with the \citet{pl}
brightest cluster galaxies and calibrated by the \citet{hr} relation.
The solid and dotted lines above the curve show the effect of cosmic scatter of
0.25 and 0.50 (respectively)
about the mean relation (derived by \citet{hr}) between
BH mass and galaxy mass.
The green curve (in the left panel) is a BH mass function predicted from the
SDSS velocity dispersion function \citep{sheth}
augmented by the \citet{bern3} high-dispersion sample and
calibrated by the  \citet{tr02}
$M_\bullet(\sigma)$ predictor for zero  cosmic scatter.  The solid and
dotted lines above the curve show the effect of a cosmic scatter of
0.15 and 0.30 in the decimal log of the BH mass about the mean
Tremaine relation.  The dark blue curve in the right panel illustrates
the same dispersion function calibrated instead by the \citet{wyithe}
$M_\bullet(\sigma)$ predictor. }
\label{fig:bh_mf}
\end{figure}

\begin{figure}
\plotone{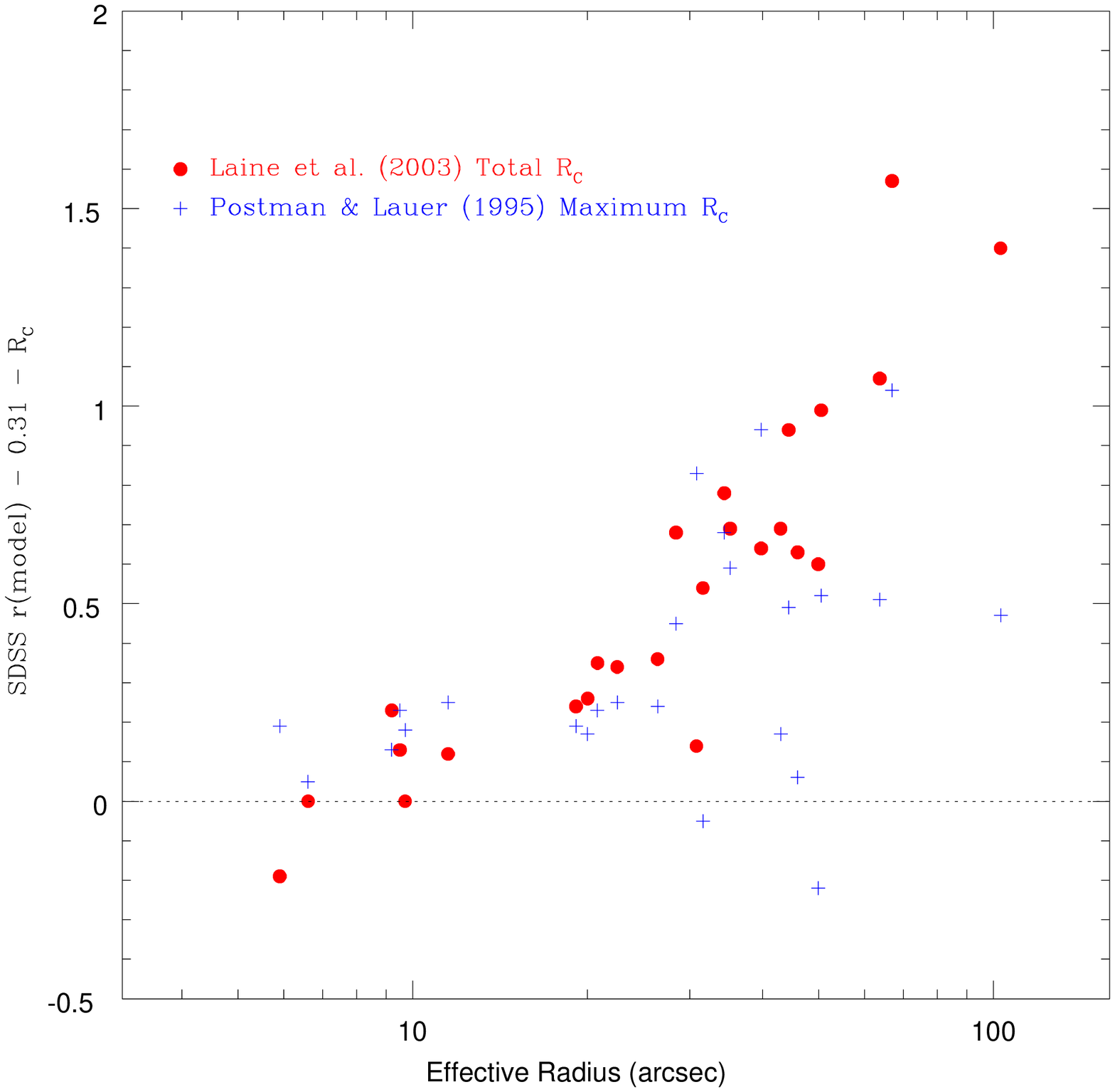}
\caption{SDSS $r$ ``model magnitudes'' are compared to two separate
$R_C$ magnitudes for the \citet{pl} BCGs inc common, where
the SDSS values are transformed assuming $R_C=r-0.31.$  Red points
show the difference between SDSS model $r$ magnitudes and the \citet{laine}
$R_C$ total BCG luminosity versus effective radius, $R_e,$ which is
derived from $r^{1/4}$ fits to the \citet{pl} surface brightness profiles.
The blue points are the same exercise, but with the \citet{pl}
maximum-aperture magnitudes used instead.
Clearly, the larger a BCG is, the more the present total luminosity
disagrees with the SDSS value.
The maximum-aperture magnitudes
are not intended to be interpreted as a total magnitude, but provide
a model-independent {\it lower} limit on its value. Since the
maximum-aperture magnitudes are brighter than the SDSS magnitudes for
most BCGs, this clearly shows that the SDSS values cannot be regarded
as total luminosities.}
\label{fig:sdss_comp}
\end{figure}

\begin{figure}
\plotone{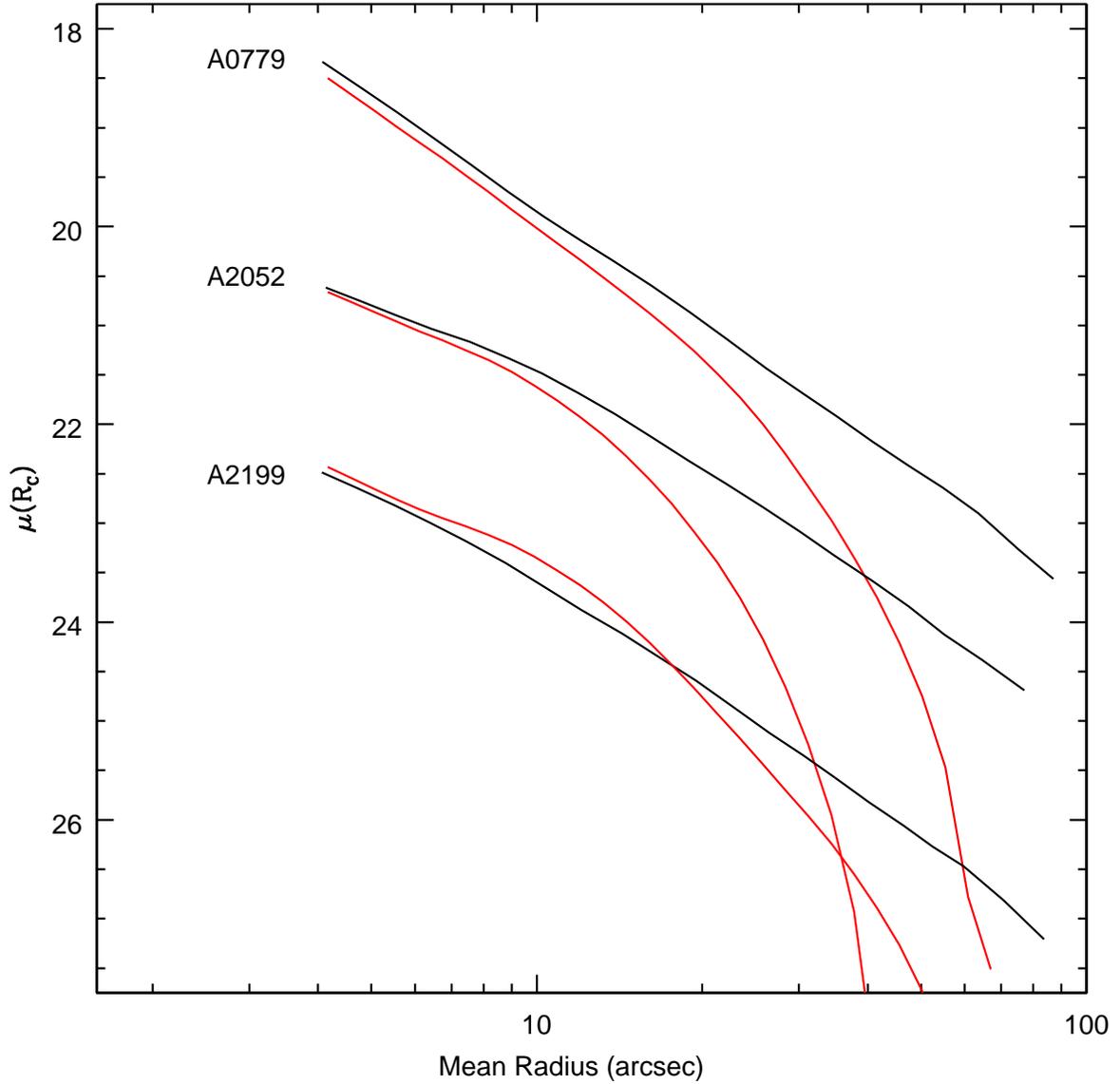}
\caption{Surface brightness profiles from \citet{pl} (black) are
compared to ``profmean" SDSS profiles (red) for three BCGs.  The
SDSS $r$ band magnitudes are transformed assuming $R_C=r-0.31.$
The SDSS profiles all fall below the \citet{pl} profiles at large
radii, consistent with excessive sky subtraction by the SDSS pipeline.}
\label{fig:profile}
\end{figure}

\begin{figure}
\plotone{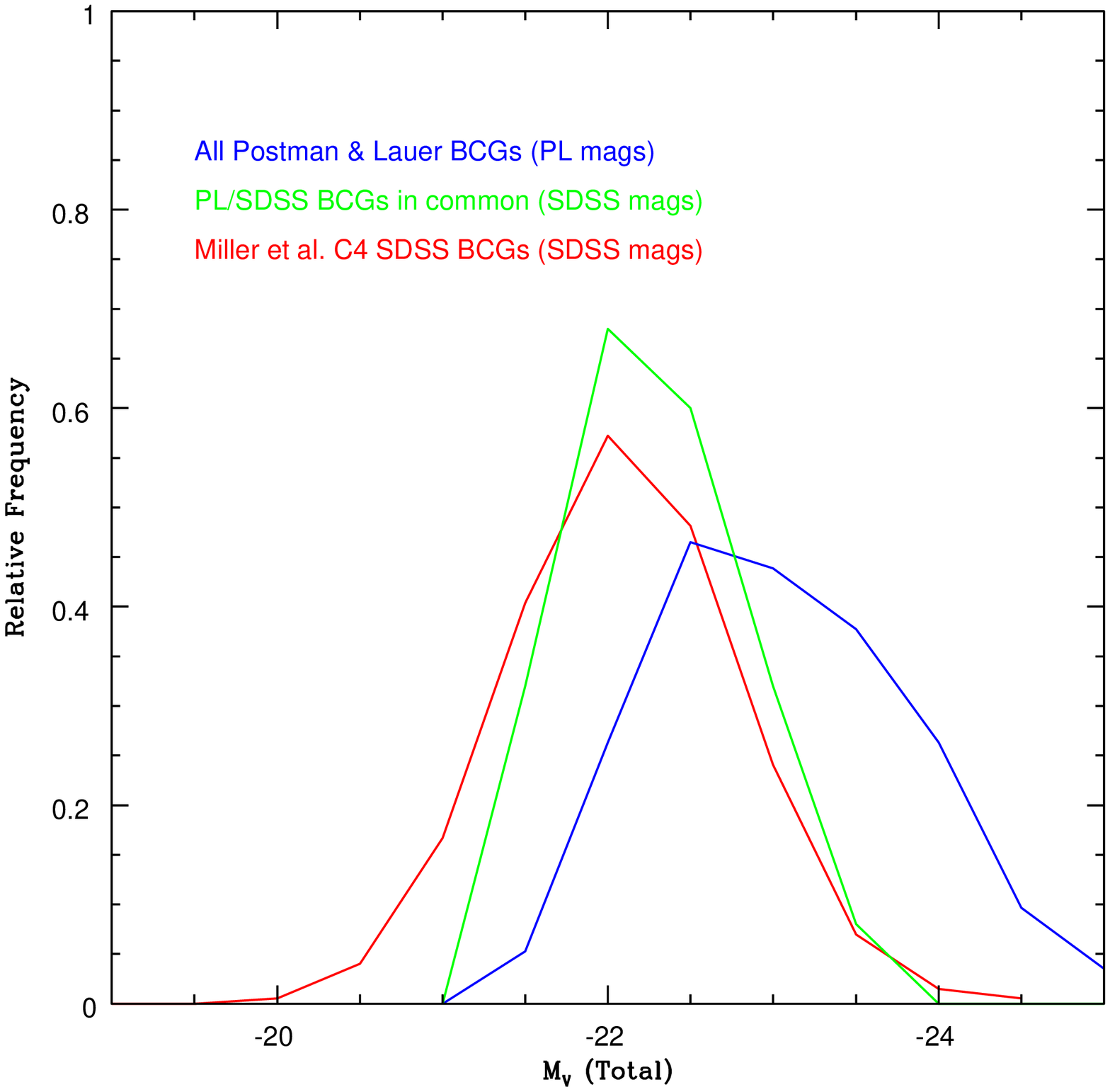}
\caption{Histograms of estimated total $M_V$ are shown for three BCG samples.
Magnitudes for the \citet{pl} volume-limited BCG sample (blue) are based
on $r^{1/4}$ fits to $R_C$-band surface photometry.  The \citet{miller}
BCG sample is based on SDSS model $r$ magnitudes, and has typical luminosities
one magnitude smaller than the \citet{pl} sample.
The histogram of the subset of \citet{pl} BCGs observed by SDSS (green)
agrees well with the \citet{miller} sample
when SDSS $r$ model magnitudes are used instead to estimate
total $M_V,$ yet we argue that
these magnitudes are strongly affected by excessive sky subtraction.
This concordance implies that the C4 BCGs, are also likely to have had
their total luminosities under-estimated.}
\label{fig:bcg_lf}
\end{figure}

\begin{figure}
\plotone{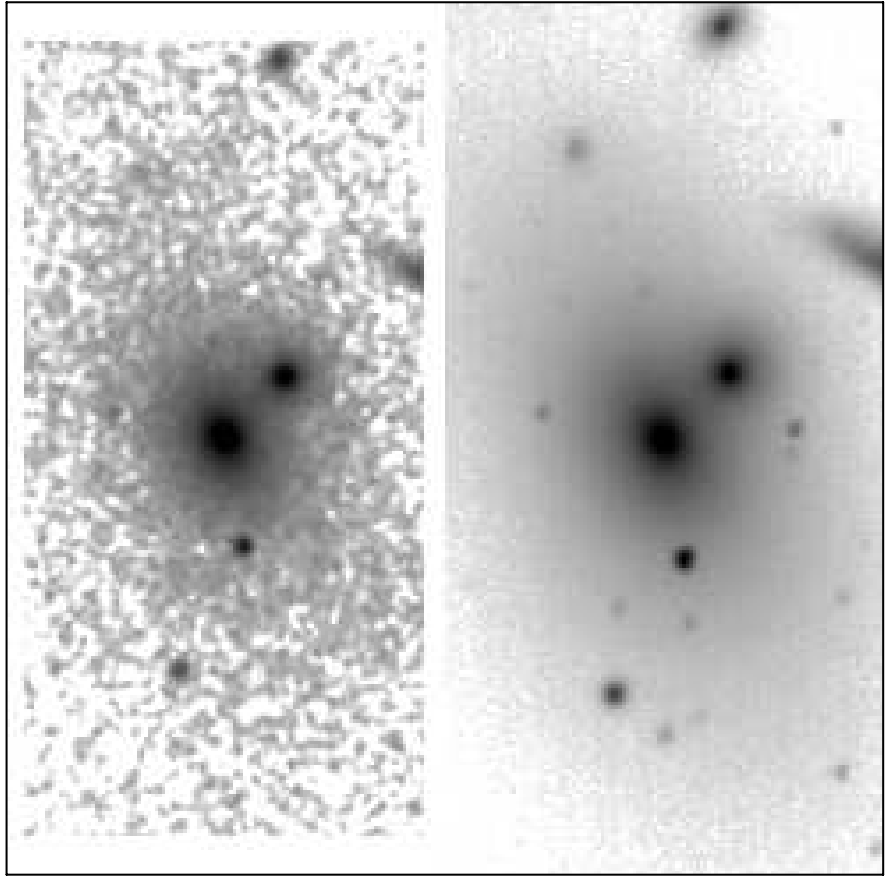}
\caption{The archived 2MASS $J$ band image of NGC 2832, the BCG in A0779,
is compared to a portion of the $R$ band image used by \citet{pl}
to derive surface photometry profile shown in the next figure.  The
stretch has been set to be the same for both images.  The $R$ band
image has been binned to a $0\asec91$ pixel scale to roughly match the
$1\asec0$ scale of the $J$ band image.  The sky level of the 2MASS
image is effectively $26\times$ brighter in $J,$ taking the observed
$R-J$ color of the galaxy into account. The $J$ band image is clearly
considerably shallower than the $R$ band image, and the envelope of the galaxy
disappears into the noise at radii where it is still clearly present
in the $R$ band.}
\label{fig:a0779i}
\end{figure}

\begin{figure}
\plotone{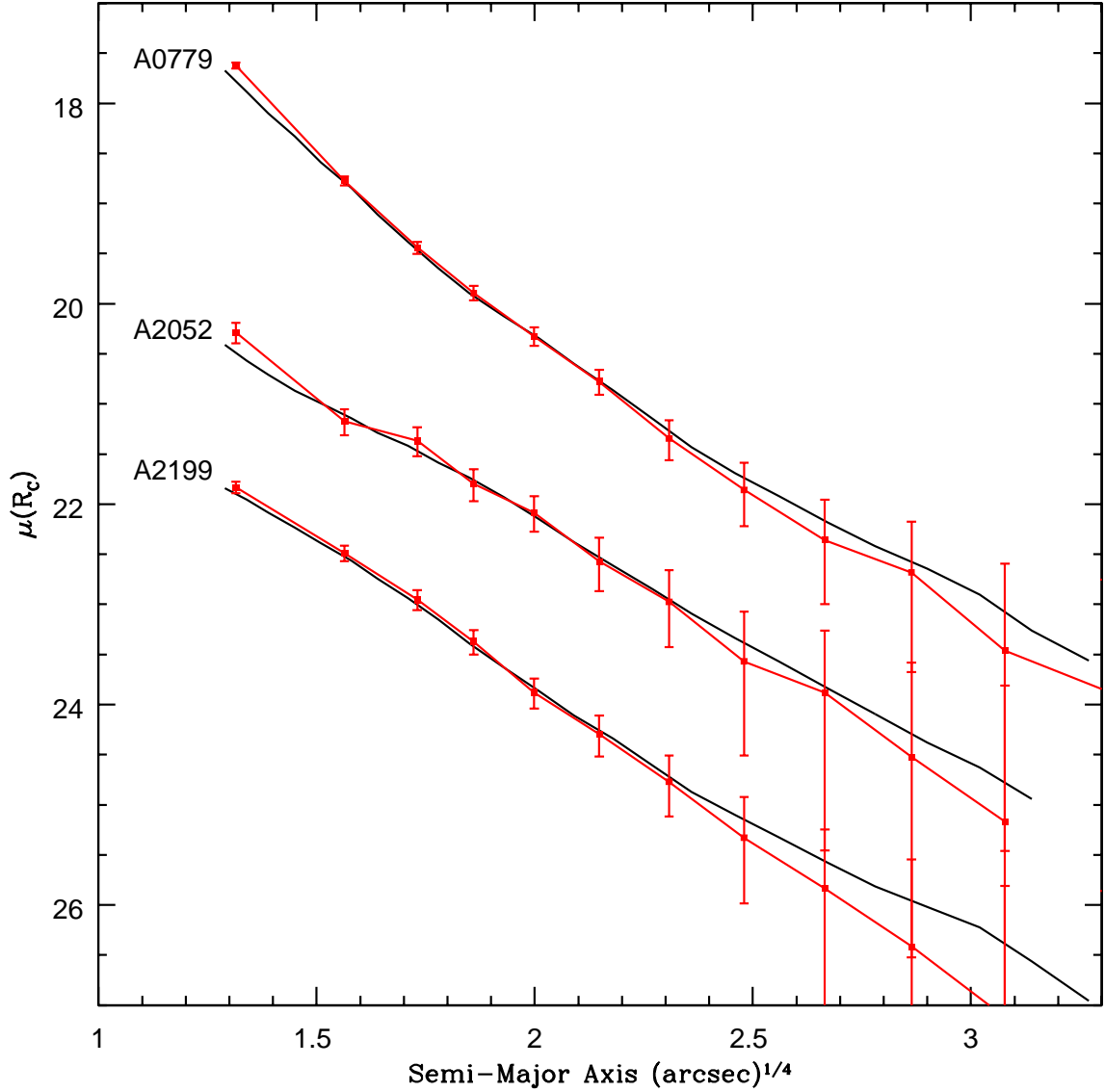}
\caption{Major axis surface $J$ photometry profiles (red) derived from
2MASS $J$ band archive images are compared to the $R$ profiles derived
by \citet{pl} (black) for the three BCGs shown in Figure \ref{fig:profile}.
The $\sim0.4\%$\ error in the 2MASS sky levels gives the large error bars.
The last $J$ band isophotes fall $\sim7$ magnitudes
below the sky, and thus are less significant than
the errors in the sky levels.
The $R$ and $J$ profiles agree within the errors, and total luminosities
estimated by $r^{1/4}$-law fits to the even the $J$ band profiles
are considerably larger than the 2MASS XSC apparent luminosities.}
\label{fig:warp_2mass}
\end{figure}

\begin{figure}
\plotone{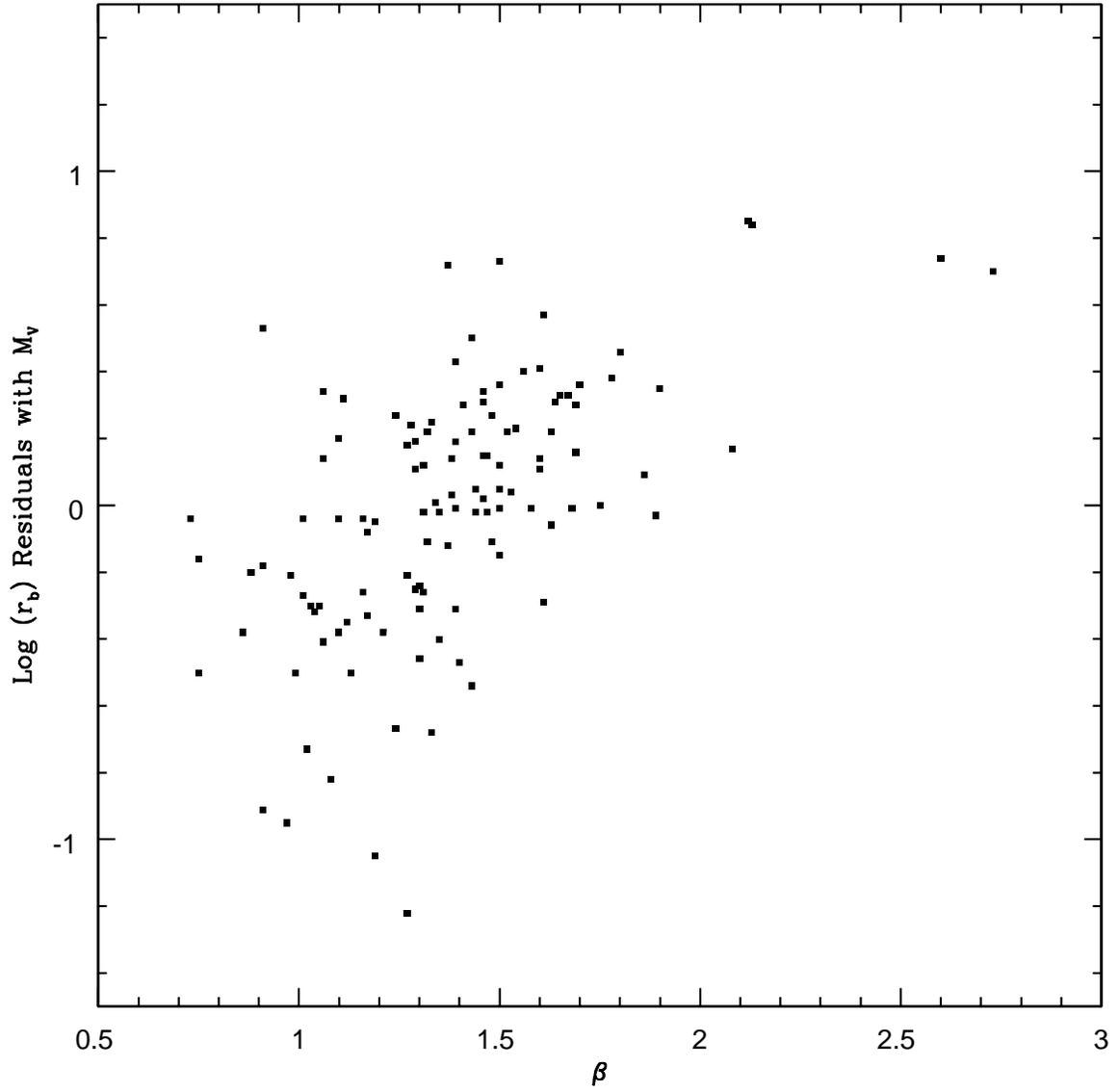}
\caption{Residuals about the mean $r_b-M_V$ relationship are plotted
for core galaxies as a function of the logarithmic envelope slope $\beta.$
At any given $M_V,$ excessively large cores (positive residuals) correspond
to higher $\beta$ (steeper envelopes)
while excessively small cores correspond to small $\beta.$}
\label{fig:rbmv_bet}
\end{figure}

\begin{figure}
\centering
\includegraphics[scale=0.60]{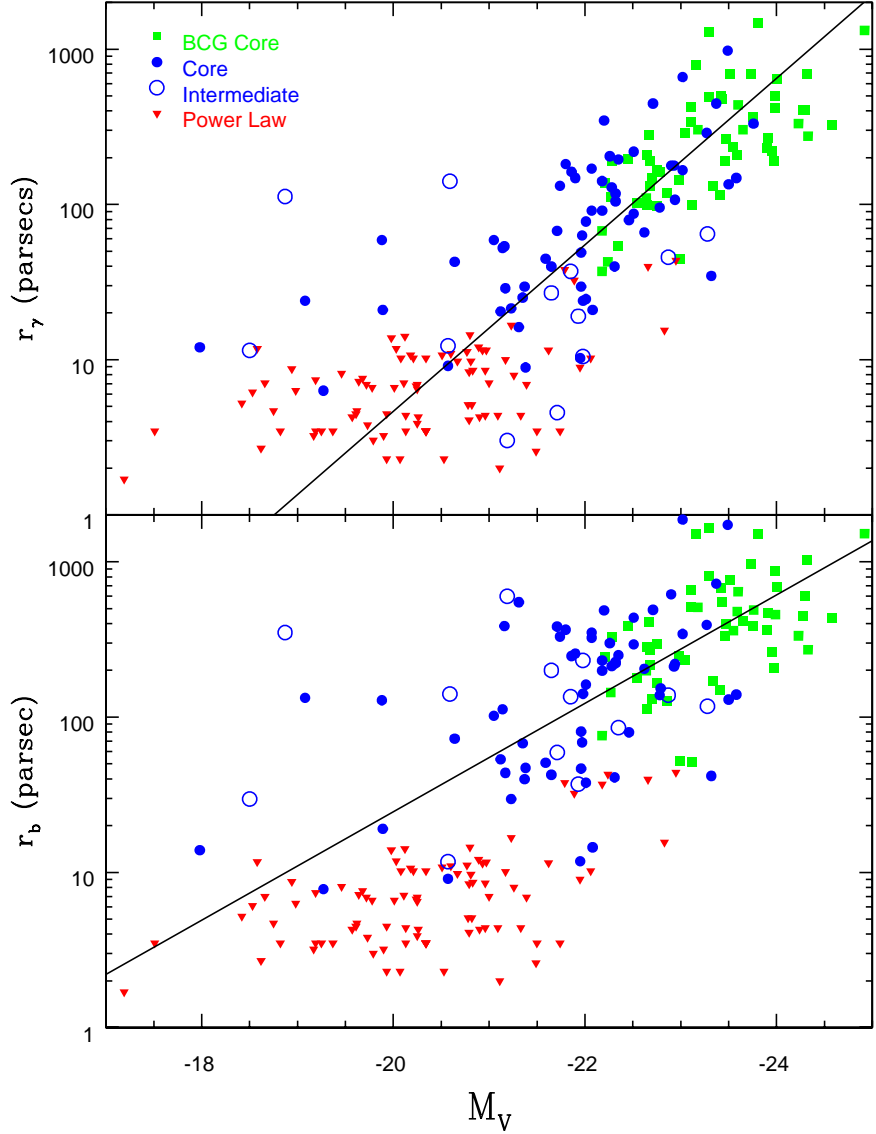}
\caption{The $r_\gamma-L$ and $r_b-L$ relationships for core
galaxies are compared.
The lines are the mean relationships for $M_V<-21.$  The $r_\gamma-L$
relationship has smaller scatter for core galaxies with $M_V<-21.$
Power-law galaxies are plotted the same in both panels, with
upper limits on $r_\gamma$ to be the same as upper limits on $r_b.$}
\label{fig:rgrb}
\end{figure}

Figure \ref{fig:rgrb} compares the
$r_\gamma-L$ relationship for $\gamma'=1/2$ (shown
earlier in Figure \ref{fig:mvrc}) and the more traditional
$r_b-L$ relationship.  A symmetrical fit to the latter yields
\begin{equation}
\log(r_b)=(-0.71\pm0.04)\left(M_V+21.64\right)/2.5+(2.05\pm0.02),
\label{eqn:mvrb}
\end{equation}
which can be compared to the $r_\gamma-L$ relation given
in equation (\ref{eqn:r_mv}).
For most galaxies $\beta>1,$ so in general,
$r_\gamma<r_b;$ on average $r_\gamma\sim0.8r_b$ for $\gamma'=1/2.$
We measured the rms scatter in the $r_\gamma-L$ relationship for core
galaxies with $M_V<-21$ as a function of $\gamma'$ over the range
$0.4<\gamma'<0.7.$
A broad minimum in the scatter of 0.31 in $\log r_\gamma$
occurs at $\gamma'\sim1/2.$
This is significantly smaller than the 0.38 scatter in
$\log r_b$ for the $r_b-L$ relationship.
The reduced scatter in the $r_\gamma-L$ relationship as compared to
the $r_b-L$ relationship is clearly evident in Figure \ref{fig:rgrb}.
We have thus chosen to use $r_\gamma$ over $r_b$ as the core scale.

Evaluation of $r_\gamma$ at $\gamma'=1/2$ also leads to a clean
separation of core and power-law galaxies.  Since power-law galaxies
have $\gamma'>0.5$ at the {\it HST} resolution limit, they
are naturally excluded from the analysis.
The upper limits for $r_\gamma$ for power-law galaxies
however have the same physical values as their $r_b$ limits;
since $r_\gamma<r_b$ for core galaxies, this may create a false
impression that the upper limits of $r_\gamma$ for power-law galaxies
are more in line with the typical $r_\gamma$ values of core galaxies.

The issue of whether or not intermediate galaxies can be included
within the class of core galaxies, or should be treated separately,
unfortunately depends on which relationship is being considered.
As is evident in Figure \ref{fig:rgrb},
the intermediate galaxies with $M_V\leq-21$ appear to be
evenly distributed about the the mean $r_b-L$ relationship;
their mean residual about the relationship is $-0.06\pm0.15$
in $\log(r_b).$ In contrast, the same galaxies fall preferentially to
the compact side of the $r_\gamma-L$ relationship, now having
a mean $\log(r_\gamma)$ residual of $-0.60\pm0.11;$
given this systematic offset, we conclude that the intermediate
galaxies should be treated separately from the core galaxies.

\clearpage

\begin{deluxetable}{lccrcccl}
\tablecolumns{8}
\tablewidth{0pt}
\tablecaption{Galaxy Parameters}
\tablehead{\colhead{ }&\colhead{ }&\colhead{ }&\colhead{$\sigma$}&\colhead{ }
&\colhead{$r_\gamma$}&\colhead{$\mu_\gamma$}&\colhead{ } \\
\colhead{Galaxy}&\colhead{Morph}&\colhead{$M_V$}&\colhead{(km/s)}
&\colhead{P}&\colhead{$\log({\rm pc})$}&\colhead{(V-Band)}&\colhead{Alt-ID}}
\startdata
NGC 0404       &S$0-$&$-17.19$&     38&$\backslash$&0.23&14.62& \\          
NGC 0474       &S0   &$-20.12$&    164&$\backslash$&1.15&15.20& \\          
NGC 0507       &S0   &$-23.02$&    311&$\cap$&2.22&16.62& \\                
NGC 0524       &S0+  &$-21.85$&    253&$\wedge$&1.57&15.24& \\              
NGC 0545       &BCG  &$-22.98$&    242&$\cap$&2.16&16.52&A0194-M1 \\        
NGC 0584       &E    &$-21.38$&    207&$\cap$&0.95&14.06& \\                
NGC 0596       &E    &$-20.90$&    152&$\backslash$&0.63&13.99& \\          
NGC 0720       &E    &$-22.20$&    242&$\cap$&2.54&17.22& \\                
NGC 0741       &E    &$-23.27$&    291&$\cap$&2.46&17.48& \\                
NGC 0821       &E    &$-21.71$&    200&$\wedge$&0.66&13.75& \\              
NGC 0910       &BCG  &$-22.79$&    257&$\cap$&2.21&16.96&A0347-M1 \\       
NGC 1016       &E    &$-22.90$&    294&$\cap$&2.25&17.01& \\                
NGC 1023       &S$0-$&$-20.53$&    204&$\backslash$&0.36&12.90& \\          
NGC 1052       &E    &$-21.17$&    208&$\cap$&1.46&14.35& \\                
NGC 1172       &E    &$-20.13$&    112&$\backslash$&0.64&14.09& \\          
NGC 1316       &E    &$-23.32$&    228&$\cap$&1.54&14.30& \\                
NGC 1331       &E    &$-18.58$&     58&$\backslash$&1.07&17.07& \\         
NGC 1351       &S$0-$&$-20.08$&    137&$\backslash$&1.01&14.21& \\          
NGC 1374       &E    &$-20.57$&    185&$\cap$&0.96&14.57& \\                
NGC 1399       &E    &$-22.07$&    342&$\cap$&2.23&16.76& \\                
NGC 1400       &S$0-$&$-21.37$&    256&$\cap$&1.47&15.25& \\                
NGC 1426       &E    &$-20.78$&    153&$\backslash$&0.71&14.28& \\          
NGC 1427       &E    &$-20.79$&    162&$\backslash$&0.61&14.11& \\          
NGC 1439       &E    &$-20.82$&    154&$\backslash$&0.71&13.85& \\          
NGC 1500       &BCG  &$-22.75$&    263&$\cap$&1.99&16.34&A3193-M1 \\       
NGC 1553       &S0   &$-22.06$&    177&$\backslash$&1.01&13.54& \\          
NGC 1600       &E    &$-23.02$&    337&$\cap$&2.82&18.17& \\                
NGC 1700       &E    &$-21.95$&    235&$\cap$&1.01&13.50& \\                
NGC 2300       &S0   &$-21.74$&    261&$\cap$&2.12&16.82& \\                
NGC 2434       &E    &$-21.33$&    188&$\backslash$&0.64&14.60& \\          
NGC 2549       &S0   &$-19.17$&    143&$\backslash$&0.51&13.98& \\          
NGC 2592       &E    &$-20.01$&    265&$\backslash$&0.82&13.76& \\          
NGC 2634       &E    &$-20.83$&    181&$\backslash$&0.93&14.57& \\          
NGC 2636       &E    &$-19.19$&     69&$\backslash$&0.87&15.22& \\          
NGC 2685       &S0+  &$-19.72$&     94&$\backslash$&0.84&14.16& \\          
NGC 2699       &E    &$-20.25$&    141&$\backslash$&0.84&14.14& \\          
NGC 2778       &E    &$-18.75$&    162&$\backslash$&0.67&13.97& \\          
NGC 2832       &BCG  &$-23.76$&    335&$\cap$&2.52&17.11&A0779-M1 \\        
NGC 2841       &Sb   &$-20.57$&    206&$\wedge$&1.09&14.54& \\              
NGC 2872       &E    &$-21.62$&    285&$\backslash$&1.06&13.65& \\          
NGC 2902       &S0   &$-20.59$&$\ldots$&$\wedge$&2.15&16.95& \\             
NGC 2907       &Sa   &$-21.23$&$\ldots$&$\backslash$&1.22&13.47& \\         
NGC 2950       &S0   &$-19.73$&    182&$\backslash$&0.58&12.99& \\          
NGC 2974       &E    &$-21.09$&    227&$\backslash$&0.64&13.77& \\          
NGC 2986       &E    &$-22.32$&    262&$\cap$&2.07&16.24& \\                
NGC 3056       &S0+  &$-18.98$&$\ldots$&$\backslash$&0.80&14.10& \\         
NGC 3065       &S0   &$-19.64$&    160&$\backslash$&0.86&13.93& \\          
NGC 3078       &E    &$-21.95$&    250&$\backslash$&0.95&13.23& \\          
NGC 3115       &S$0-$&$-21.11$&    252&$\backslash$&0.30&12.65& \\          
NGC 3193       &E    &$-21.98$&    194&$\cap$&1.38&14.70& \\                
NGC 3266       &S0   &$-20.11$&$\ldots$&$\backslash$&0.85&14.95& \\         
NGC 3348       &E    &$-22.18$&    238&$\cap$&1.96&16.05& \\                
NGC 3377       &E    &$-20.07$&    139&$\backslash$&0.36&12.24& \\          
NGC 3379       &E    &$-21.14$&    207&$\cap$&1.72&15.59& \\                
NGC 3384       &S$0-$&$-19.93$&    148&$\backslash$&0.36&13.03& \\          
NGC 3414       &S0   &$-20.25$&    237&$\backslash$&0.81&13.56& \\          
NGC 3551       &BCG  &$-23.55$&    268&$\cap$&2.37&17.35&A1177-M1 \\       
NGC 3585       &E    &$-21.93$&    207&$\wedge$&1.28&14.29& \\              
NGC 3595       &E    &$-20.96$&$\ldots$&$\backslash$&0.93&14.67& \\         
NGC 3599       &S0   &$-19.93$&     85&$\backslash$&0.65&14.64& \\          
NGC 3605       &E    &$-19.61$&     92&$\backslash$&0.65&14.96& \\          
NGC 3607       &S0   &$-19.88$&    224&$\cap$&1.77&16.26& \\                
NGC 3608       &E    &$-21.12$&    193&$\cap$&1.31&15.05& \\                
NGC 3610       &E    &$-20.96$&    162&$\backslash$&0.64&12.86& \\          
NGC 3613       &E    &$-21.59$&    210&$\cap$&1.65&15.65& \\                
NGC 3640       &E    &$-21.96$&    182&$\cap$&1.47&15.39& \\                
NGC 3706       &S$0-$&$-22.31$&    270&$\cap$&1.60&14.16& \\                
NGC 3842       &BCG  &$-23.18$&    314&$\cap$&2.48&17.40&A1367-M1 \\        
NGC 3900       &S0+  &$-20.80$&    118&$\backslash$&1.16&14.25& \\          
NGC 3945       &S0+  &$-20.25$&    174&$\backslash$&0.59&14.19& \\          
NGC 4026       &S0   &$-19.79$&    178&$\backslash$&0.48&12.96& \\          
NGC 4073       &E    &$-23.50$&    278&$\cap$&2.13&16.55& \\                
NGC 4121       &E    &$-18.53$&     86&$\backslash$&0.79&14.55& \\          
NGC 4128       &S0   &$-20.79$&    203&$\backslash$&0.92&13.62& \\          
NGC 4143       &S0   &$-19.68$&    214&$\backslash$&0.88&13.98& \\          
NGC 4150       &S0   &$-18.66$&     85&$\backslash$&0.85&13.87& \\          
NGC 4168       &E    &$-21.80$&    184&$\cap$&2.26&17.58& \\                
NGC 4239       &E    &$-18.50$&     62&$\wedge$&1.06&16.82& \\              
NGC 4261       &E    &$-22.26$&    309&$\cap$&2.31&16.09& \\                
NGC 4278       &E    &$-21.05$&    238&$\cap$&1.77&15.82& \\                
NGC 4291       &E    &$-20.64$&    285&$\cap$&1.63&15.29& \\                
NGC 4365       &E    &$-22.18$&    256&$\cap$&2.15&16.53& \\                
NGC 4374       &E    &$-22.28$&    282&$\cap$&2.11&15.67& \\                
NGC 4382       &S0+  &$-21.96$&    179&$\cap$&1.69&15.34& \\                
NGC 4387       &E    &$-19.25$&    104&$\backslash$&0.54&15.13& \\          
NGC 4406       &E    &$-22.46$&    235&$\cap$&1.90&16.03& \\                
NGC 4417       &S0   &$-18.94$&    131&$\backslash$&0.94&13.96& \\          
NGC 4434       &E    &$-19.19$&    120&$\backslash$&0.54&14.44& \\          
NGC 4458       &E    &$-19.27$&    103&$\cap$&0.80&13.57& \\                
NGC 4464       &E    &$-18.82$&    127&$\backslash$&0.54&13.92& \\          
NGC 4467       &E    &$-17.51$&     68&$\backslash$&0.54&15.07& \\          
NGC 4472       &E    &$-22.93$&    291&$\cap$&2.25&16.53& \\                
NGC 4473       &E    &$-21.16$&    179&$\cap$&1.73&15.40& \\                
NGC 4474       &S0   &$-18.42$&     87&$\backslash$&0.72&14.74& \\         
NGC 4478       &E    &$-19.89$&    138&$\cap$&1.32&15.50& \\                
NGC 4482       &E    &$-18.87$&     26&$\wedge$&2.05&19.52& \\             
NGC 4486       &E    &$-22.71$&    332&$\cap$&2.65&17.25& \\                
NGC 4486B      &cE   &$-17.98$&    170&$\cap$&1.08&14.44& \\                
NGC 4494       &E    &$-21.50$&    150&$\backslash$&0.54&13.40& \\          
NGC 4503       &S$0-$&$-19.57$&    111&$\backslash$&0.63&14.42& \\          
NGC 4551       &E    &$-19.37$&    108&$\backslash$&0.54&14.86& \\          
NGC 4552       &E    &$-21.65$&    253&$\cap$&1.60&15.17& \\                
NGC 4564       &E    &$-20.26$&    157&$\backslash$&0.63&13.43& \\          
NGC 4589       &E    &$-21.35$&    224&$\cap$&1.40&15.41& \\                
NGC 4621       &E    &$-21.74$&    225&$\backslash$&0.54&12.43& \\          
NGC 4636       &E    &$-21.86$&    203&$\cap$&2.21&16.76& \\                
NGC 4648       &E    &$-20.24$&    220&$\backslash$&0.83&13.34& \\          
NGC 4649       &E    &$-22.51$&    336&$\cap$&2.34&16.77& \\                
NGC 4660       &E    &$-20.13$&    188&$\backslash$&0.54&12.53& \\          
NGC 4696       &BCG  &$-24.33$&    254&$\cap$&2.44&17.77&A3526-M1 \\        
NGC 4697       &E    &$-21.49$&    174&$\backslash$&0.41&14.13& \\          
NGC 4709       &E    &$-22.32$&    242&$\cap$&2.02&16.91& \\                
NGC 4742       &E    &$-19.90$&    109&$\backslash$&0.51&12.43& \\          
NGC 4874       &E    &$-23.49$&    278&$\cap$&2.99&18.98& \\                
NGC 4889       &BCG  &$-23.73$&    401&$\cap$&2.84&17.80&A1656-M1 \\        
NGC 5017       &E    &$-20.67$&    184&$\backslash$&0.99&13.30& \\          
NGC 5061       &E    &$-22.01$&    186&$\cap$&1.39&14.06& \\                
NGC 5077       &E    &$-22.07$&    256&$\cap$&1.96&16.07& \\                
NGC 5198       &E    &$-21.23$&    196&$\cap$&1.33&15.19& \\                
NGC 5308       &S$0-$&$-21.26$&    211&$\backslash$&0.90&13.15& \\          
NGC 5370       &S0   &$-20.60$&    133&$\backslash$&1.04&15.34& \\          
NGC 5419       &E    &$-23.37$&    333&$\cap$&2.65&17.53& \\                
NGC 5557       &E    &$-22.62$&    254&$\cap$&1.82&15.58& \\                
NGC 5576       &E    &$-21.31$&    183&$\cap$&1.21&14.38& \\                
NGC 5796       &E    &$-21.98$&    288&$\wedge$&1.02&14.40& \\              
NGC 5812       &E    &$-21.39$&    200&$\backslash$&0.84&14.27& \\          
NGC 5813       &E    &$-22.01$&    239&$\cap$&1.89&16.32& \\                
NGC 5831       &E    &$-21.00$&    164&$\backslash$&0.85&14.41& \\          
NGC 5838       &S$0-$&$-20.51$&    266&$\backslash$&1.03&13.61& \\          
NGC 5845       &E    &$-19.98$&    234&$\backslash$&1.14&13.81& \\          
NGC 5898       &E    &$-21.65$&    218&$\wedge$&1.43&15.41& \\              
NGC 5903       &E    &$-21.90$&    198&$\cap$&2.17&17.07& \\                
NGC 5982       &E    &$-21.97$&    240&$\cap$&1.80&15.62& \\                
NGC 6086       &BCG  &$-23.11$&    336&$\cap$&2.53&17.26&A2162-M1 \\        
NGC 6166       &BCG  &$-23.80$&    310&$\cap$&3.17&19.32&A2199-M1 \\        
NGC 6173       &BCG  &$-23.59$&    278&$\cap$&2.32&16.72&A2197-M1 \\        
NGC 6278       &S0   &$-20.81$&    150&$\backslash$&0.99&13.97& \\          
NGC 6340       &S0   &$-19.46$&    144&$\backslash$&0.91&14.54& \\          
NGC 6849       &S0   &$-22.78$&    216&$\cap$&1.98&16.81& \\                
NGC 6876       &E    &$-23.58$&    234&$\cap$&2.17&17.02& \\                
NGC 7014       &BCG  &$-22.18$&    263&$\cap$&1.83&15.54& \\                
NGC 7052       &E    &$-22.35$&    271&$\cap$&2.29&16.19& \\                
NGC 7213       &Sa   &$-21.71$&    182&$\cap$&1.83&15.88& \\                
NGC 7332       &S0   &$-19.62$&    124&$\backslash$&0.67&12.78& \\          
NGC 7457       &S$0-$&$-18.62$&     69&$\backslash$&0.43&15.86& \\          
NGC 7578B      &BCG  &$-23.41$&    214&$\cap$&2.06&16.19&A2572-M1 \\        
NGC 7619       &E    &$-22.94$&    322&$\cap$&2.03&15.90& \\                
NGC 7626       &E    &$-22.87$&    276&$\wedge$&1.66&14.98& \\              
NGC 7647       &BCG  &$-23.97$&    282&$\cap$&2.28&17.14&A2589-M1 \\        
NGC 7727       &Sa   &$-21.19$&    196&$\wedge$&0.48&14.11& \\              
NGC 7743       &S0+  &$-20.18$&     84&$\backslash$&1.03&14.07& \\          
NGC 7785       &E    &$-22.08$&    245&$\cap$&1.32&15.28& \\                
IC 0115        &BCG  &$-22.67$&$\ldots$&$\cap$&2.45&17.24&A0195-M1 \\       
IC 0613        &BCG  &$-22.27$&    262&$\cap$&2.05&16.25&A1016-M1 \\       
IC 0664        &BCG  &$-22.86$&    336&$\cap$&2.07&15.81&A1142-M1 \\       
IC 0712        &BCG  &$-23.29$&    345&$\cap$&2.69&17.68&A1314-M1 \\       
IC 0875        &S0   &$-20.21$&$\ldots$&$\backslash$&1.01&13.45& \\         
IC 1459        &E    &$-22.51$&    306&$\cap$&1.94&15.39& \\                
IC 1565        &BCG  &$-22.99$&    303&$\cap$&1.65&16.86&A0076-M1 \\        
IC 1633        &BCG  &$-23.91$&    355&$\cap$&2.43&16.66&A2877-M1 \\       
IC 1695        &BCG  &$-23.90$&    364&$\cap$&2.36&16.68&A0193-M1 \\        
IC 1733        &BCG  &$-23.43$&    301&$\cap$&2.68&17.63&A0260-M1 \\        
IC 2738        &BCG  &$-22.18$&    275&$\backslash$&1.57&16.15&A1228-M1 \\   
IC 4329        &BCG  &$-23.95$&    275&$\cap$&2.34&16.26&A3574-M1 \\        
IC 4931        &BCG  &$-23.47$&    273&$\cap$&2.42&16.86&A3656-M1 \\       
IC 5353        &BCG  &$-22.64$&    262&$\cap$&2.04&16.37&A4038-M1 \\        
UGC 4551       &S0   &$-19.78$&    167&$\backslash$&0.82&15.00& \\          
UGC 4587       &S0   &$-20.77$&$\ldots$&$\backslash$&1.05&14.79& \\         
UGC 6062       &S0   &$-20.34$&$\ldots$&$\backslash$&1.01&14.66& \\         
VCC 1199       &E    &$-20.34$&$\ldots$&$\backslash$&0.54&15.42& \\         
VCC 1440       &E    &$-20.34$&$\ldots$&$\backslash$&0.54&15.62& \\         
VCC 1545       &E    &$-20.34$&$\ldots$&$\backslash$&0.54&17.49& \\         
VCC 1627       &E    &$-20.34$&$\ldots$&$\backslash$&0.54&15.68& \\         
ESO 378-20     &S0   &$-20.97$&$\ldots$&$\backslash$&1.06&13.11& \\         
ESO 443-39     &S0   &$-20.93$&$\ldots$&$\backslash$&1.06&14.97& \\         
ESO 447-30     &S0   &$-21.17$&$\ldots$&$\backslash$&1.00&13.99& \\         
ESO 462-15     &E    &$-22.83$&    305&$\backslash$&1.19&14.75& \\          
ESO 507-27     &S0   &$-20.89$&    203&$\backslash$&1.08&14.25& \\         
ESO 507-45     &S0   &$-23.28$&    311&$\wedge$&1.81&14.95& \\              
MCG 11-14-25A  &E    &$-19.08$&$\ldots$&$\cap$&1.38&15.77& \\               
MCG 08-27-18   &S0   &$-20.03$&     89&$\backslash$&1.07&15.09& \\         
A0119-M1       &BCG  &$-24.01$&    294&$\cap$&2.81&18.52& \\                
A0168-M1       &BCG  &$-23.12$&    345&$\cap$&2.00&16.90& \\                
A0189-M1       &BCG  &$-21.89$&    230&$\backslash$&1.51&17.56& \\          
A0261-M1       &BCG  &$-22.95$&$\ldots$&$\backslash$&1.64&18.22& \\         
A0295-M1       &BCG  &$-23.11$&$\ldots$&$\cap$&2.63&17.81& \\               
A0376-M1       &BCG  &$-23.60$&    276&$\cap$&2.64&18.10& \\                
A0397-M1       &BCG  &$-23.42$&    289&$\cap$&2.70&17.80& \\                
A0419-M1       &BCG  &$-21.79$&$\ldots$&$\backslash$&1.58&17.08& \\         
A0496-M1       &BCG  &$-24.28$&    273&$\cap$&2.61&18.14& \\                
A0533-M1       &BCG  &$-22.68$&$\ldots$&$\cap$&2.28&17.08& \\               
A0548-M1       &BCG  &$-22.75$&    220&$\cap$&2.22&17.12& \\                
A0634-M1       &BCG  &$-22.70$&    245&$\cap$&2.17&17.16& \\                
A0912-M1       &BCG  &$-22.24$&$\ldots$&$\cap$&1.63&16.36& \\               
A0999-M1       &BCG  &$-22.45$&    272&$\cap$&2.29&17.03& \\               
A1020-M1       &BCG  &$-22.65$&    345&$\cap$&2.32&16.87& \\                
A1631-M1       &BCG  &$-23.34$&    249&$\cap$&2.12&16.49& \\                
A1831-M1       &BCG  &$-23.51$&$\ldots$&$\cap$&2.84&18.58& \\               
A1983-M1       &BCG  &$-22.35$&    270&$\wedge$&1.73&15.50& \\                
A2040-M1       &BCG  &$-23.46$&    223&$\cap$&2.28&17.41& \\                
A2052-M1       &BCG  &$-23.04$&    216&$\cap$&2.46&18.53& \\                
A2147-M1       &BCG  &$-23.16$&    278&$\cap$&2.90&19.03& \\                
A2247-M1       &BCG  &$-22.66$&    209&$\backslash$&1.60&20.06& \\         
A3144-M1       &BCG  &$-22.28$&$\ldots$&$\cap$&2.28&16.79& \\               
A3376-M1       &BCG  &$-23.29$&    300&$\cap$&3.11&18.93& \\                
A3395-M1       &BCG  &$-24.23$&    276&$\cap$&2.52&18.09& \\               
A3528-M1       &BCG  &$-24.30$&    434&$\cap$&2.61&17.54& \\               
A3532-M1       &BCG  &$-24.58$&$\ldots$&$\cap$&2.51&17.49& \\               
A3554-M1       &BCG  &$-23.99$&$\ldots$&$\cap$&2.62&18.36& \\               
A3556-M1       &BCG  &$-23.65$&$\ldots$&$\cap$&2.48&16.95& \\               
A3558-M1       &BCG  &$-24.92$&    275&$\cap$&3.12&19.29& \\               
A3562-M1       &BCG  &$-24.32$&    236&$\cap$&2.84&18.87& \\                
A3564-M1       &BCG  &$-22.68$&$\ldots$&$\cap$&2.12&16.63& \\               
A3570-M1       &BCG  &$-22.54$&    268&$\cap$&2.01&16.19& \\               
A3571-M1       &BCG  &$-25.30$&    325&$\cap$&3.03&19.39& \\               
A3677-M1       &BCG  &$-22.21$&$\ldots$&$\cap$&2.14&16.49& \\               
A3716-M1       &BCG  &$-23.75$&    247&$\cap$&2.56&17.99& \\               
A3736-M1       &BCG  &$-23.98$&$\ldots$&$\cap$&2.70&17.93& \\               
A3747-M1       &BCG  &$-22.65$&    232&$\cap$&2.00&15.97& \\                
\enddata
\label{tab:glob}
\tablecomments{Derivation of the parameters listed are presented
in \citet{l06}.  Bulge luminosities are given for S0 and spiral galaxies.
Velocity dispersion are provided by the ``Hyperleda'' database.
The profile type, P, is $\backslash=$ power-law,
$\wedge=$ intermediate form, and $\cap=$ core.  BCG identifications are
referred to their hosting Abell clusters; see \citet{pl} for details.}
\end{deluxetable}

\begin{deluxetable}{lccccl}
\tablecolumns{6}
\tablewidth{0pt}
\tablecaption{Core Galaxies with Measured Black Hole Masses}
\tablehead{
\colhead{Galaxy}&\colhead{$M_V$}&\colhead{$\log(r_\gamma/\rm{pc})$}
&\colhead{$I_\gamma$ (V)}&\colhead{$\log(M_\bullet/M_\odot)$}
&\colhead{$M_\bullet$ Source}}
\startdata
NGC 1399   &     $-22.07$&  2.23&  16.76& 8.95 & 1, 2\\
NGC 3379   &     $-21.14$&  1.72&  15.59& 8.00 & 3\\
NGC 3608   &     $-21.12$&  1.31&  15.05& 8.28 & 4\\
NGC 4261   &     $-22.26$&  2.31&  16.09& 8.72 & 5\\
NGC 4291   &     $-20.64$&  1.63&  15.29& 8.49 & 4\\
NGC 4374   &     $-22.28$&  2.11&  15.67& 9.00 & 6, 7\\
NGC 4473   &     $-21.16$&  1.73&  15.40& 8.04 & 4\\
NGC 4486   &     $-22.71$&  2.65&  17.25& 9.48 & 8 \\
NGC 4649   &     $-22.51$&  2.34&  16.77& 9.30 & 4\\
NGC 7052   &     $-22.35$&  2.29&  16.19& 8.52 & 9\\
IC  1459   &     $-22.51$&  1.94&  15.39& 9.18 & 10, 11\\
\enddata
\label{tab:bhrc}
\tablecomments{Black hole mass references are 1) \citet{n1399}, 2)
\citet{g06}, 3) \citet{gbh}, 4) \citet{g03}, 5) \citet{n4261},
6) \citet{n4374}, 7) \citet{mac}, 8) \citet{m87bh}, 9) \citet{n7052},
10) \citet{ver}, and 11) \citet{cap}.
For galaxies with two references, the black hole mass is
an average value.}
\end{deluxetable}
\clearpage

\clearpage

\begin{thebibliography}{}

\bibitem[Adelberger \& Steidel (2005)]{adel05} Adelberger, K. L \& 
Steidel, C. C. 2005 \apj, 630, 50
\bibitem[Batcheldor et al.(2006)]{bat} Batcheldor, D., 
Marconi, A., Merritt, D., \& Axon, D.~J.\ 2006, ArXiv Astrophysics 
e-prints, arXiv:astro-ph/0610264
\bibitem[Bechtold et al.(2003)]{bechtold} Bechtold, J., et al.\ 
2003, \apj, 588, 119
\bibitem[Begelman(2006)]{begel06} Begelman, M.~C.\ 2006, \apj, 643, 1065 
\bibitem[Begelman et al.(1980)]{bbr} Begelman, M. C., Blandford, R. D., \&
Rees, M. J. 1980, \nat, 287, 307
\bibitem[Bender et al.(1992)]{bend92} Bender, R., Burstein, 
D., \& Faber, S.~M.\ 1992, \apj, 399, 462
\bibitem[Bernardi et al.(2006b)]{b06b} Bernardi, M., Hyde, 
J.~B., Sheth, R.~K., Miller, C.~J., \& Nichol, R.~C.\ 2006b, ArXiv 
Astrophysics e-prints, arXiv:astro-ph/0607117
\bibitem[Bernardi et al.(2006c)]{b06c} Bernardi, M., Sheth, 
R.~K., Tundo, E., \& Hyde, J.~B.\ 2006c, ArXiv Astrophysics e-prints, 
arXiv:astro-ph/0609300
\bibitem[Bernardi et al.(2003)]{bern2} Bernardi, M., et al.\ 
2003, \aj, 125, 1849
\bibitem[Bernardi et al.(2006a)]{bern3} Bernardi, M., et al.\ 
2006a, \aj, 131, 2018
\bibitem[Binney \& Tabor(1995)]{bt} Binney, J., \& Tabor, 
G.\ 1995, \mnras, 276, 663
\bibitem[Birnboim \& Dekel(2003)]{birn} Birnboim, Y., \& 
Dekel, A.\ 2003, \mnras, 345, 349
\bibitem[Blanton et al.(2003)]{blanton} Blanton, M.~R., et al.\ 
2003, \apj, 592, 819
\bibitem[Blumenthal et al.(1984)]{cdm} Blumenthal, G.~R., 
Faber, S.~M., Primack, J.~R., \& Rees, M.~J.\ 1984, \nat, 311, 517
\bibitem[Bower et al.(2000)]{n3031} Bower, G.~A., Wilson, 
A.~S., Heckman, T.~M., Magorrian, J., Gebhardt, K., Richstone, D.~O., 
Peterson, B.~M., \& Green, R.~F.\ 2000, \baas, 32, 1566
\bibitem[Bower et al.(1998)]{n4374} Bower, G.~A., et al.\ 1998, \apjl, 492, L111
\bibitem[Boylan-Kolchin et al.(2006)]{bmq} Boylan-Kolchin, 
M., Ma, C.-P., \& Quataert, E.\ 2006, \mnras, 369, 1081
\bibitem[Cappellari et al.(2002)]{cap} Cappellari, M., 
Verolme, E.~K., van der Marel, R.~P., Kleijn, G.~A.~V., Illingworth, G.~D., 
Franx, M., Carollo, C.~M., \& de Zeeuw, P.~T.\ 2002, \apj, 578, 787 
\bibitem[Carollo et al.(1997)]{carollo} Carollo, C. M., Franx, M.,
Illingworth, G. D., \& Forbes, D. 1997, \apj, 481, 710
\bibitem[Churazov et al.(2002)]{chur} Churazov, E., Sunyaev, 
R., Forman, W., B\" ohringer, H.\ 2002, \mnras, 332, 729 
\bibitem[de Vaucouleurs et al.(1991)]{rc3} de Vaucouleurs,
G., de Vaucouleurs, A., Corwin, H.~G., Buta, R.~J., Paturel, G., \& Fouque,
P.\ 1991, Volume 1-3, XII, 2069 pp.~7 figs..~ Springer-Verlag Berlin
\bibitem[Djorgovski \& Davis(1987)]{dd} Djorgovski, S., \& 
Davis, M.\ 1987, \apj, 313, 59
\bibitem[Dressler(1989)]{d89} Dressler, A.\ 1989, IAU 
Symp.~134: Active Galactic Nuclei, 134, 217
\bibitem[Dressler et al.(1987)]{d87} Dressler, A., 
Lynden-Bell, D., Burstein, D., Davies, R.~L., Faber, S.~M., Terlevich, R., 
\& Wegner, G.\ 1987, \apj, 313, 42
\bibitem[Ebisuzaki et al.(1991)]{ebi} Ebisuzaki, T., Makino, J., \&
Okumura, S. K. 1991, \nat, 354, 212
\bibitem[Faber \& Jackson(1976)]{fj} Faber, S.~M., \&
Jackson, R.~E.\ 1976, \apj, 204, 668
\bibitem[Faber et al.(1997)]{f97} Faber, S. M., Tremaine, S., Ajhar,
E. A., Byun, Y., Dressler, A., Gebhardt, K., Grillmair, C., Kormendy, J.,
Lauer, T. R., \& Richstone, D.  1997, \aj, 114, 1771
\bibitem[Faber et al.(1989)]{f89} Faber, S. M., Wegner, G., Burstein, D.,
Davies, R. L., Dressler, A., Lynden-Bell, D., \& Terlevich, R. J.,
1989, \apjs, 69, 763
\bibitem[Fabian et al.(2002)]{fab} Fabian, A.~C., Voigt, 
L.~M., \& Morris, R.~G.\ 2002, \mnras, 335, L71
\bibitem[Ferrarese et al.(1996)]{n4261} Ferrarese, L., Ford, 
H.~C., \& Jaffe, W.\ 1996, \apj, 470, 444
\bibitem[Ferrarese \& Merritt(2000)]{fm} Ferrarese, L., \& Merritt, D.\
2000, \apj, 539, L9
\bibitem[Ferrarese et al.(2006)]{f06} Ferrarese, L., et 
al.\ 2006, \apjs, 164, 334
\bibitem[Fukugita et al.(1995)]{fuku} Fukugita, M., 
Shimasaku, K., \& Ichikawa, T.\ 1995, \pasp, 107, 945
\bibitem[Gebhardt et al.(2000a)]{g00}Gebhardt, K., et al.\ 2000a, \apj, 539, L13
\bibitem[Gebhardt et al.(2000b)]{gbh} Gebhardt, K., et al.\ 
2000b, \aj, 119, 1157
\bibitem[Gebhardt et al.(2003)]{g03}Gebhardt, K., et al.\ 2003, \apj, 583, 92 
\bibitem[Gebhardt et al.(2007)]{g06}Gebhardt, K., et al.\ 2007, in preparation
\bibitem[Gonzalez et al.(2005)]{gzz} Gonzalez, A.~H., 
Zabludoff, A.~I., \& Zaritsky, D.\ 2005, \apj, 618, 195
\bibitem[Graham et al.(1996)]{graham} Graham, A., Lauer, 
T.~R., Colless, M., \& Postman, M.\ 1996, \apj, 465, 534
\bibitem[Greenhill et al.(1997)]{n4945} Greenhill, L.~J., 
Moran, J.~M., \& Herrnstein, J.~R.\ 1997, \apjl, 481, L23
\bibitem[H\"aring \& Rix(2004)]{hr}H\"aring, N. \& Rix, H. 2004, \apj, 604, L89
\bibitem[Harms et al.(1994)]{m87} Harms, R.~J., et al.\ 1994, \apjl, 435, L35
\bibitem[Hogg et al.(2004)]{hogg} Hogg, D.~W., et al.\ 2004, \apjl, 601, L29
\bibitem[Hopkins et al.(2006)]{hop06}Hopkins, P. F., Hernquist, L.,
Cox, T. J., Di Matteo, T., Robertson, B., \& Springel, V. 2006, \apjs, 163, 1   
\bibitem[Houghton et al.(2006)]{n1399} Hougton, R. C. W., Magorrian, J.,
Sarzi, M., Thatte, N., Davies, R. L., Krajnovic, D. 2006, \mnras, in press
\bibitem[Jarrett et al.(2000)]{xsc} Jarrett, T.~H., 
Chester, T., Cutri, R., Schneider, S., Skrutskie, M., \& Huchra, J.~P.\ 
2000, \aj, 119, 2498
\bibitem[Jarrett et al.(2003)]{lga} Jarrett, T.~H., 
Chester, T., Cutri, R., Schneider, S.~E., \& Huchra, J.~P.\ 2003, \aj, 125, 
525
\bibitem[Kormendy(1985)]{k85} Kormendy, J.\ 1985, \apj, 295, 73
\bibitem[Kormendy(1993)]{k93} Kormendy, J.\ 1993, Coleccion 
Nuevas Tendencias (The nearest active galaxies), Proceedings of the meeting 
on The nearest active galaxies, held in Madrid in May 1992, Madrid: Consejo 
Superior de Investigaciones Cientificas, |c1993, Edited by J.~Beckman, 
L.~Colina and H.~Netzer, p.~197-218., 197
\bibitem[Kormendy et al.(2007)]{k06} Kormendy, J., Fisher, D. B., Cornell,
M. E., \& Bender, R.\ 2007, \apj, submitted
\bibitem[Kormendy \& Richstone(1995)]{kr} Kormendy, J., \& 
Richstone, D.\ 1995, \araa, 33, 581
\bibitem[Kormendy et al.(1997)]{n4486b} Kormendy, J., et al.\ 
1997, \apjl, 482, L139
\bibitem[Laine et al.(2002)]{laine}Laine, S., van der Marel, R. P.,
Lauer, T. R., Postman, M., O'Dea, C. P., \& Owen, F. N. 2003, \aj, 125, 478
\bibitem[Lauer(1985)]{l85} Lauer, T.~R.\ 1985, \apj, 292, 104
\bibitem[Lauer et al.(1995)]{l95}Lauer, T.~R., et al.\ 1995, \aj, 110, 2622
\bibitem[Lauer et al.(2005)]{l05}Lauer, T.~R., et al.\ 2005, \aj, 129, 2138
\bibitem[Lauer et al.(2007a)]{l06}Lauer, T.~R., et al.\ 2007a, \apj, submitted
\bibitem[Lauer et al.(2007b)]{l07b}Lauer, T.~R., et al.\ 2007b, in preparation
\bibitem[Macchetto et al.(1997)]{m87bh} Macchetto, F., 
Marconi, A., Axon, D.~J., Capetti, A., Sparks, W., \& Crane, P.\ 1997, 
\apj, 489, 579
\bibitem[Maciejewski \& Binney(2001)]{mac} Maciejewski, W., 
\& Binney, J.\ 2001, \mnras, 323, 831
\bibitem[Magorrian et al.(1998)]{mag} Magorrian, J., et al.\ 1998, \aj,
115, 2285
\bibitem[Marconi et al.(2001)]{n5128} Marconi, A., Capetti, 
A., Axon, D.~J., Koekemoer, A., Macchetto, D., \& Schreier, E.~J.\ 2001, 
\apj, 549, 915 
\bibitem[Marconi et al.(2004)]{marc} Marconi, A., Risaliti, 
G., Gilli, R., Hunt, L.~K., Maiolino, R., \& Salvati, M.\ 2004, \mnras, 
351, 169
\bibitem[McLure et al.(2004)]{mclure} McLure, R.~J., Willott, 
C.~J., Jarvis, M.~J., Rawlings, S., Hill, G.~J., Mitchell, E., Dunlop, 
J.~S., \& Wold, M.\ 2004, \mnras, 351, 347
\bibitem[Merritt(2006)]{m06} Merritt, D.\ 2006, \apj, 648, 
976
\bibitem[Miller et al.(2005)]{miller} Miller, C.~J., et al.\ 
2005, \aj, 130, 968
\bibitem[Milosavljevi\'c \& Merritt(2001)]{mnm}
Milosavljevi\'c, M., \& Merritt, D. 2001, \apj, 563, 34
\bibitem[Naab et al.(2006)]{naab} Naab, T., Khochfar, S., \& 
Burkert, A.\ 2006, \apjl, 636, L81
\bibitem[Nelson et al.(2000)]{n7332} Nelson, C.~H., Weistrop, 
D., Bower, G.~A., Green, R.~F., \& STIS Team 2000, \baas, 32, 701
\bibitem[Netzer(2003)]{net} Netzer, H.\ 2003, \apjl, 583, L5
\bibitem[Novak et al.(2006)]{novak} Novak, G.~S., Faber, 
S.~M., \& Dekel, A.\ 2006, \apj, 637, 96
\bibitem[Oegerle \& Hoessel(1991)]{ho} Oegerle, W.~R., \&
Hoessel, J.~G.\ 1991, \apj, 375, 15
\bibitem[Patel et al.(2006)]{patel} Patel, P., Maddox, S., 
Pearce, F.~R., Arag{\'o}n-Salamanca, A., \& Conway, E.\ 2006, \mnras, 370, 851
\bibitem[Postman \& Lauer(1995)]{pl}Postman, M., \& Lauer, T. R. 1995,
\apj, 440, 28
\bibitem[Press et al.(1992)]{press}Press, W.~H., Teukolsky, S.~A.,
Vetterling, W.~T., \& Flannery, B.~P.\ 1992, Numerical Recipes
(2d ed.; Cambridge: Cambridge Univ. Press)
\bibitem[Prugniel \& Simien(1996)]{ps} Prugniel, P.,~\&
Simien, F.\ 1996, \aap, 309, 749
\bibitem[Quillen et al.(2000)]{quil}Quillen, A. C., Bower, G. A.,
\& Stritzinger, M. 2000, \apjs, 128, 85
\bibitem[Quinlan(1996)]{q06} Quinlan, G.~D.\ 1996, New 
Astronomy, 1, 35
\bibitem[Quinlan \& Hernquist(1997)]{qh} Quinlan, G.~D., 
\& Hernquist, L.\ 1997, New Astronomy, 2, 533
\bibitem[Ravindranath et al.(2001)]{rav} Ravindranath, S., Ho, L. C.,
Peng, C. Y., Filippenko, A. V., \& Sargent, W. L.  W. 2001, \aj, 122, 653
\bibitem[Rest et al.(2001)]{rest} Rest, A., van den Bosch, F. C., Jaffe, W.,
Tran, H., Tsvetanov, Z., Ford, H. C., Davies, J., \& Schafer, J. 2001,
\aj, 121, 2431
\bibitem[Richards et al.(2005)]{rich05} Richards, G.~T., et 
al.\ 2005, \mnras, 360, 839
\bibitem[Richstone et al.(1998)]{rinuc98} Richstone, D., et 
al.\ 1998, \nat, 395, A14 
\bibitem[Robertson et al.(2006)]{rob06} Robertson, B., 
Hernquist, L., Cox, T.~J., Di Matteo, T., Hopkins, P.~F., Martini, P., \& 
Springel, V.\ 2006, \apj, 641, 90 
\bibitem[Salpeter(1964)]{sal} Salpeter, E.~E.\ 1964, \apj, 140, 796
\bibitem[Schechter(1976)]{schechter} Schechter, P.\ 1976, \apj, 203, 297
\bibitem[Schlegel et al.(1998)]{sfd} Schlegel, D.~J., 
Finkbeiner, D.~P., \& Davis, M.\ 1998, \apj, 500, 525
\bibitem[S\' ersic(1968)]{sersic} S\' ersic, J.~L.\ 1968, Cordoba,
Argentina: Observatorio Astronomico, 1968
\bibitem[Sheth et al.(2003)]{sheth} Sheth, R.~K., et al.\ 
2003, \apj, 594, 225
\bibitem[Small \& Blandford(1992)]{smbl} Small, T.~A., \& 
Blandford, R.~D.\ 1992, \mnras, 259, 725 
\bibitem[So\l tan(1982)]{soltan} So\l tan, A.\ 1982, \mnras, 200, 115 
\bibitem[Springel et al.(2005)]{spr05} Springel, V., Di 
Matteo, T., \& Hernquist, L.\ 2005, \mnras, 361, 776 
\bibitem[Steidel et al.(2002)]{ste02}Steidel, C. . et al. 
2002, \apj 576, 653  
\bibitem[Tadhunter et al.(2003)]{cyga} Tadhunter, C., 
Marconi, A., Axon, D., Wills, K., Robinson, T.~G., \& Jackson, N.\ 2003, 
\mnras, 342, 861
\bibitem[Tonry et al.(2001)]{ton}Tonry, J. L., Dressler, A., Blakeslee, J. P.,
Ajhar, E. A., Fletcher, A. B., Luppino, G. A., Metzger, M. R., \& Moore, C. B.
2001, \apj, 546, 681
\bibitem[Tremaine et al.(2002)]{tr02}Tremaine, S. et al. 2002, \apj, 574, 740
\bibitem[Tundo et al.(2006)]{tundo} Tundo, E., Bernardi, M., 
Hyde, J.~B., Sheth, R.~K., \& Pizzella, A.\ 2006, ArXiv Astrophysics 
e-prints, arXiv:astro-ph/0609297 
\bibitem[van der Marel \& van den Bosch(1998)]{n7052} van der 
Marel, R.~P., \& van den Bosch, F.~C.\ 1998, \aj, 116, 2220
\bibitem[Verdoes Kleijn et al.(2000)]{ver} Verdoes Kleijn, 
G.~A., van der Marel, R.~P., Carollo, C.~M., \& de Zeeuw, P.~T.\ 2000, \aj, 
120, 1221 
\bibitem[Vestergaard(2004)]{vester1} Vestergaard, M.\ 2004, \apj, 601, 676
\bibitem[Voit \& Donahue(2005)]{vd} Voit, G.~M., \& 
Donahue, M.\ 2005, \apj, 634, 955
\bibitem[Wyithe(2006)]{wyithe} Wyithe, J.~S.~B.\ 2006, \mnras, 365, 1082
\bibitem[Wyithe \& Loeb(2005)]{wl} Wyithe, J.~S.~B., \& 
Loeb, A.\ 2005, \apj, 634, 910
\bibitem[Yu \& Lu(2004)]{yulu} Yu, Q., \& Lu, Y.\ 2004, 
\apj, 610, 93
\bibitem[Yu \& Tremaine(2002)]{yu02} Yu, Q., \& Tremaine, 
S.\ 2002, \mnras, 335, 965
\bibitem[Zibetti et al.(2005)]{zib} Zibetti, S., White, 
S.~D.~M., Schneider, D.~P., \& Brinkmann, J.\ 2005, \mnras, 358, 949
\end{thebibliography}
\end{document}